\documentclass[fleqn,usenatbib]{mnras}
\usepackage[T1]{fontenc}
\usepackage{ae,aecompl}
\usepackage{hyperref}
\usepackage{graphicx}	
\usepackage{amsmath}	
\usepackage{amssymb}	
\usepackage{color}
\usepackage{mathtools}
\usepackage[thinc]{esdiff}
\usepackage{bm}
\usepackage{newtxtext,newtxmath}
\usepackage{ulem}
\usepackage{xparse}
\usepackage{verbatim}
\usepackage{xcolor}
\definecolor{bostonuniversityred}{rgb}{0.8, 0.0, 0.0}
\definecolor{agsgreen}{rgb}{0, 0.7, 0.0}

\let\oldhat\hat
\renewcommand{\vec}[1]{\mathbf{#1}}
\renewcommand{\hat}[1]{\oldhat{\mathbf{#1}}}

\let\vec\mathbf

\title[Redshift-space distortions with split densities]{Redshift-space distortions with split densities}

 \author[E. Paillas et al.]
 {\parbox{\textwidth}{
     Enrique Paillas$^{1,2}$\thanks{E-mail : epaillas@astro.puc.cl},
     Yan-Chuan Cai$^{3}$,
     Nelson Padilla$^{1,2}$
     and Ariel~G. S\'anchez$^{4}$
     }
 \vspace{.2cm}\\
 $^{1}$Instituto de Astrof\'isica, Pontificia Universidad Cat\'olica de Chile, Av. Vicu\~na Mackenna 4860, Santiago, Chile \\
 $^{2}$Centro de Astro-Ingenier\'ia, Pontificia Universidad Cat\'olica de Chile, Av. Vicu\~na Mackenna 4860, Santiago, Chile \\
 $^{3}$Institute for Astronomy, University of Edinburgh, Royal Observatory, Edinburgh EH9 3HJ, UK \\
 $^{4}$Max-Planck-Institut f\"ur Extraterrestrische Physik, Postfach 1312, Giessenbachstr., D-85741 Garching, Germany
 }

\date{Accepted XXX. Received YYY; in original form ZZZ}

\pubyear{2020}

\begin{document}
\label{firstpage}
\pagerange{\pageref{firstpage}--\pageref{lastpage}}
\maketitle

\begin{abstract}
Accurate modelling of redshift-space distortions (RSD) is challenging in the non-linear regime for two-point statistics e.g. the two-point correlation function (2PCF). We take a different perspective to split the galaxy density field according to the local density, and cross-correlate those densities with the entire galaxy field. Using mock galaxies, we demonstrate that combining a series of cross-correlation functions (CCFs) offers improvements over the 2PCF as follows: 1. The distribution of peculiar velocities in each split density is nearly Gaussian. This allows the Gaussian streaming model for RSD to perform accurately within the statistical errors of a ($1.5\,h^{-1}$Gpc)$^3$ volume for almost all scales and all split densities. 2. The PDF of the density field at small scales is non-Gaussian, but the CCFs of split densities capture the non-Gaussianity, leading to improved cosmological constraints over the 2PCF. We can obtain unbiased constraints on the growth parameter $f\sigma_{12}$ at the per-cent level, and Alcock-Paczynski (AP) parameters at the sub-per-cent level with the minimal scale of $15\,h^{-1}{\rm Mpc}$. This is a $\sim$30 per cent and $\sim$6 times improvement over the 2PCF, respectively. The diverse and steep slopes of the CCFs at small scales are likely to be responsible for the improved constraints of AP parameters. 3. Baryon acoustic oscillations (BAO) are contained in all CCFs of split densities. Including BAO scales helps to break the degeneracy between the line-of-sight and transverse AP parameters, allowing independent constraints on them. We discuss and compare models for RSD around spherical densities.
\end{abstract}

\begin{keywords}
cosmological parameters, large-scale structure of Universe
\end{keywords}

\section{Introduction}

In the modern era of cosmology, galaxy redshift surveys have granted us the information to achieve percent-level constraints on the parameters of our cosmological paradigm, the $\Lambda$ cold dark matter ($\Lambda$CDM) model. This has allowed us to better understand how the Universe evolved from a very homogeneous initial state, to the highly complex and rich structures we observe nowadays. Under this model, the present-day \textit{cosmic web} assembled in a hierarchical structure formation scenario, where small-scale quantum fluctuations from the inflationary epoch grew over time due to gravitational instabilities. The rate of growth of these overdensities is always at contest with the expansion of the Universe \citep{Linder2019}. As such, precise measurements of the growth rate of structure constitute not only valuable tests for $\Lambda$CDM, but also for alternative cosmological models, such as modified gravity theories, which can predict rather different growth histories under the same expansion history \citep{Copeland2006, Linder2007, Jennings2011, Joyce2014, Koyama2016}.

A powerful method for constraining the growth rate of cosmic structure is through the analysis of redshift-space distortions \citep[RSD,][]{Jackson1972, Kaiser1987}. This effect arises when converting redshifts to distances, where the cosmological redshifts are perturbed by peculiar velocities. This produces distortions in the redshift-space clustering pattern of galaxies. Since peculiar velocities are induced by gravity, an accurate modelling of this effect grants us the possibility to estimate the rate at which structure is assembling \citep{Kaiser1987}.

Success has been achieved by using RSD to measure the growth of structure and testing theories of gravity with galaxies from large surveys such as BOSS \citep{Alam2017}, eBOSS \citep{eboss2020}, VIPERS \citep{Pezzotta2017}, GAMA \citep{Blake2013}, the 6 degree Field Galaxy Survey 6dFGS \citep{Beutler2012}, the Subaru FMOS galaxy redshift survey \citep{Okumura2016} and the WiggleZ Dark Energy Survey \citep{Blake2011}, achieving 5-10 percent level of accuracy down to the scale of $\sim 20\,h^{-1}{\rm Mpc}$. The next generation of spectroscopic redshift surveys such 
the Dark Energy Spectroscopic Instrument \citep[DESI,][]{Levi2019} and \textit{Euclid} \citep{euclid} promise percent-level of accuracy for parameter constraints, demanding even higher precision for the modelling.

To best extract cosmological information from these surveys, great efforts have been made to improve the modelling of RSD in terms of measurements of the two-point correlation function \citep[2PCF, e.g.][]{Sheth2001a, Reid2011, Bianchi2016} or the power spectrum \citep[e.g.][]{Scoccimarro2004, Seljak2011, Chen2020}. These two-point statistics characterising the variance of the density field are able to capture all information if the field is Gaussian, which is the case in the linear regime at early times. In the non-linear regime, however, the density field becomes non-Gaussian, so the variance of the field becomes an incomplete statistic. In a non-Gaussian field, the 2PCF will not be able to extract all the information. This is a major limitation for two-point statistics.

Beyond perturbation theory, the streaming model \citep{Peebles1980, Fisher1995} has been widely used for modelling RSD in configuration space. It provides a mapping between {the} real and redshift-space correlation functions. A crucial ingredient for the model to work accurately is the full distribution function of the pairwise velocities, which is highly non-Gaussian at the quasi-linear and non-linear scales, making it challenging to model \citep[e.g.][]{Scoccimarro2004}. Extensive effort has been made to model the pairwise velocity distribution \citep{Kuruvilla2018, Cuesta-Lazaro2020}. \citet{Reid2011} have shown that, under the assumption of a Gaussian pairwise velocity distribution, 
taking the true real-space 2PCF, and the mean streaming velocity and velocity dispersion profiles from simulations, the predictions of the streaming model agree with direct redshift-space measurements at a few percent level at $s \sim 25\,h^{-1}{\rm Mpc}$, and fail at smaller scales. \cite{Kuruvilla2018} have shown that dropping the Gaussian assumption, with the full distribution function, the model works perfectly for the two-dimensional redshift-space correlation function. 
Alternatively, \cite{Tinker2007} showed that the skewed PDF of the pairwise  velocity arises when combining halo pairs from environments with different densities \citep[see also a more recent related study by][]{Shirasaki2021}. By splitting a sample of haloes into different quantiles according to their local number density, they verified that the pairwise velocities at fixed density are approximately Gaussian, and this helps to improve the accuracy of 
the streaming model. Following this spirit, we will generalise the idea to include underdense environments, and instead of the 2PCF, we will focus on the cross-correlation function (CCF) between environments of different densities with the entire galaxy field. We will show that a Gaussian distribution remains a good approximation for the PDF of velocities in underdense regions, and hence it works for all density environments.

Alternative approaches for overcoming the limitation of two-point measurements include higher-order statistics, such as the three-point correlation function and the bispectrum \citep[e.g.][]{Sefusatti2006, Gil-Marin2012, Slepian2015, Slepian2017} [and indeed, the streaming model has recently been generalised to three-point correlation function \citep{Kuruvilla2020}], non-linear transformation to re-Gaussianise the density field (\citealt{Neyrinck2009, Neyrinck2011a, Neyrinck2011b, Wang2011}), density split statistics for weak lensing analysis \citep{Gruen2015,Gruen2018, Friedrich2018}, counts-in-cells statistics \citep{Szapudi2004, Klypin2018,Jamieson2020, Mandal2020}, and using the concept of separate Universe to model density-dependent two-point statistics (\citealt{Wagner2015, Chiang2015}), which, in essence, corresponds to a three-point quantity. These approaches usually provide complementary cosmological constraints by accessing information from the non-Gaussian field.

More recently, there has been increasing attention in using RSD around cosmic voids, taking the advantage of the milder density contrasts around them, which may be better described by linear dynamics \citep{Hamaus2014,Pollina2017}. \cite{Paz2013} employed the Gaussian streaming model for extracting velocity profiles around voids; \cite{Hamaus2015} applied the Gaussian streaming model for voids to constrain growth parameters, with the assumption that the density and streaming velocity are linearly coupled. \cite{Cai2016} wrote down how the redshift-space correlation function is mapped to its real-space version when only streaming velocities are accounted for. When expanding it to the linear order, it works well for small density perturbations, but by definition does not work when $\delta$ is relatively large, i.e. near void centres. \cite{Nadathur2019a} derived expressions for RSD around voids and convolved them with a Gaussian velocity distribution, showing that it helps to improve the performance of the model, and stressed that the Gaussian streaming model provides a poor fit to their data. These models, though somewhat disputed at the details, have been applied to observational data, and in some cases, have led to significant improvement for the constraints of AP and growth parameters \citep[e.g.][]{Hamaus2017, Achitouv2019, Hawken2020, Correa2020}. Another goal of our study is to clarify some of these disputes in the model. We will present some model comparisons in Sec.~\ref{sec:model_comparison}. Nevertheless, like other beyond-two-point statistics, RSD around voids are an interesting development, but there is no obvious reason to include only voids in the analysis.

In this work, we build upon the idea of separate universes, counts-in-cells and density split statistics for weak lensing. We will model RSD around different density environments. We will first split random positions of the spherically smoothed galaxy density field in different quantiles, and compute CCFs between these positions and the entire galaxy field in redshift space. This is in essence density split in three dimensions (DS). The CCF between regions in each quantile with the galaxy field corresponds to the stacked galaxy number density around environments of different depths. DS can also be seen as a generalisation of the void-galaxy CCF, as it naturally includes voids, clusters and intermediate-density regions in a general framework to exploit their combined constraining power on cosmology. With this setup, {\it the main question we wish to ask is: does the combination of a series of CCFs contain a different amount of cosmological information than the conventional 2PCF?} A closely related question is: does DS make the modelling more accurate for RSD than the standard 2PCF?

We will show that with DS, the distribution of the velocities in each density quantile are well-fit by a Gaussian function. This allows the Gaussian streaming model to perform accurately at almost all scales. Perhaps more importantly, the CCFs of split 3D density naturally captures the non-Gaussian distribution of the density, and this leads to improved cosmological constraints over two-point statistics.

The outline of the manuscript is as follows. Sec.~\ref{sec:model} introduces the theory of dynamical and geometrical distortions, and describes the Gaussian streaming model, which sets the basis for our theoretical predictions. Sec.~\ref{sec:rsd_with_ds} describes the algorithm for splitting the galaxy density field. Sec.~\ref{sec:performance_of_rsd_models} shows the performance of the RSD model using mock galaxy catalogues. Sec.~\ref{sec:constraints} presents the cosmological constraints obtained with the splitting density method. Finally, we discuss and summarize our main conclusions in Sec.~\ref{sec:conclusions}.

\section{Models for redshift-space distortions} \label{sec:model}

To compare the power of constraining cosmology between the 2PCF and DS using RSD, we need to have adequate dynamical distortion models for both the 2PCF and DS. We will employ the Gaussian streaming model \citep{Fisher1995} for both of them. In addition, we also need to account for the geometrical distortions, the Alcock-Paczynski effect \citep{Alcock1979}. We introduce these two distortion effects in this section.

\subsection{Dynamical distortions} \label{sec:dynamical_distortions}
To the linear order in velocity, the observed redshift of a distant galaxy is the sum of two components, the cosmological redshift and the redshift due to its peculiar velocity sourced by gravity. The observed redshift-space distance of a galaxy is then
\begin{equation}
\label{eq:mapping}
    \vec{s} = \vec{r} + \frac{v_{\parallel}}{aH} \hat{z},
\end{equation}
where $\bf r$ and $\bf s$ are real- and redshift-space distance vectors, $v_{\parallel}$ is the peculiar velocity along the line-of-sight direction $\hat {z}$; $a$ is the scale factor of the Universe, $H$ is the Hubble parameter at $a$. With mass conservation, we have
\begin{equation}
    \left[ 1 + \xi^s(\vec{s}) \right] \mathrm{d}^3\vec{s} = \left[ 1 + \xi(\vec{r}) \right] \mathrm{d}^3\vec{r}\ ,
\end{equation}
 where $\xi(\vec{r})$ and $\xi^s(\vec{s})$ denote the real and redshift-space correlation functions, which measure the excess probability of finding a galaxy pair separated by a given scale. The streaming model \citep{Peebles1980} provides a mapping between the real-space correlation function to the redshift-space anisotropic correlation function:
\begin{equation} \label{eq:GSM_integral}
    1 + \xi^s(s_{\perp}, s_{\parallel}) = \int \left[1 + \xi(r)\right]
     \mathcal{P}(v_{\parallel}, \vec{r}) \mathrm{d} v_{\parallel},
\end{equation}
where $r^2 = r_{\parallel}^2 + r_{\perp}^2$ is the real-space separation and $v_{\parallel} = aH (s_{\parallel} - r_{\parallel})$ is the pairwise line-of-sight velocity, which has a probability distribution  $\mathcal{P}(v_{\parallel}, \vec{r})$. Note that for a given $r$, the PDF for the line-of-sight velocities depends on the subtended angle from the line of sight $\theta$, as it has the contribution from both the radial and tangential components. Assuming a Gaussian form for $\mathcal{P}(v_{\parallel}, \vec{r})$, and that the density field is also Gaussian, neglecting higher order terms, the mapping becomes:
\begin{align} 
\label{eq:GSM} \nonumber
1 + \xi^s(s_{\perp}, s_{\parallel}) = &\int \left(1 + \xi(r)\right) \frac{1}{\sqrt{2\pi \sigma_{\parallel}^2(r, \mu)}}\\
&\exp\left\{- \frac{\left[v_{\parallel} - v_{r}(r) \mu\right]^2}{2\sigma_{\parallel}^2(r,\mu)}\right\} \mathrm{d}v_{\parallel},
\end{align}
where $\mu = r_{\parallel}/r=\cos\theta$, and $v_r(r)$ is the pairwise velocity along the radial direction, also known as the mean streaming velocity. This was first derived by \cite{Fisher1995}, and is usually referred to as the Gaussian streaming model (GSM)  \citep[see also the derivations by][in Fourier space]{Scoccimarro2004,Vlah2019}. 
Note that the velocity dispersion $\sigma_{\parallel}(r, \mu)$ depends on both $r$ and $\mu$.  Eq.~(\ref{eq:GSM_integral}) can be used to predict the redshift-space correlation function, as long as the full distribution of pairwise velocities and the real-space correlation function are fully known. However, it is challenging to predict the distribution of pairwise velocities from first principles. Eq.~(\ref{eq:GSM}) may work but only if the Gaussian assumption holds. Exploring the validity of this assumption is a focus for this study.

Another key ingredient for the GSM is the pairwise streaming velocity $v_{r}(r)$. In linear theory, it can be expressed in terms of the correlation function \citep{Peebles1980,Sheth2001b}
\begin{equation}
\label{eq:vel_den_coupling_0}
 v_{r}(r) = - \frac{2}{3} \frac{\beta aHr \bar{\xi}(r)}{[1 + \xi(r)]} .
\end{equation}
Here $\xi(r)$ is the galaxy real-space correlation function, $\beta=f/b$ where $f = {\rm d} \ln D/{\rm d}\ln a$ is the linear growth rate, $D$ is the growth factor, $b$ is the linear galaxy bias, and
\begin{equation}
\label{eq:xi}
    \bar{\xi}(r) = \frac{3}{r^3} \int_0^r \xi(x) x^2 \mathrm{d}x. 
\end{equation}
The linear coupling between density and velocity is not adequate for the quasi-nonlinear regime. From this point and throughout the paper, our notations will not use the usual subscript $g$ to refer to quantities related to galaxies. We will use $\xi(r)$ and $\bar{\xi}(r)$ to refer to the {\it galaxy} correlation function and its cumulative version. We will use the subscript $m$ to refer to quantities about dark matter.

Modelling the exact coupling is another crucial step towards accurate modelling for RSD with the GSM. We adopt the empirical function introduced in \cite{Juszkiewicz1999}, with one free parameter to model the coupling between the galaxy density and velocity field:
\begin{align} \label{eq:vel_den_coupling_1}
    v_{r}(r) = -\frac{2}{3} aHr\beta \overline{\overline{\xi}}(r)\left[1 + \nu \overline{\overline{\xi}}(r)\right] \ .
\end{align}
In the expression above, $\nu$ is a free parameter. When $\nu=0$, the above goes back to the linear model. $\overline{\overline{\xi}}$ is defined by the following relation:
\begin{align} \label{eq:xi_bar}
    \overline{\xi}(r) \equiv
\overline{\overline{\xi}}(r) [1 + \xi(r)] .
\end{align}
As we will describe in the following sections, we are also interested in the cross-correlation between randomly positioned centres ranked by their local number density of galaxies, and the entire galaxy number density field. In such a case, the equations for the streaming model outlined above still apply, but as there are no peculiar motions for the random centres themselves, the pairwise velocity profile becomes the stacked velocity profile, and the linear density-velocity coupling becomes
\begin{align} \label{eq:vel_den_coupling_1.5}
    v_r(r) = - \frac{1}{3} a H r \beta \overline{\xi}(r), \
\end{align}
where $\bar{\xi}(r)$ from Eq.~(\ref{eq:xi}) now refers to the cumulative cross-correlation function between fixed positions (e.g. voids or clusters) with the galaxy field, i.e. the stacked density profile, and so it is the same as the cumulative {\it galaxy} density contrast $\Delta(r)$ within $r$. This is commonly adopted in void RSD studies (e.g. \citealt{Hamaus2014, Cai2016,Achitouv2017a,Hawken2020, Hamaus2017,Nadathur2020}). Again, we will go beyond the linear coupling by introducing
\begin{align} \label{eq:vel_den_coupling_2}
    v_r(r) = - \frac{1}{3} \frac{a H r \beta \overline{\xi}(r)}{1 + \nu \xi(r)}, \,
\end{align}
where $\nu$ is a free parameter and $\nu=0$ is the linear model.  As we will see later, this empirical expression will allow us to not only fit the velocity profiles around voids, but also around high density regions. 

Without causing confusion, we will use the same notations [$v_{\parallel}$, $v_r$, $\sigma_{\parallel}, \xi, \bar\xi$] to refer to variables for both two-point correlation function (2PCF) and cross-correlation function (CCF) throughout the paper.

\subsection{Geometrical distortions}
In observations, when converting observed redshifts to distances using a cosmology that is different from the true underlying cosmology of the Universe, we artificially induce geometrical distortions in the clustering of galaxies, an effect that is also known as Alcock-Paczynski (AP) distortions \citep{Alcock1979}. We can parametrise these distortions by rescaling the transverse and line-of-sight separation vectors \citep{Ballinger1996}:
\begin{align}
    s_{\perp} = q_{\perp} s_{\perp}' \\
    s_{\parallel} = q_{\parallel} s_{\parallel}' \ ,
\end{align}
where the primes represent quantities in the fiducial cosmology. The scaling factors are related to cosmological parameters via
\begin{align}
    q_{\perp} = \frac{D_M}{D_{M}'} \\
    q_{\parallel} = \frac{H'}{H} \ ,
\end{align}
where $D_M$ and $H$ are the comoving angular diameter distance and the Hubble parameter at $a$, respectively. The redshift-space correlation function can then be rescaled as
\begin{align} 
\label{eq:AP_rescaling}
    \xi^s(s_{\perp}, s_{\parallel}) = \xi^s(q_{\perp}s_{\perp}', q_{\parallel}s_{\parallel}') \ .
\end{align}
The dynamical and geometrical distortions act at the same time on the observed redshift-space correlation function -- the only observable at our disposal in this context. Adjusting the dynamical distortion parameters (RSD) and geometrical distortion parameters (AP) to fit for the observed redshift space clustering will in turn allow us to constrain those parameters of interests \citep[e.g.][]{Sanchez2016, Beutler2017, Hou2018, Hamaus2020, Bautista2021}.

\begin{figure}
    \centering
    \includegraphics[width=\columnwidth]{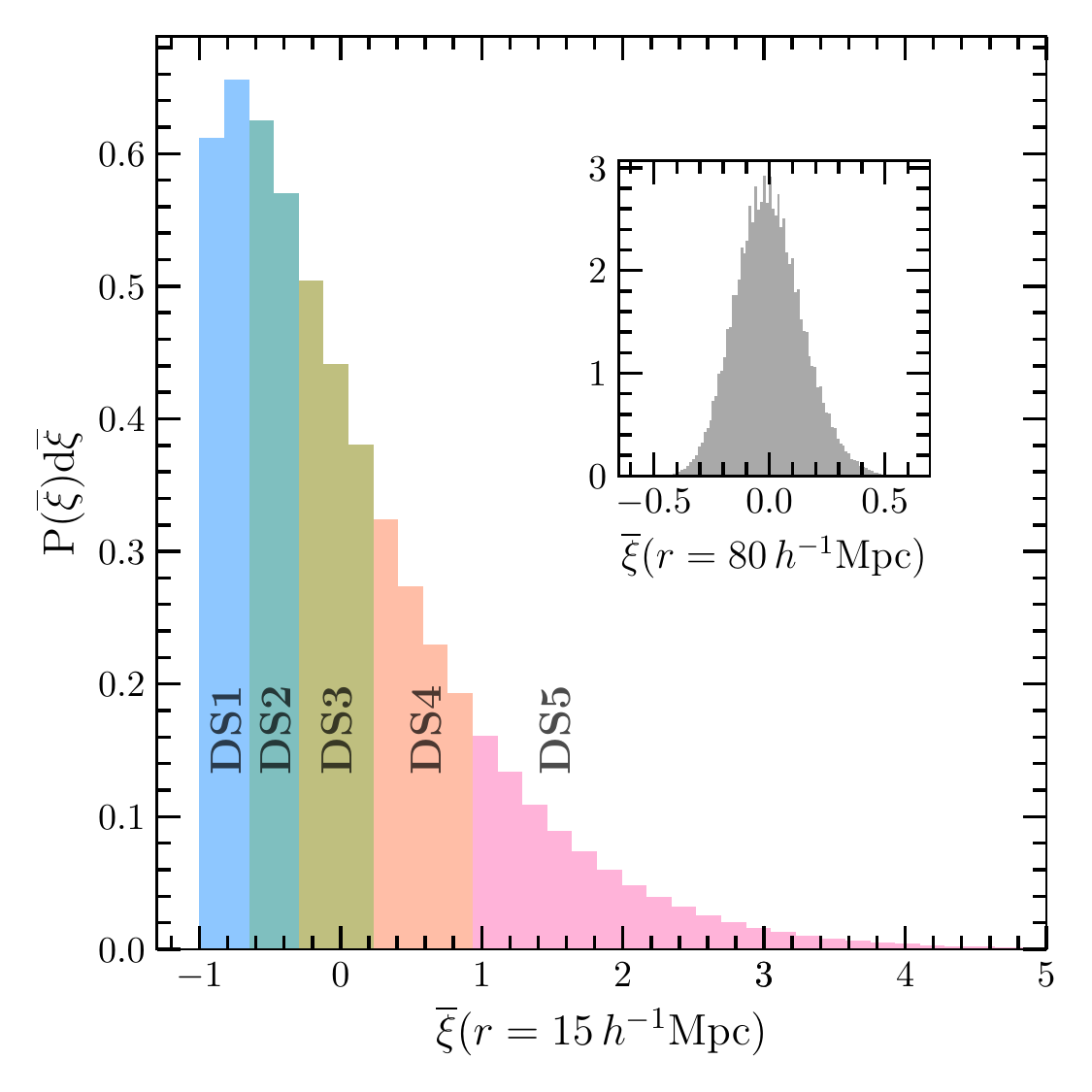}
    \caption{The probability  distribution for the integrated galaxy density contrast $\bar\xi$, smoothed by a top-hat window function of $15\,h^{-1}{\rm Mpc}$. The different colors delimit the split of the PDF into different quintiles, according to the value of $\bar{\xi}$, ranging from underdensities (DS1-2) to overdensities (DS4-5). The positions in each quintile will be cross-correlated with the entire galaxy field in redshift space. These are the main observables we will employ in this paper. The inner subplot shows the PDF at a scale of $80\,h^{-1}{\rm Mpc}$. }
    \label{fig:Delta_PDF}
\end{figure}

\begin{figure*}
    \centering
    \includegraphics[width=1.\textwidth]{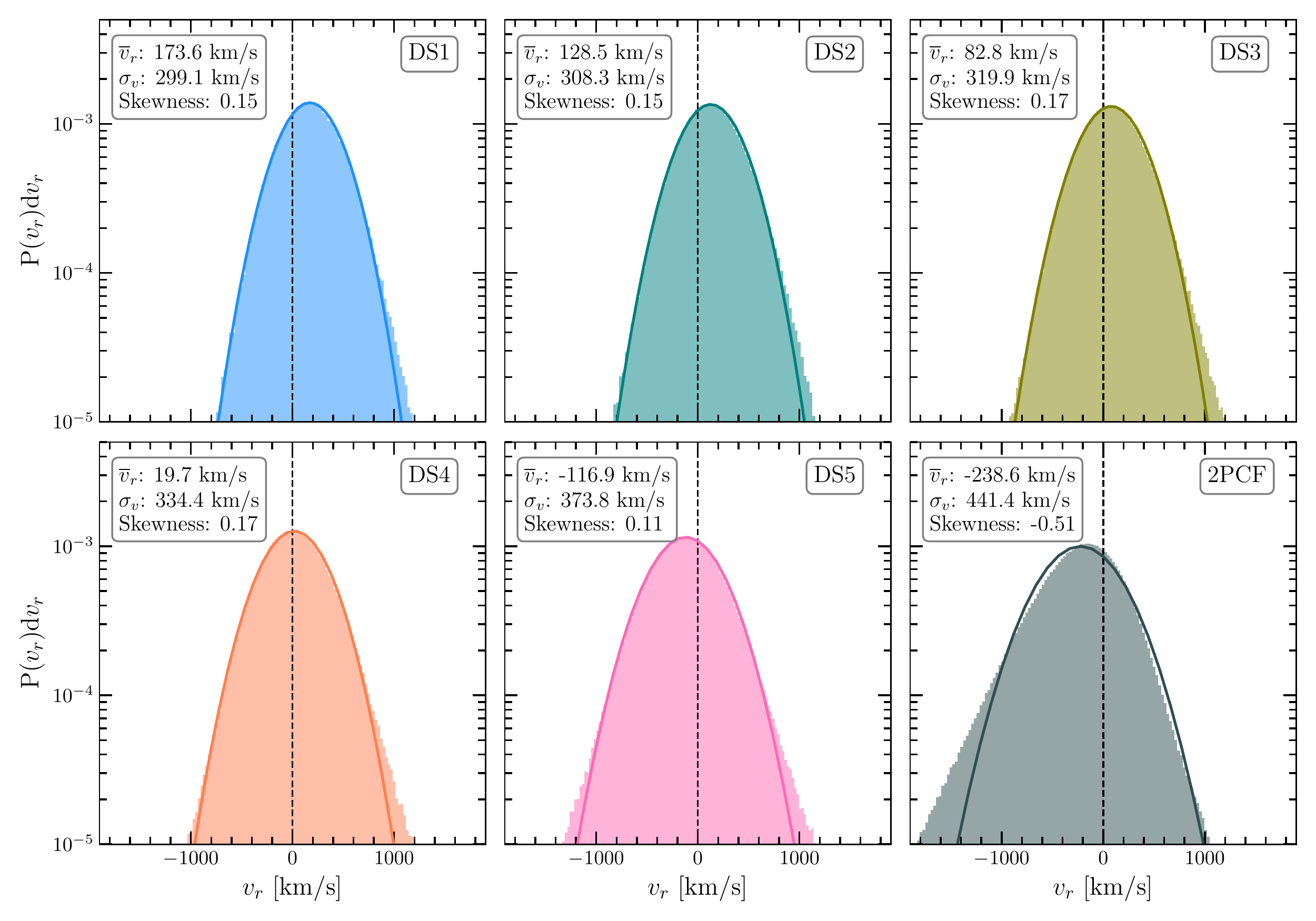}
    \caption{Distributions of radial velocities in density splitting (DS) quintiles and pairwise velocities for the two-point correlation function (2PCF) (different panels as labelled), measured from one of the mock realisations, at a scale of $r=15\,h^{-1}{\rm Mpc}$. For DS, the colours and legends (DS1-5) are matched to those in Fig.~\ref{fig:Delta_PDF} to represent densities of different depths, varying from voids (DS1) to clusters (DS5). The mean $\bar v_r$, standard deviation $\sigma_v$ and skewness of the distributions are shown on the upper left corner of each panel. The skewness is the third standardised moment of a distribution,  characterising the asymmetry of the distribution about its mean. The solid lines show a Gaussian distribution with the same mean and standard deviation as the data, after applying 3$\sigma$ clippings. The PDFs for velocities in each DS are well-fit by a Gaussian profile, with their mean $\bar{v}_{r}$ varying from positive in DS1 to negative in DS5, corresponding to outflow and infall around those quintiles, respectively. The PDF for the pairwise velocities (2PCF) is significantly skewed towards negative values, strongly deviating from the Gaussian distribution.}
    \label{fig:velocity_PDF}
\end{figure*}

\begin{figure*}
    \centering
    \includegraphics[width=1.0\textwidth]{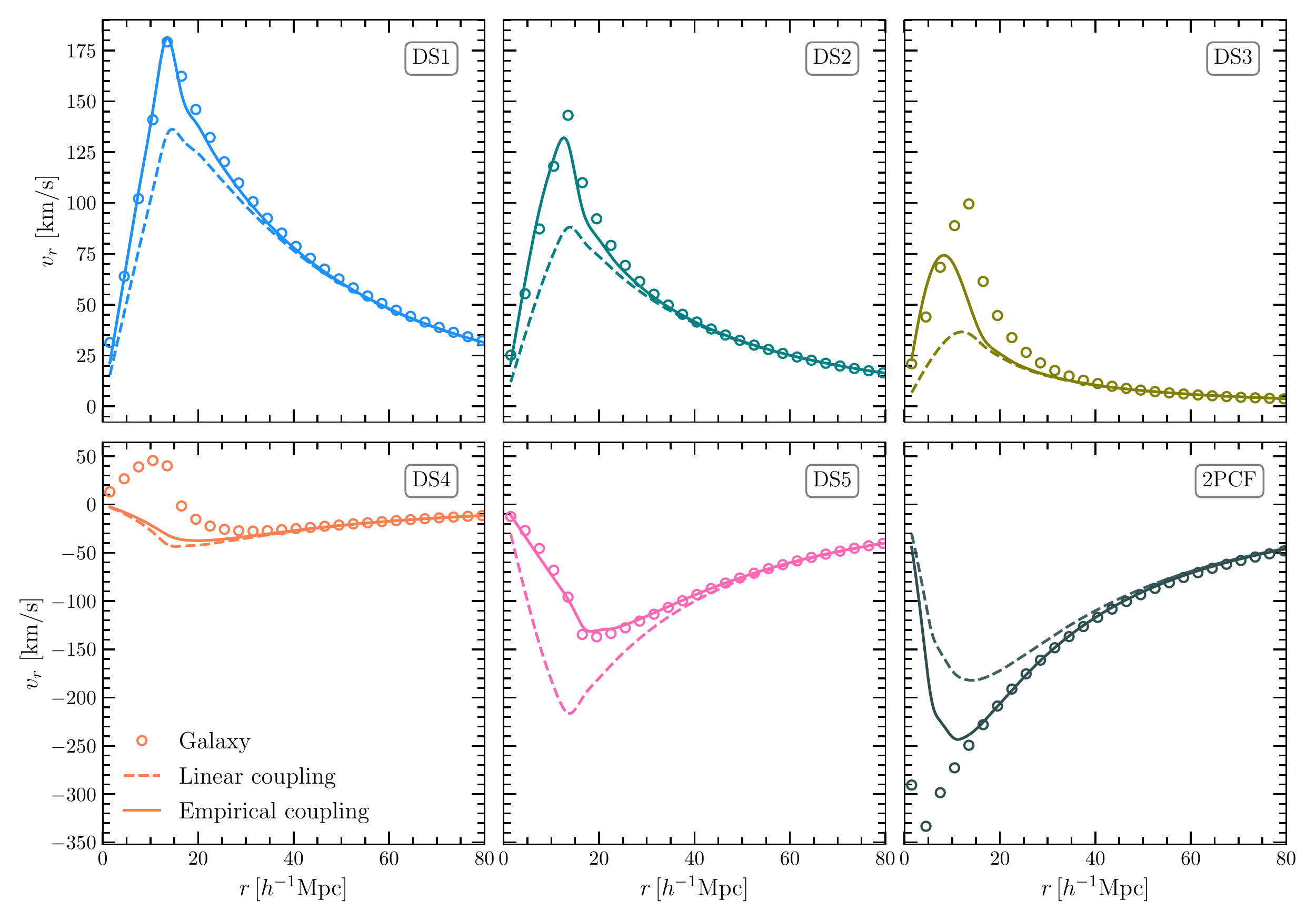}
    \caption{Galaxy radial velocity profiles around regions corresponding to different density quintiles (DS1-5) and the pairwise radial velocity profiles (2PCF), as indicated in each panel. The circles show measured velocities from the simulations. The dashed lines show the predictions from linear theory (Eqs.~\ref{eq:vel_den_coupling_0} \& \ref{eq:vel_den_coupling_1.5}), whereas the solid lines show the best-fit results from our empirical model (Eqs.~\ref{eq:vel_den_coupling_1} \& \ref{eq:vel_den_coupling_2}) using the redshift-space galaxy correlation functions as observables. The linear coupling model fails to reproduce the observed velocities on small scales for all cases, where non-Gaussianity becomes important. The empirical model accurately captures density-velocity velocity coupling for DS1 and DS5 (voids and clusters), but it underestimates the velocity profiles for intermediate quintiles and for the pairwise velocity profiles at small scales.}
    \label{fig:velocity_profiles}
\end{figure*}

\begin{figure}
    \centering
    \includegraphics[width=\columnwidth]{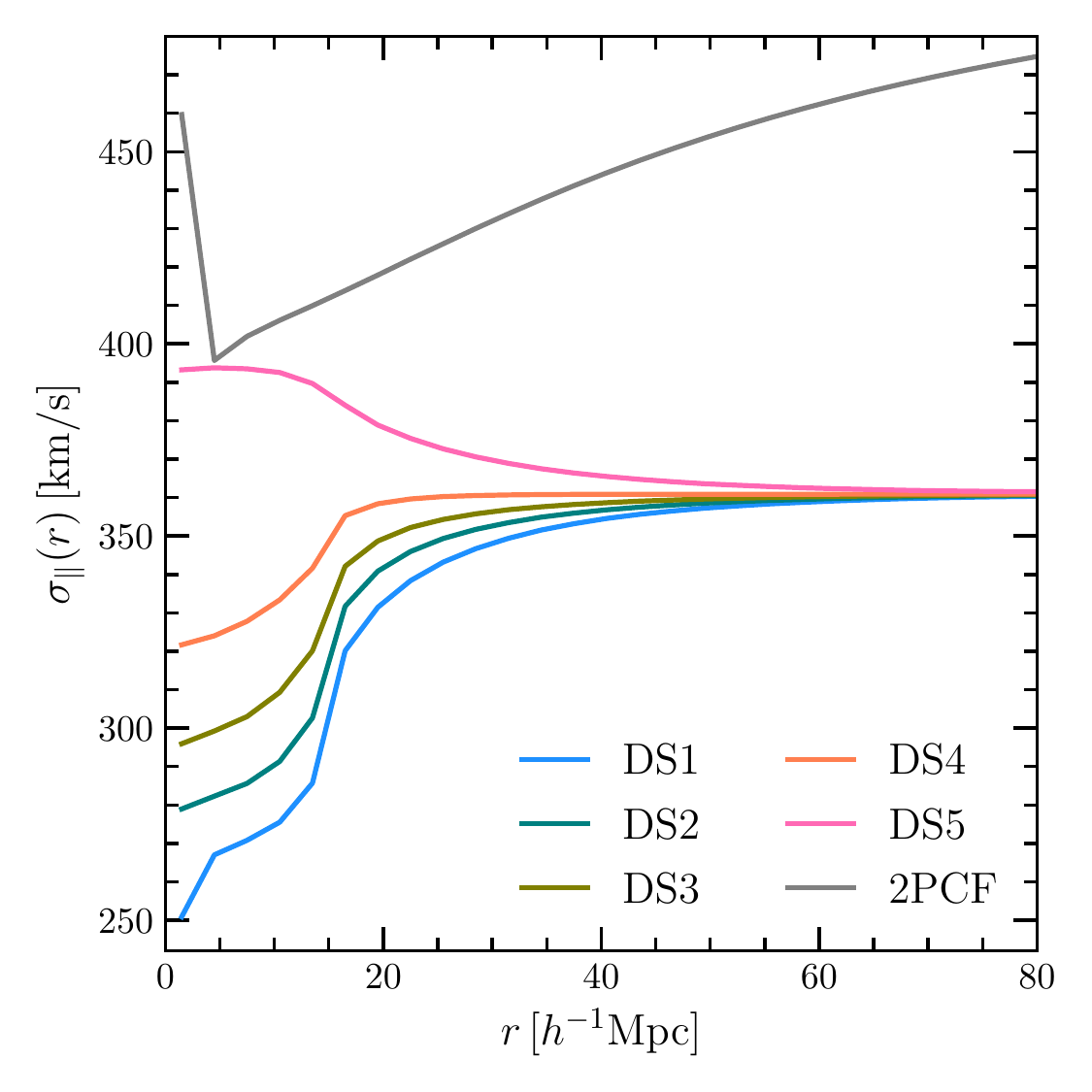}
    \caption{The line-of-sight velocity dispersion as a function of $r$ (after integrating over the $\mu$ dependence). These are direct measurements from the mock galaxies. Different colours correspond to the different DS quintiles and to the pairwise galaxy velocities, as indicated in the legend.}
    \label{fig:sigma_los_vs_r}
\end{figure}

\section{RSD with splitting densities} \label{sec:rsd_with_ds}
A crucial assumption for the Gaussian streaming model (Eq.~\ref{eq:GSM}) to work is that the PDF of the pairwise velocity needs to be Gaussian. This is only true at high-$z$ or in the linear regime. \cite{Reid2011} have shown that this assumption can already lead to a $2$ percent level bias for the quadrupole at $s=25\,h^{-1}{\rm Mpc}$.
To improve the performance of the model, one obvious way is to go beyond the Gaussian assumption by modelling the full distribution of pairwise velocity. This is non-trivial from first principles and extra degrees of freedom are usually introduced (\citealt{Kuruvilla2018,Cuesta-Lazaro2020}).

We take an alternative approach to analyse the data by splitting the galaxy field into different density environments. The assumption is that the non-Gaussian PDF of the pairwise velocity at small scales can be decomposed into many Gaussian PDFs of different widths. This was elucidated in \cite{Tinker2007}, where it was shown that the PDF of pairwise velocities at a specific density environment is indeed close to Gaussian. The halo model was also adopted for the modelling in \cite{Tinker2007}. Examples for the velocity PDFs for several overdense environments were shown, but not for underdense environments. We will generalise this to low-density regions. We will use cross-correlations instead of auto-correlation functions. In this case, the relevant statistic is the distribution function of the velocity field, rather than the pairwise velocities. The main steps for our method are summarised as follows:

1. start from a galaxy sample in real space and apply a spherical top-hat filtering with a filter radius $R$ on random positions.

2. rank order the filtered density contrasts $\Delta(r=R)$ (the same as $\bar{\xi}(R)$ as noted in Sec.~\ref{sec:model}) and split them into $n$ density bins. For this study, we adopt $R=15\,h^{-1}{\rm Mpc}$ and $n=5$. 

3. cross-correlate the positions in each density bin, or quintile, with the entire galaxy sample in redshift space to obtain the CCF $\xi^{i}(s,\mu | \Delta^i)$, where $i=1,2,3...n$\footnote{Note that we have used $\bf s$ to denote the redshift-space distance vector in the earlier part of the paper. Here we use the same symbol to denote the distance between the DS centres (in real space) to galaxies (in redshift space). Their meanings are technically different.}. These are in essence a series of conditioned correlation functions, with the condition being $\Delta^i(r=R)$ satisfying our density splitting criteria. We will write $\Delta^i$ instead of $\Delta^i(r=R)$ without loss of clarity. 

 The CCFs $\xi^i(r,\mu | \Delta^i)$ are the same as $\delta^i(r,\mu | \Delta^i)$ i.e. the stacked number densities of galaxies around the centres of spheres within the $i$-th density bin. Therefore, the CCF for the lowest density bin is similar to the void-galaxy cross-correlation function, and the highest density bin is similar to the cluster-galaxy cross-correlation function. The notation $\Delta(r)$ is also the same as $\bar\xi (r)$ from Eq.~(\ref{eq:xi}), with $\xi(r)=\delta(r)$, as mentioned in Sec.~\ref{sec:model}.

Instead of the 2PCF, the series of CCFs will be our main observable to be used to constrain cosmology. The streaming model and its Gaussian version can naturally be applied to model these void-galaxy, cluster-galaxy cross-correlations, and in general cross-correlations of different local densities with the galaxy field. In GSM (Eq.~\ref{eq:GSM}), one can simply replace $\xi(r)$ by the real-space CCF (or the stacked density profile), and the pairwise velocity profile $v_r(r)$ by the stacked velocity profiles around different local densities. This can be seen by keeping the positions of the spherical regions fixed in real-space, i.e. one of the two points in the pair has zero velocity and thus the pairwise velocity becomes the radial velocity of a galaxy relative to its real-space spherical centre. Note that although the mathematical form is the same, the two-point correlation function is, fundamentally, a two-point statistic, while the cross-correlation between a set of centres with the entire galaxy field is a first-moment measurement.

The number of density quantiles chosen in this work is a balance between different factors. Firstly, a larger number of quantiles ensures that the distribution of densities in each bin is narrow, which results in a distribution of radial velocities that is more Gaussian, and thus a better performance for the GSM. Secondly, increasing the number of bins also increases the size of the covariance matrix that is needed for our likelihood analysis (see Sec.~\ref{sec:constraints}). A bigger covariance matrix will also require a larger number of mock realizations in order to be accurately estimated. Since we have 300 mocks at our disposal, 5 quantiles is a good compromise between these two factors.

\section{Performance of RSD models} \label{sec:performance_of_rsd_models}
Following the above steps, we will use mock galaxies to measure the series of $\xi^i(s,\mu | \Delta^i)$. We will also measure the 2PCF $\xi(s,\mu)$. We then use the GSM to extract cosmological information from the above two measurements from the same simulations. We will focus on asking: does the combination of the whole series of CCFs, i.e. all $\xi^i(s,\mu | \Delta^i)$, contain any different amount of cosmological information than the conventional 2PCF? Before doing this, it is crucial to validate the performance of RSD modelling with the Gaussian streaming model for CCFs from split densities and the 2PCF. This is the focus of this section. We will describe the mock galaxies catalogue (Sec.~\ref{sec:simulations}), followed by splitting densities with the mock galaxies (Sec.~\ref{sec:density_split}). We then analyse the velocity distribution functions with the split densities (Sec.~\ref{sec:velocity_PDF}), and cross-correlate each split density quintile with the entire galaxy field and compare them with the Gaussian streaming model (Sec.~\ref{sec:GSM_DS}). We review  and compare other RSD models for cross-correlations in Sec.~\ref{sec:model_comparison}.

\subsection{Mock galaxies} \label{sec:simulations}
We will use the {\sc Minerva} simulation for our analysis \citep{Grieb2016, Lippich2019}. It consists of a set of 300 N-body simulations that represent different realisations of the same cosmology, which corresponds to the best-fitting flat $\Lambda$CDM model to the combination of CMB data \citep{Planck2016} and SDSS DR9  CMASS wedges, presented in \citet{Sanchez2013}. The model is characterised by a matter density parameter of $\Omega_\mathrm{m} = 0.285$, a baryon physical density of $\omega_\mathrm{b} = 0.02224$, a present-day Hubble rate of $H_0 = 69.5\ $kms$^{-1}$Mpc$^{-1}$, an amplitude of density fluctuations of $\sigma_8 = 0.828$ and an scalar spectral index of $n_\mathrm{s} = 0.968$. For each box, a total of $1000^3$ dark matter particles were evolved with {\sc GADGET} \citep{Springel2005} in a cosmological box of $1.5\ h^{-1}\mathrm{Gpc}$ aside. The volume for each box is therefore 3.375$(h^{-1}\mathrm{Gpc})^3$. This will be the default volume that sets the statistical errors for our predictions. Initial conditions were generated using second order-Lagrangian perturbation theory \citep{Jenkins2010}, starting from $z=63$.

Dark matter haloes and their associated substructures are identified using \textsc{subfind} \citep{Springel2001}. These haloes were populated at the $z = 0.57$ snapshots of the simulations using the Halo Occupation Distribution method (HOD; \citealt{Peacock2000,Benson2000,Berlind2002,Kravtsov2004}). We adopt the HOD functional form presented in \cite{Zheng2007}. The HOD parameters were calibrated to reproduce the clustering properties of the BOSS CMASS DR9 galaxy sample for the same cosmology, as in \cite{Manera2013}.

For each HOD catalogue we also construct a redshift-space analog by shifting the positions of galaxies along the line of sight with Eq.~(\ref{eq:mapping}) (taken to be the $z$-axis of the simulation) using their peculiar velocities. In the remainder of the text, we will refer to each of these 300 simulations populated with HOD galaxies as mock realisations.

\subsection{Splitting densities with mock galaxies} \label{sec:density_split}

For each mock realisation, we place $1.3 \times 10^6$ (which is roughly the same as the number of mock galaxies) random points in the cosmological volume of the simulation, and measure the real-space integrated galaxy number densities $\bar \xi$ within spheres of radius $r=15\,h^{-1}{\rm Mpc}$ centred on those points. We then rank the centres according to the value of $\overline{\xi} (r = 15~h^{-1}{\rm Mpc})$ in increasing order, and split them into 5 bins with equal number of centres, that we refer to as \textit{Density Split} quintiles, labelling them as DS1 to DS5 from low to high density. The coloured histogram in Fig.~\ref{fig:Delta_PDF} shows the distribution of  $\overline{\xi}(r=15~h^{-1}{\rm Mpc})$ around the random positions in one of the mock realisations. DS1 is the lowest density quintile and has a negative $\overline{\xi}$, corresponding to voids. DS2-3 are also underdense, although their density contrasts are shallower than voids. DS4 marks the transition towards overdense regions, having a positive $\overline{\xi}$. DS5 has the largest positive amplitude, corresponding to cluster-like regions. As has been shown by previous studies, the full PDF of $\overline{\xi} (r = 15\,h^{-1}{\rm Mpc})$ is highly non-Gaussian, with its peak corresponding to a negative density, i.e. a major fraction of the volume is in voids, and a high-density tail that extends to large positive values. It follows closely a lognormal distribution \citep[e.g.][]{Coles1991,Colombi1994, Bernardeau1995,Scoccimarro2004,Uhlemann2016, Repp2018, Einasto2020}. These features are expected, and are the result of the growth of structures via gravitational evolution. While initial overdensities collapsed due to gravitational attraction into small and highly overdense structures, initially underdense regions expanded and ended up covering a large volume, but could only reach moderate underdensities, as the density contrast is bound from below (i.e. $\delta_m = -1$). The PDF and its evolution in the non-linear regime follow spherical dynamics closely, and can be predicted accurately \citep{Uhlemann2016, Repp2018, Jamieson2020}. On large scales where the density evolution is linear, the density PDF is expected to be Gaussian. An example is shown by the inset of Fig. \ref{fig:Delta_PDF}. At the smoothing scale of $r= 80\,h^{-1}{\rm Mpc}$, the distribution is much more symmetric and can be well fitted by a Gaussian PDF.

\begin{figure*}
    \centering
    \includegraphics[width=1.\textwidth]{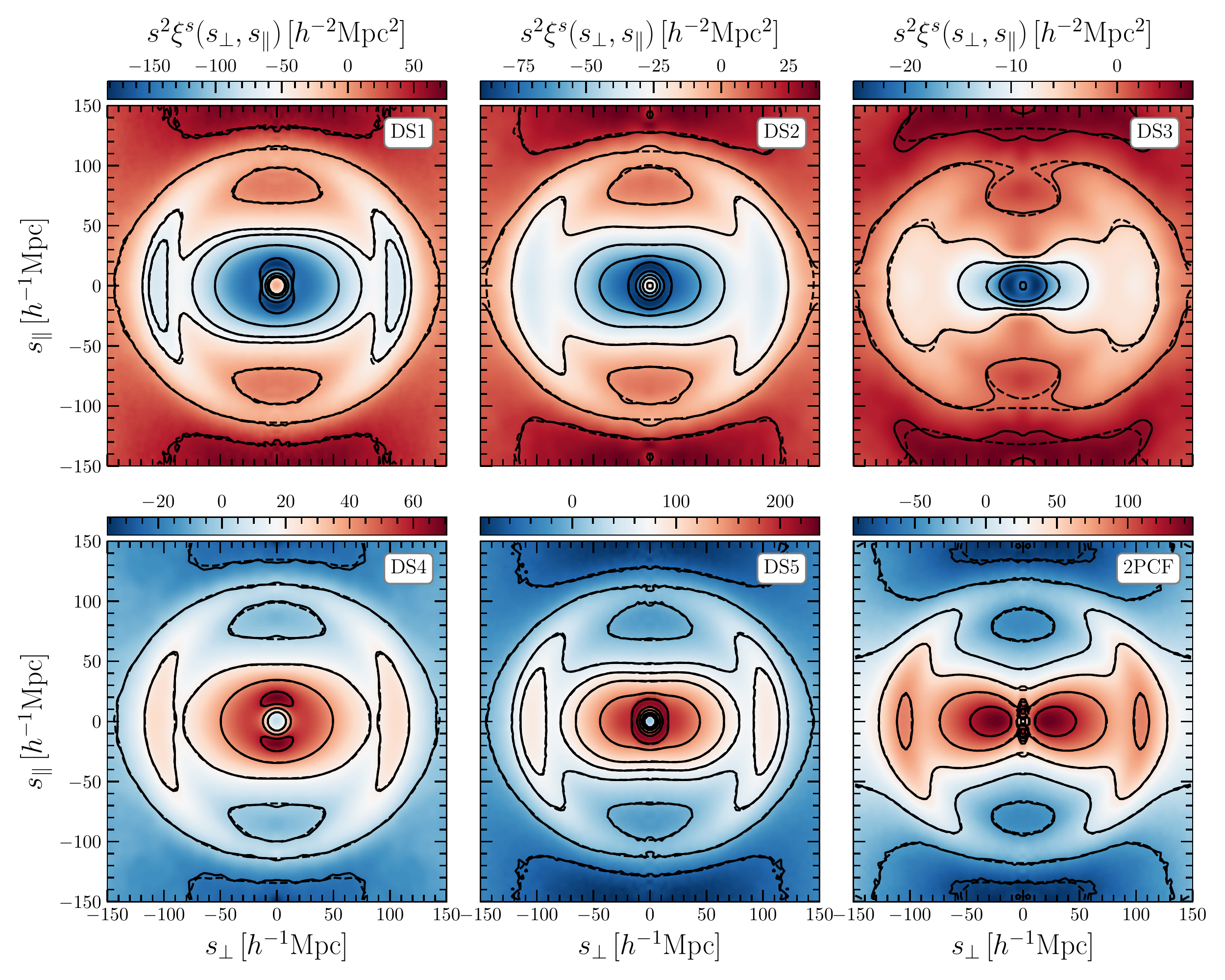}
    \caption{Cross-correlations between centres of density quintiles in DS with the entire galaxy sample in redshift space (labels as DS1-5), and the galaxy two point correlation function (2PCF), averaged over the 300 mock galaxy catalogues. The colourbars indicate the clustering amplitude, scaled by a factor of $s^2$ to highlight large-scale features. The solid lines show contours of constant amplitude in the simulation. The predictions from the Gaussian streaming model (Eq.~\ref{eq:GSM}), with all its ingredients measured from simulations, are shown in dashed lines. Contours for DS quantiles 2, 3 and 4 have been smoothed using a Gaussian kernel with a Gaussian $\sigma$ of 1, 2 and 1 pixels, respectively.}
    \label{fig:sigma_pi_largescales}
\end{figure*}

\begin{figure*}
    \centering
    \includegraphics[width=\textwidth]{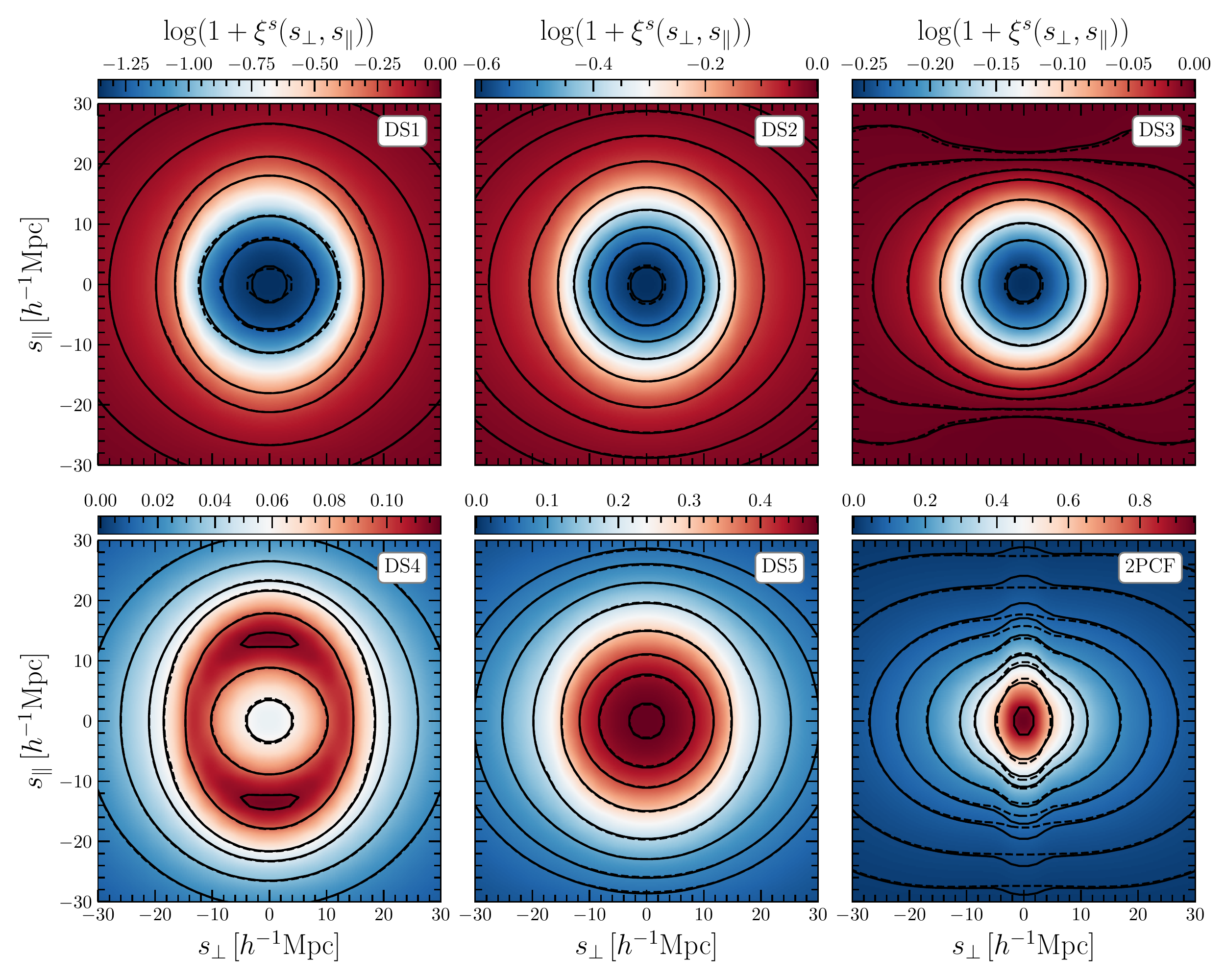}
    \caption{Similar to Fig. \ref{fig:sigma_pi_largescales}, but showing clustering at small scales. The Gaussian streaming model (in dashed contours) lays on top of the measurements from mock galaxies (solid contours) for CCFs from all density quintiles (DS1-5). All the CCFs show no sign of the Fingers-of-God effect. Deviations for the 2PCF from the same model can be seen close to the line-of-sight direction, mainly due to the strong Fingers-of-God effect. At these relatively small scales, linear theory is not as accurate, but its prediction agrees reasonably well for DS1 and DS2 (underdense environments), as we have checked.}
    \label{fig:sigma_pi_smallscales}
\end{figure*}

\begin{figure*}
     \centering
     \begin{tabular}{ccc}
     \hspace{-0.7 cm}
        \includegraphics[width=0.35\textwidth]{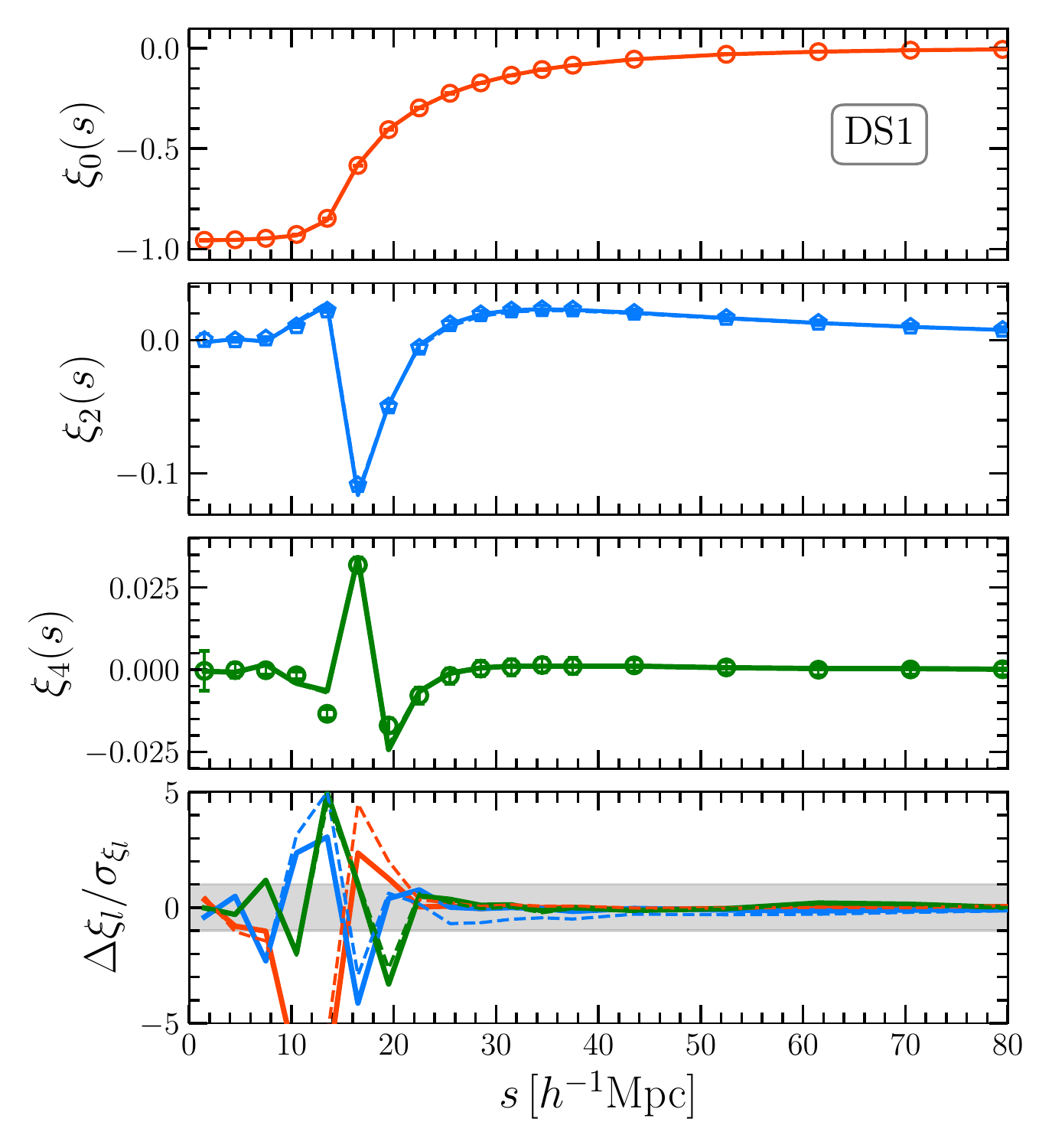} &
         \hspace{-0.6 cm}
        \includegraphics[width=0.35\textwidth]{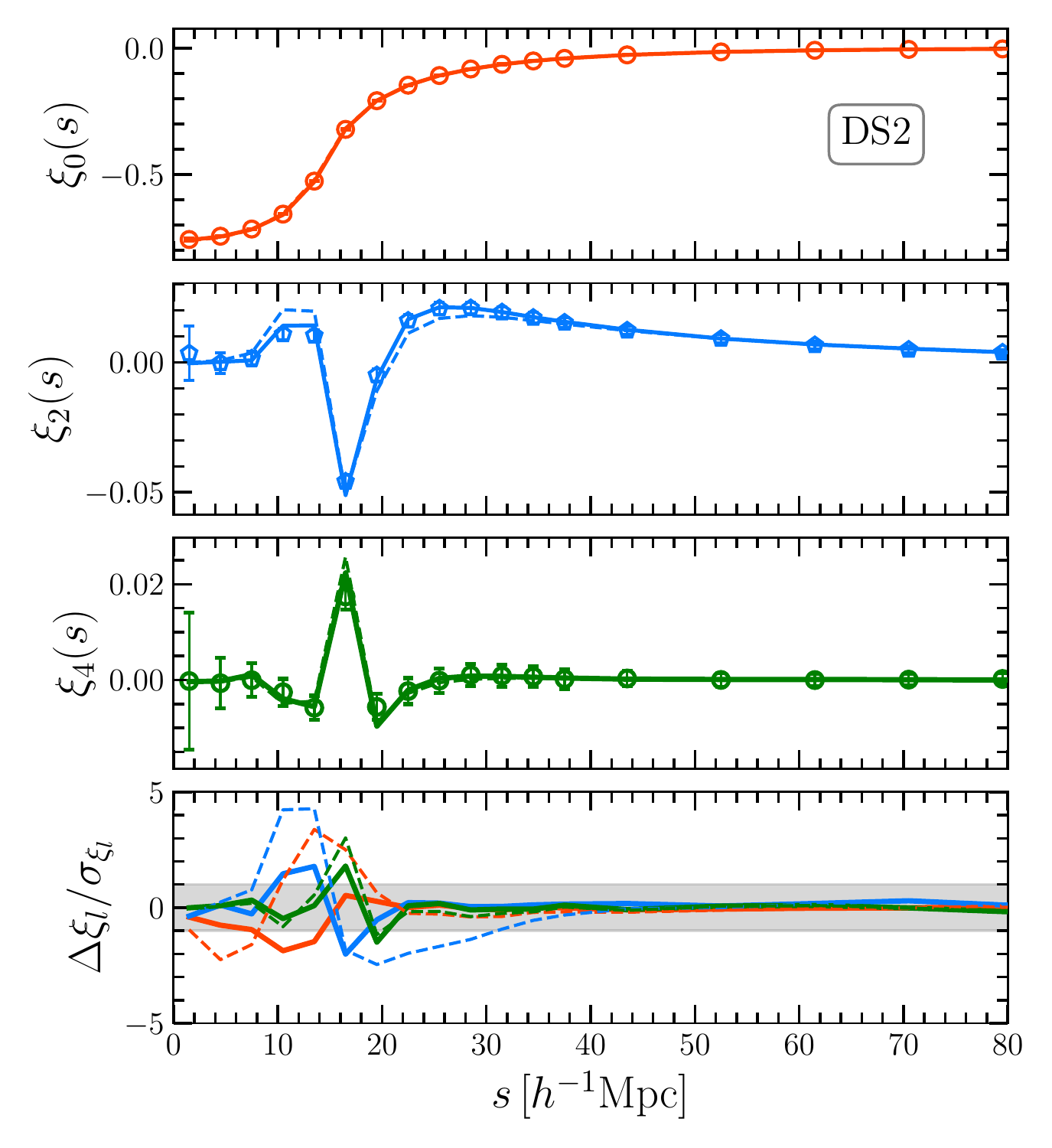} &
         \hspace{-0.6 cm}
        \includegraphics[width=0.35\textwidth]{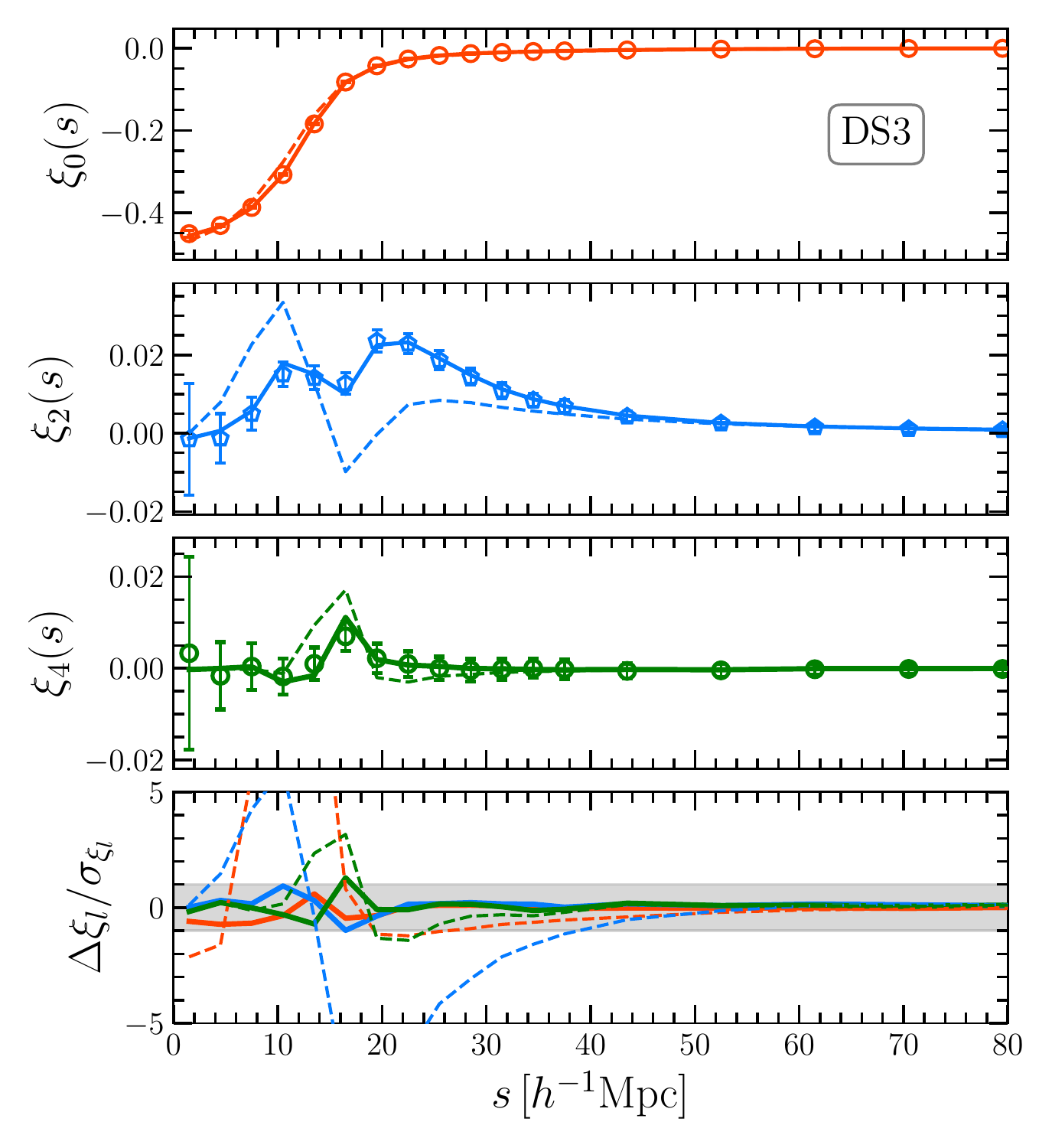} \\
        \hspace{-0.7 cm}
        \includegraphics[width=0.35\textwidth]{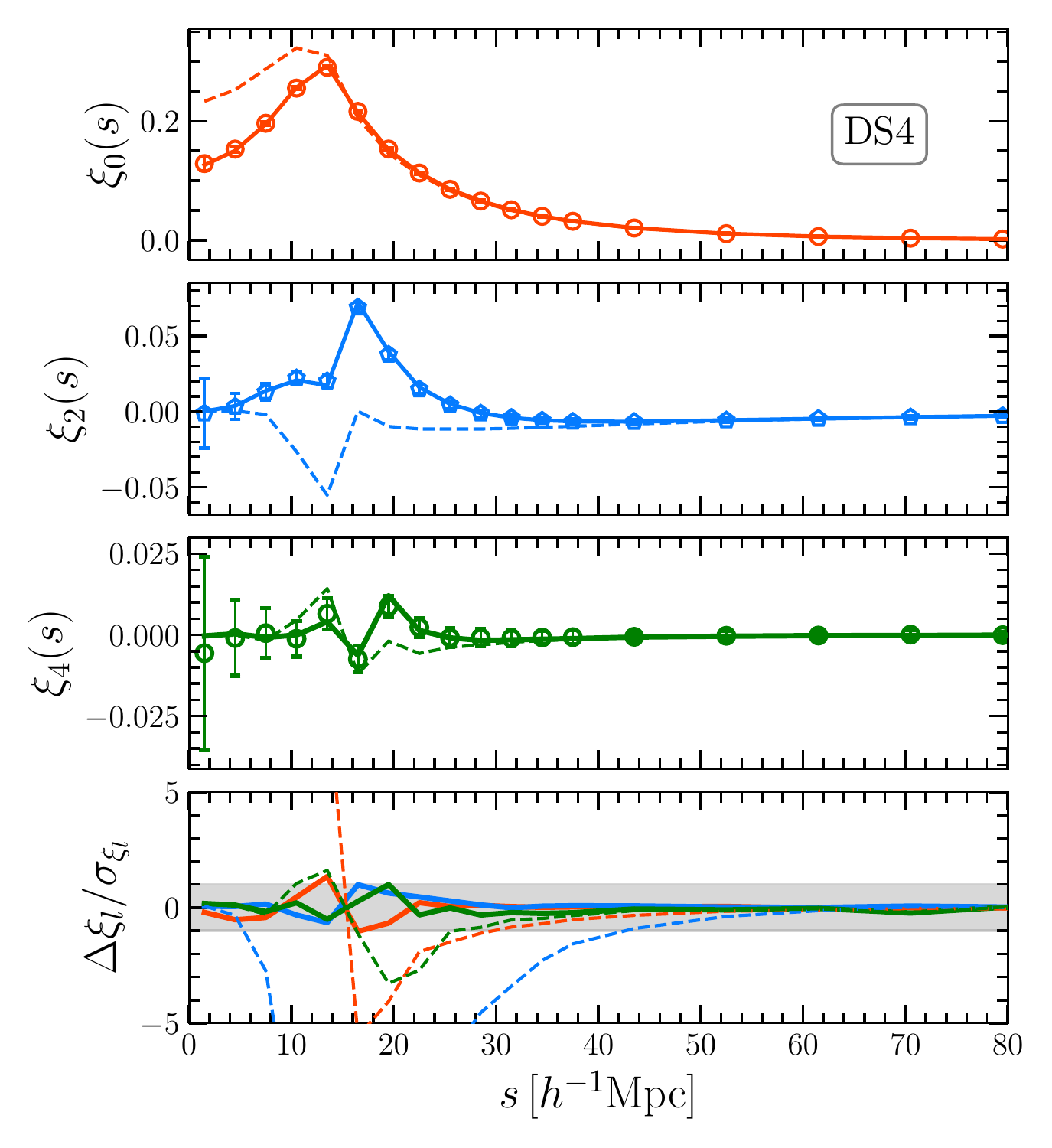} &
         \hspace{-0.6 cm}
        \includegraphics[width=0.35\textwidth]{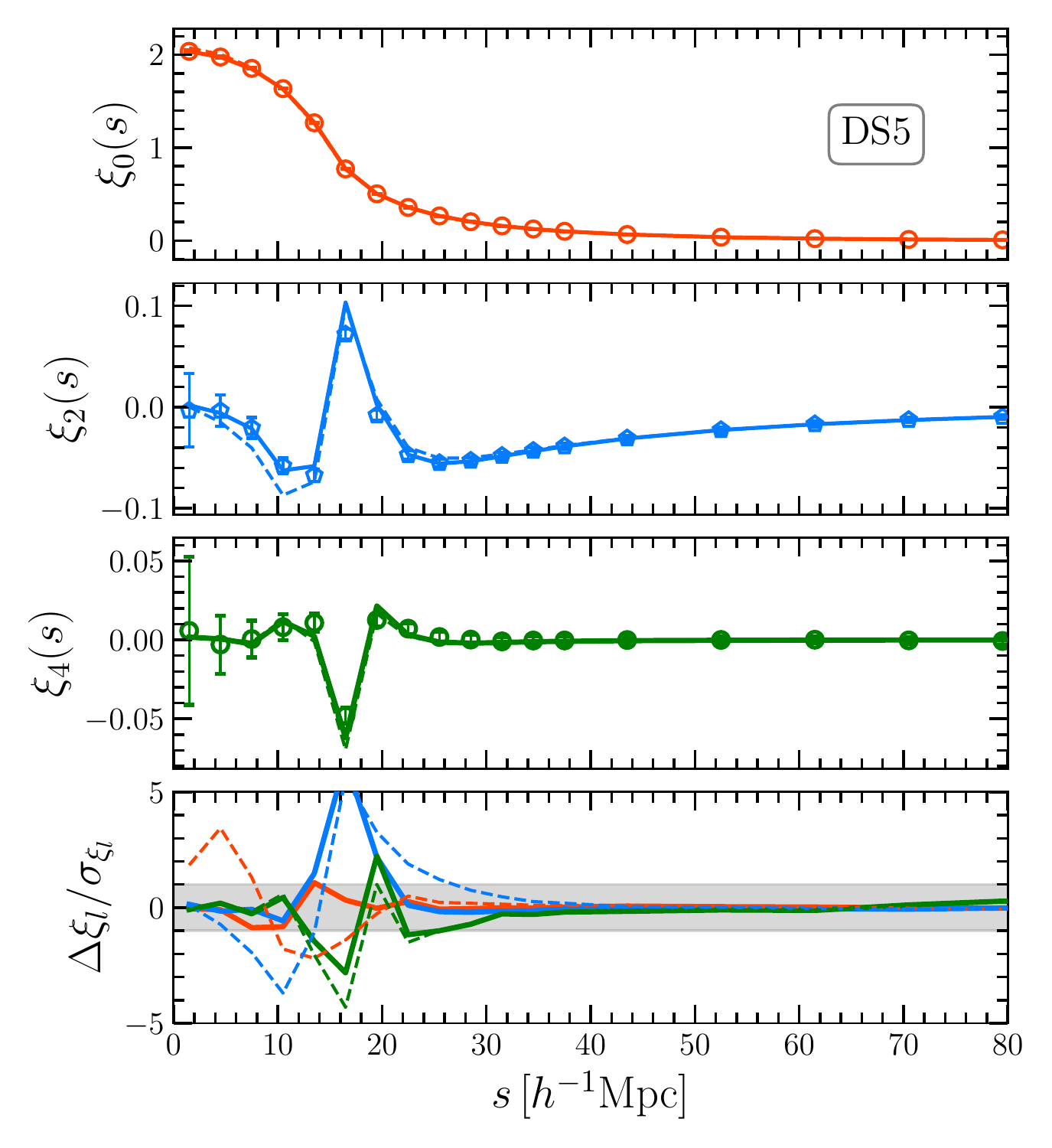} &
         \hspace{-0.6 cm}
        \includegraphics[width=0.35\textwidth]{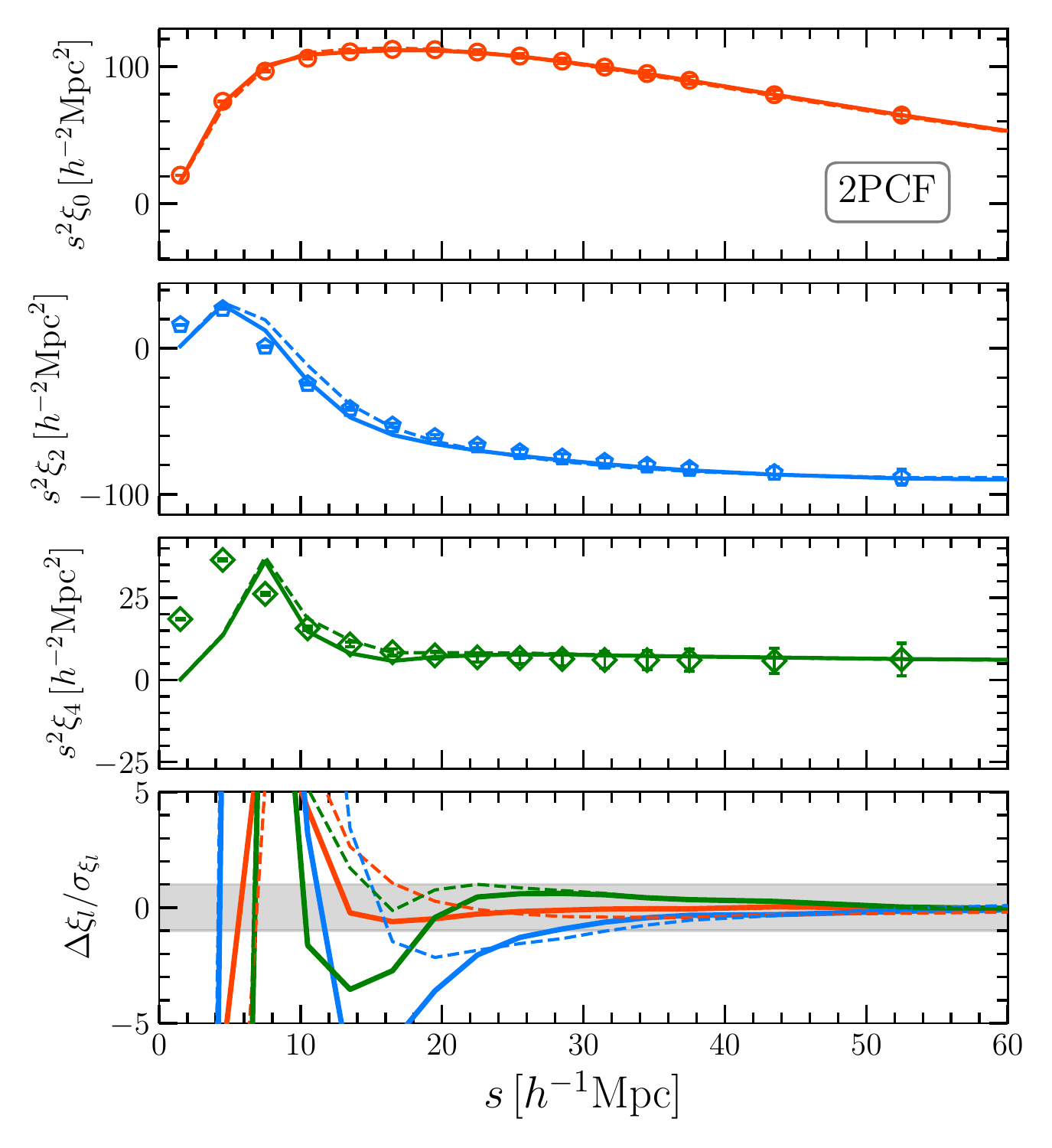}
     \end{tabular}
     \caption{Comparison of monopoles ($\xi_0$, orange), quadrupoles ($\xi_2$, blue) and hexadecapole ($\xi_4$, green) between models (lines) and simulations (data points with errors). The first five main panels (DS1-5) show the CCF multipoles of each DS quintile, whereas the last panel at the bottom right (labeled 2PCF) shows the multipoles for the galaxy two point correlation function- Solid lines represent model predictions with the Gaussian streaming model with all its ingredients measured from simulation. Dashed lines shows our best-fit model with the free parameters that account for the density-velocity coupling and velocity dispersion. The error bars represent 1$\sigma$ standard deviations for 1 simulation box of ($1.5\,h^{-1}{\rm Gpc}$)$^3$ from the 300 mock realisations. The bottom sub-panels show the deviation between the model and the measurements from simulations, in units of the dispersion in the simulation. The grey-shaded area shows the $1\sigma$ region. The amplitudes of the galaxy 2PCF multipoles have been scaled by a factor of $s^2$ for clarity.}
     \label{fig:multipoles}
\end{figure*}

\begin{figure}
    \centering
    \includegraphics[width=\columnwidth]{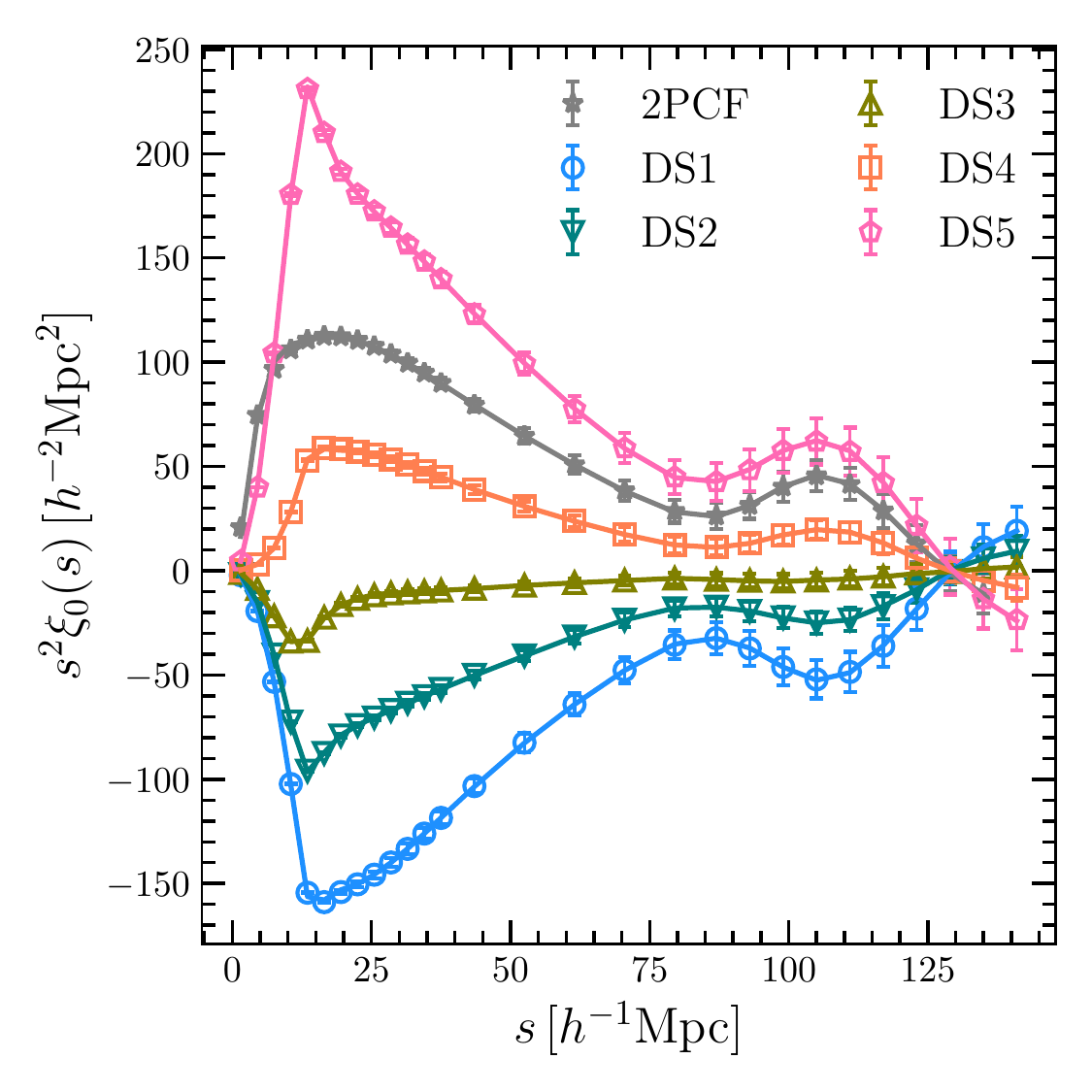}
    \caption{Monopoles $s^2\xi_0$ of the redshift-space cross-correlation functions between positions in different density quintiles (labeled as DS1-5) and the entire galaxy sample. The BAO feature is seen as prominent peaks or dips around $\sim 105\,h^{-1}{\rm Mpc}$ for all quintiles. The case for the two-point correlation function is labeled 2PCF. Data points with errors are measurements from simulations. Solid lines are predictions from the Gaussian streaming model, with all its ingredients measured from the simulation (see also Fig~\ref{fig:sigma_pi_largescales} for their 2D versions).}
    \label{fig:monopoles_s2}
\end{figure}

\subsection{Velocity distribution functions in different environments}
\label{sec:velocity_PDF}

With the above setup, we can measure the velocity distribution function at different scales\footnote{The relevant quantity that enters the streaming model equation is the line-of-sight velocity distribution, which receives contributions from both the radial and tangential components, and that is what we have used to plug into the GSM. For this part of the paper, we choose to show the velocity PDFs for the radial component to make them better correspond to the streaming velocity profiles that we will show later. We have checked that line-of-sight velocity PDFs are qualitatively similar to the PDFs of the radial components.}.

On large scales, the PDFs of these radial velocities are expected to be  Gaussian, but it is the most interesting at small scales where non-linear evolution is prominent. An example at $r=15\,h^{-1}{\rm Mpc}$ is presented in Fig.~\ref{fig:velocity_PDF}. We can see that the PDFs for each quintile are well fitted by a Gaussian function of different widths (DS1-5 of Fig.~\ref{fig:velocity_PDF}). The mean of the Gaussian, $\bar v_r$, is usually offset from zero. It monotonically decreases from DS1 ($\bar v_r=$~131.2 km/s) to DS5 ($\bar v_r=-63.9$~km/s). This indicates that the radial velocity is turning from outflow at DS1 to infall at DS5. This is expected, because the local density increases from DS1 to DS5 i.e. DS1 is underdense, similar to the case of voids; DS5 is overdense, corresponding to clusters; and DS2-4 are their transition phases. This is clearly seen in the full radial velocity profiles around each density quintile Fig.~\ref{fig:velocity_profiles} (data points), where the outflow for DS1-2, and the infall for DS5 continue up to large radii.

In contrary, at the same scale, the full pairwise velocity distribution, the quantity relevant for modelling the redshift-space 2PCF, is highly skewed (bottom-right panel of Fig.~\ref{fig:velocity_PDF}). The mean pairwise velocity is also highly negative, indicating that the pairs are predominantly sampling the high density regions. This is expected, as pair-counting at this small scale is heavily weighted by high density regions. This leads to a large velocity dispersion as the dominant feature of the pairwise velocity distribution, which manifests itself as Fingers-of-God \citep[FoG,][]{Jackson1972} at those scales in the redshift-space 2PCF. This is naturally avoided in DS, as we will see in the next sub-section.

The non-Gaussianity of the PDF for the pairwise velocity violates the Gaussian assumption necessary for modelling the 2PCF with the GSM. It is therefore unsurprising that the model becomes inaccurate for the 2PCF at small scales \citep[e.g.][]{Reid2011}. The Gaussian nature of the velocity PDFs for DS1-DS5 ensures that such key condition for the GSM is met, thus promises a better performance, as we will show in the next sub-section.

Perhaps more importantly, by splitting the non-Gaussian density PDF, and cross-correlating local densities with the entire galaxy field, we will naturally capture the non-Gaussianity of the density field. This lays the foundation for a possible gain of cosmological information over standard two-point statistics, as we will demonstrate in Sec.~\ref{sec:constraints}.

For completeness, we show the velocity dispersion profiles in Fig.~\ref{fig:sigma_los_vs_r}. We note that $\sigma_{\parallel}(r, \mu)$ is necessarily a function of $r$ and $\mu$. Its variation with $\mu$ is relatively strong for the pairwise velocities in the 2PCF and should not be neglected for the modelling. After averaging over $\mu$, all the $\sigma_{\parallel}$'s in DS appear to be flattened on large scales and converge to the same value. This offers possibilities to use a single free parameter to capture their amplitudes on large scales (see Sec.~\ref{sec:constraints}). The dispersion decreases with scale for DS1-4 as these are underdensities at small scales. Such a trend is the opposite for DS5, which corresponds to over-densities. For the 2PCF, the dispersion decreases with scale until the 1-halo term kicks in at the smallest scales ($\sim 1\,h^{-1}$Mpc), which causes the velocity dispersion to increase sharply. This is the main drive for the FoG for the 2PCF. The velocity distribution function is expected to be highly non-Gaussian, as already seen at $r=15~h^{-1}$Mpc in Fig.~\ref{fig:velocity_PDF}.

\subsection{Performance of the GSM for splitting densities \& the 2PCF}
\label{sec:GSM_DS}

With the promising behavior of the velocity PDFs in the different DS quintiles, we will test the performance of the GSM in the simulation in this section. At this stage, we will take all the ingredients of the model [i.e. $\xi(r |\Delta^i)$, $v_r(r|\Delta^i )$, $\sigma_{\parallel}(r, \mu,|\Delta^i)$ for DS and $\xi(r)$, $v_r(r)$, $\sigma_{\parallel}(r, \mu)$ for the 2PCF] from our mock galaxies. This allows us to focus on testing the validity of the Gaussian assumption. We will relax some of these model conditions in the next section for cosmological constraints.

In simulations with periodic boundary conditions, the cross-correlation functions can be estimated without the 
use of a random catalogue as
\begin{equation}
    \xi(s, \mu |\Delta^i) = \frac{D_1 D_2(s, \mu)}{N_1 N_2}\frac{V_{\rm box}}{\delta V(s, \mu)} -1,
\end{equation}
where $D_1 D_2(s, \mu)$ are the pair counts between DS centres and galaxies, $N_1$ and $N_2$ are the total number of DS centres and galaxies, respectively, $V_{\rm box}$ is the volume of the simulation box, and $\delta V(s, \mu)$ is the volume of a bin centred on $(s, \mu)$ with radial thickness $\mathrm{d}s$, which can be calculated analytically as
\begin{equation}
    \delta V = \frac{4\pi}{3} \frac{(s + \mathrm{d}s/2)^3 - (s - \mathrm{d}s/2)^3}{N_{\mu}} ,
\end{equation}
where $N_{\mu}$ is the total number of $\mu$ bins. For each DS quintile, we measure the real-space quantities $\xi(r |\Delta^i)$, $v_r(r|\Delta^i )$, $\sigma_{\parallel}(r, \mu,|\Delta^i)$, as well as the redshift-space CCF $\xi^s(s,\mu|\Delta^i)$ from our mock galaxies, taking the average over the 300 simulation boxes. We use radial bins of $3\,h^{-1}{\rm Mpc}$, $9\,h^{-1}{\rm Mpc}$ and $6\,h^{-1}{\rm Mpc}$ widths at scales of $0-39\,h^{-1}{\rm Mpc}$, $39-84\,h^{-1}{\rm Mpc}$ and $84-150\,h^{-1}{\rm Mpc}$, respectively, as well as $\mu$ bins of a constant width of 0.02 between -1 and 1. We try to minimise the number of radial bins while being able to sample important features in the correlation function such as the BAO. This is necessary to make sure that our data vector is sufficiently smaller than number of independent mocks used to construct our covariance matrix (see also the discussions about covariance matrix in Sec. 5). We plug the model ingredients into Eq.~(\ref{eq:GSM_integral}) to obtain the predicted redshift-space CCF. These are compared with direct measurements in redshift space in Figs.~\ref{fig:sigma_pi_largescales} \& \ref{fig:sigma_pi_smallscales}. The CCF for DS1 resembles the void-galaxy CCF \citep[see e.g.][]{Paz2013, Paillas2017, Correa2019}. This is not a surprise, since the algorithms designed for identification of spherical underdensities in the cited works use a top-hat filter for the definition of the voids, in a similar fashion as we have done by using a top-hat filter for the density splitting. DS5 shows an overdensity of galaxies near the centre, similar to the case of cluster-galaxy CCF \citep[e.g.][]{Seldner1977, Lilje1988, Lilje1989, Croft1999,Yang2005,Zu2013, Mohammad2016}.

It is striking to see that despite the big variation of distortion patterns for DS across different density quintiles and scales, the agreement between the model (dashed lines) and simulation (solid lines) is nearly perfect at all scales. In contrary, for the 2PCF, deviations between the model and simulation are obvious even at  scales of a few tens of $h^{-1}$Mpc when $\mu$ is near unity, i.e. along the line of sight direction (Fig.~\ref{fig:sigma_pi_smallscales}) . 

The distortion patterns are similar between the CCFs and the 2PCF on large scales, where both appear to be flattened. At scales $s\simeq 30\ h^{-1}$Mpc, however, the CCF becomes elongated for DS1 and DS2, while the 2PCF remains flattened at the same scale, with the exception of the FoG feature near $\mu=1$, which causes deviations between the model and the simulation results. At even smaller scales, the distortion of CCF for DS1 and DS5 becomes very weak, while the 2PCF is dominated by the elongated FoG. The GSM clearly fails to capture the FoG feature (bottom-right panel of Fig.~\ref{fig:sigma_pi_smallscales}). This again indicates that the dispersion is insufficient to describe the non-Gaussian PDF of the pairwise velocities. In contrast, there is no FoG feature in the CCF of the different DS quintiles, and this allows the GSM to perform better at the same scale. This is expected, as the centres of the CCFs do not usually correspond to a galaxy. The selection of centres based on the top-hat smoothed density field at $15\ h^{-1}$Mpc naturally averages out the FoG. This is also consistent with the fact that the velocity PDF for each quintile of DS is very close to Gaussian (Fig.~\ref{fig:velocity_PDF}).

To characterize the performance of the GSM more quantitatively, we extracted the  monopole, quadrupole, and hexadecapole [$\xi_{0,2,4}(s)$] of the correlation functions inferred from our simulations following
\begin{equation} \label{eq:multipoles}
    \xi_{\ell}(s) \equiv \frac{2\ell + 1}{2} \int_{-1}^{1} L_\ell(\mu)\xi^s(s, \mu) \mathrm{d}\mu \ ,
\end{equation}
where $L_\ell(\mu)$ is the Legendre polynomial of order $\ell$. Fig.~\ref{fig:multipoles} shows a comparison of the 
simulation results (symbols) and the corresponding model predictions (solid lines). The agreement for $\xi_{0, 2, 4}(s)$ is excellent at all scales for the DS results, with small deviations near the smoothing scale where the slopes of the multipoles are steep. Instead, the multipoles of the standard 2PCF show noticeable deviations from the theory predictions at scales below 20 $h^{-1}$Mpc. This comparison becomes clearer when we quantify the agreement by taking the fractional difference between the model predictions and the simulations in terms of the 1-$\sigma$ errors  corresponding to a volume of (1.5$h^{-1}$ Gpc)$^3$. We can see that the ratios at around $\sim 15\ h^{-1}$Mpc for DS may sometimes fluctuate beyond the 1-$\sigma$ level, but overall stay within the errors. For 2PCF however, the differences between the model and the results from the simulations are noticeable at scales $s < 30\,h^{-1}{\rm Mpc}$ for $\xi_2$, and become stronger at smaller scales. 

These results for the 2PCF are similar to those reported in \cite{Reid2011} with similar mocks, where a few percent accuracy was achieved at $\sim 25\ h^{-1}$Mpc for the quadrupole  \citep[see also][]{Chen2020}. The performance of the GSM is better for most density quintiles, and for almost all scales. The exceptions are: 1. for the quadrupole in DS1 and DS5 at $s\sim 15\ h^{-1}$Mpc (solid blue lines in the lower panels corresponding to DS1 and DS5 in Fig.~\ref{fig:multipoles}). This is likely due to taking ratios with respect to $\xi_2(s)$ which crosses zero at those scales; 2. the monopole in DS1 also fluctuates outside the 1$\sigma$ error at $s<20\ h^{-1}$Mpc (solid orange line at bottom panel corresponding to DS1). This is possibly due to the sparsity of galaxies for DS1 at those scales.

As can be seen in Fig.~\ref{fig:sigma_pi_largescales}, the BAO feature is clearly present in the CCF of all quintiles at $s\sim 105\ h^{-1}$Mpc, especially when $\mu \sim 0$, i.e. the direction perpendicular to the line of sight. To better visualise this large-scale feature, Fig.~\ref{fig:monopoles_s2} shows the monopoles of the CCFs of all DS and the standard 2PCF, rescaled by $s^2$. The BAO feature is clearly seen for all quintiles as a peak (DS4-5) or dip (DS1-3), with an amplitude proportional to that of the overall CCF. As we will see later, accessing this information  significantly improves the constraints on the 
geometrical (AP) distortion parameters. While the information from the BAO feature has been commonly used in RSD analyses of the galaxy 2PCF (e.g., \citealt{Sanchez2016, Beutler2017, Bautista2021}), it has often been ignored in void RSD studies 
\citep[but see][for a demonstration of how voids can help to optimise the BAO measurement in a galaxy sample]{Zhao2020}. 
This is partly due to the fact that many authors choose to re-scale the void-galaxy profiles by the void radius \citep[e.g.][]{Cai2016, Hamaus2020}. {This rescaling helps to boost the signal from the void ridge, which is useful for gravitational lensing analyses around voids \citep[e.g.][]{Higuchi2013,Krause2013, Melchior2014, Clampitt2015, Cautun2017, Cai2017, Sanchez2017, Raghunathan2020}. It is also convenient when applying the Alcock-Paczynski test for voids \citep{Sutter2012a}. However, it effectively washes out any BAO signal present on large scales.}

\subsection{Other RSD models suitable for split densities}
\label{sec:model_comparison}

\begin{figure*}
     \centering
     \begin{tabular}{cc}
        \includegraphics[width=0.45\textwidth]{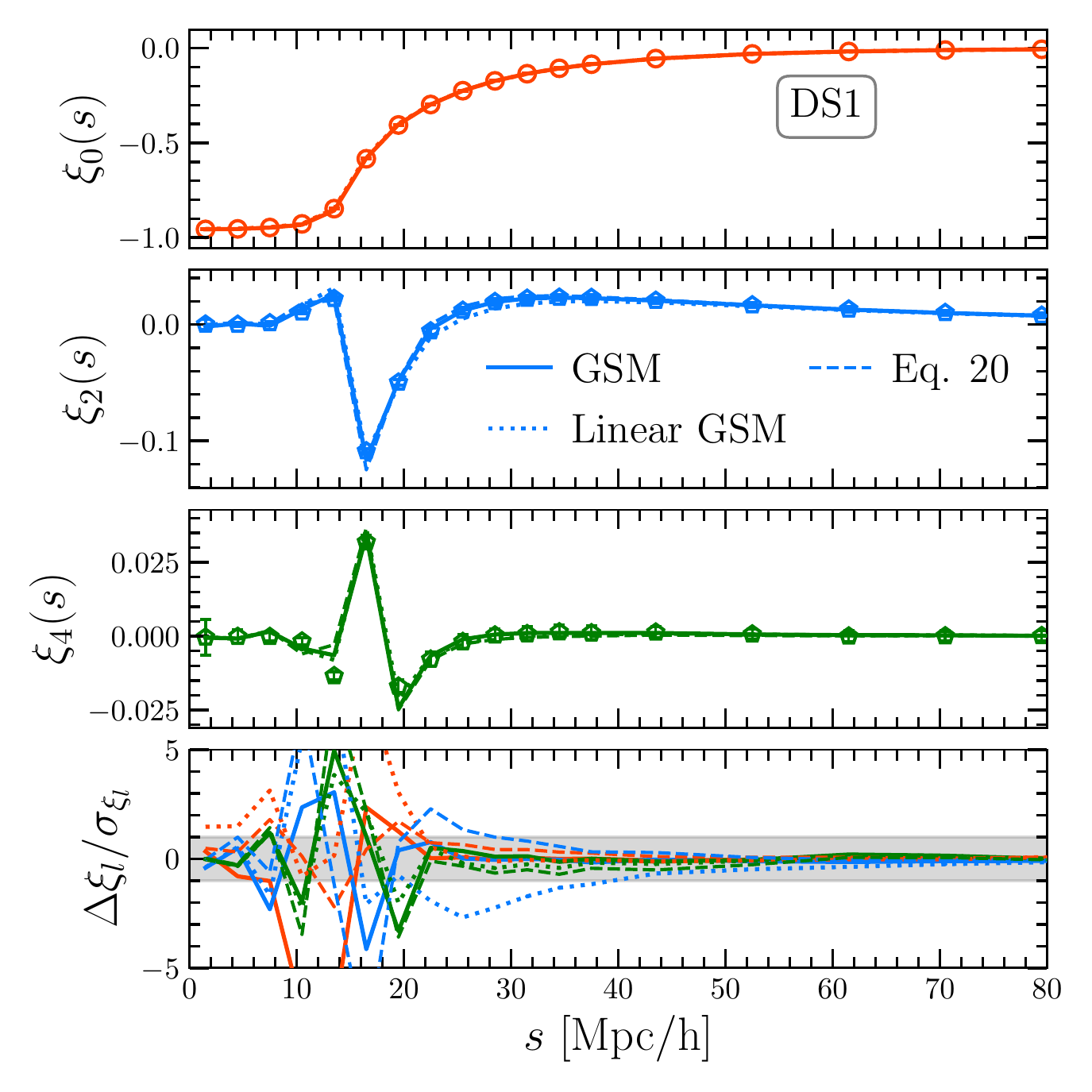} &
        \includegraphics[width=0.45\textwidth]{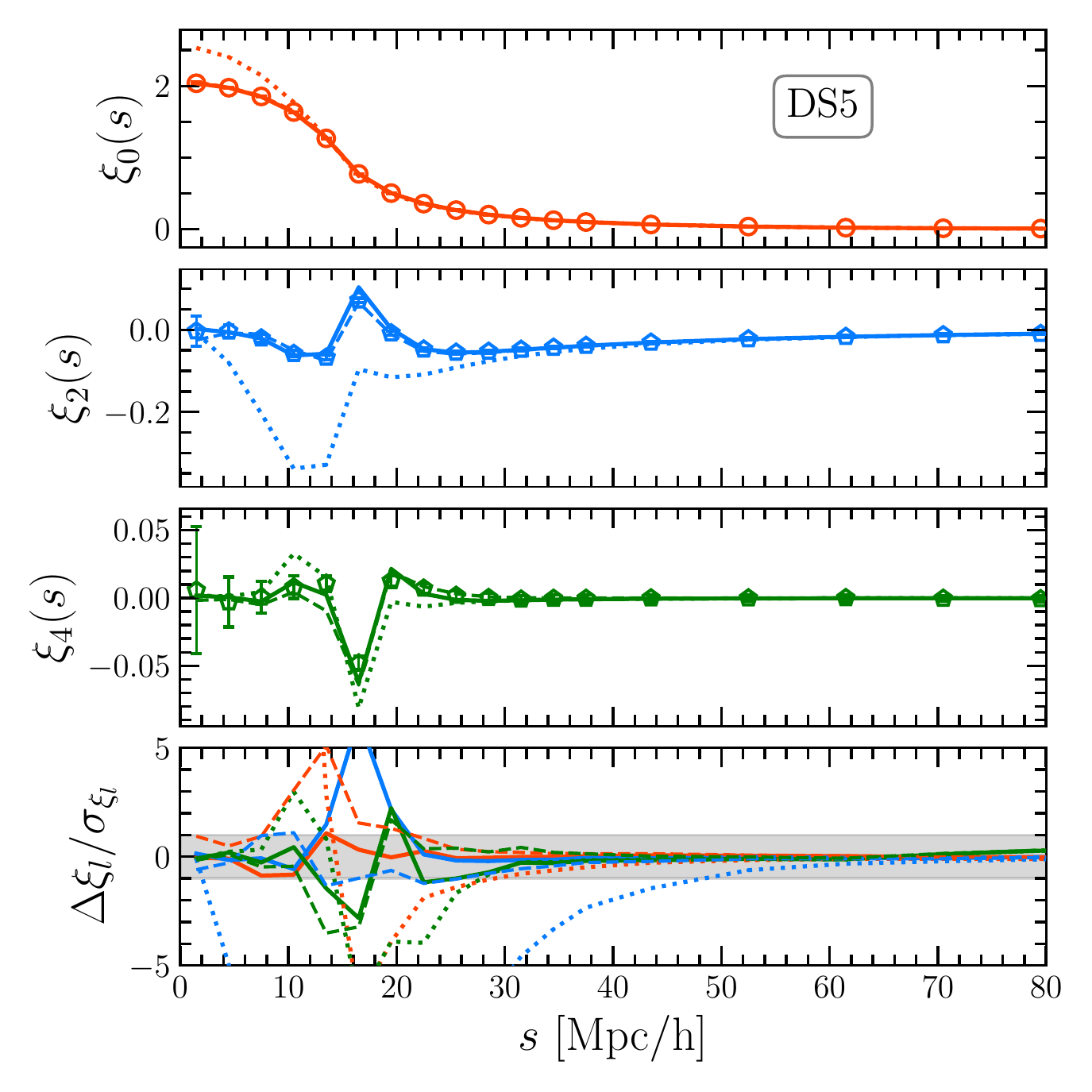}
     \end{tabular}
     \caption{Similar to Fig.~\ref{fig:multipoles}, showing multipoles for the cross-correlation functions in the extreme DS quintiles (voids on the left and clusters on the right-hand side panels, respectively). The data points and errors are measurements from the mock galaxy catalogues. The solid and dashed lines show predictions from the Gaussian streaming model (Eq.~\ref{eq:GSM_integral}) and Eq.~(\ref{eq:GSM_NP}) (with Eqs.~\ref{eq:full_Jacobian} \& \ref{eq:dispersion_only}), where all model ingredients are fully known. The dotted lines show predictions from the Gaussian streaming model but with the linear density-velocity coupling assumption (Eq.~\ref{eq:vel_den_coupling_1.5}).}
     \label{fig:multipoles_GSM_vs_NP19}
\end{figure*}

The modelling of RSD around different density quantiles has no fundamental difference from the modelling of the void-galaxy or cluster-galaxy CCFs \citep{Lilje1991, Croft1999, Zu2013, Mohammad2016}. \cite{Hamaus2015} employed the Gaussian streaming model for modelling RSD around voids and, together with the assumption that the density and streaming velocity are linearly coupled, they found that the model is biased at the $\sim$10 percent level for both AP and growth parameters, but their application of the model to BOSS CMASS mock galaxies and real data seems to be consistent with their fiducial cosmology \citep{Hamaus2016}. Accounting only for the streaming velocity, \cite{Cai2016} derived the full expression for RSD around voids, and for spherical density profiles in general
 (Eqs.~1-4 of their paper), which we summarise here \footnote{\citep{Cai2016} noted $\xi$ as $\delta$ because the cross-correlation function is indeed the same as the stacked density profile, which is usually written as $\delta(r)$. There was a typo where they should have noted the redshift-space coordinate as ${\bf s}$, rather than $\bf r$, which was pointed out in \cite{Nadathur2019a}.}: 
\begin{equation}
\label{eq:full_Jacobian}
1+\xi^s({\bf s})=[1+\xi^r( r)]\left[1+(1-\mu^2)\frac{1}{aH}\frac{v(r)}{r}+\mu^2\frac{1}{aH}\frac{\partial v}{\partial r}\right]^{-1}.
\end{equation}
The above can be applied to map between the real and redshift-space CCFs around spherical regions. When expanding to the linear order, we have 
\begin{equation}
\xi^s({\bf s})=\xi( r)-(1-\mu^2)\frac{1}{aH}\frac{v(r)}{r}-\mu^2\frac{1}{aH}\frac{\partial v}{\partial r}. 
\end{equation}
This takes the same form as expanding the Gaussian streaming model to the linear order, keeping only the first derivative, i.e. neglecting terms that are related to the velocity dispersion [e.g. Page 289 of \cite{Peebles1980}, Eqs.~24-25 of \cite{Fisher1995},  Eqs.~16-17 of \cite{Reid2011}]. When further assuming linear coupling between the density and velocity, i.e. Eq.~(\ref{eq:vel_den_coupling_1.5}), we have
\begin{equation}
\xi^s({\bf s})=\xi( r)+ \frac{1}{3}f\bar{\xi}(r)+f\mu^2[\xi(r)-\bar{\xi}(r)].
\end{equation}
This is the linear theory expression for CCFs \citep{Kaiser1987, Cai2016}\footnote{Note that, although \cite{Nadathur2019a} assume the linear coupling of Eq.~(\ref{eq:vel_den_coupling_1.5}), they expand Eq.~(\ref{eq:full_Jacobian}) and keep terms such as $\xi\bar{\xi}$ and $\xi{\xi}$ ($\xi \Delta$ and $\xi\delta$ in their notation). This to us is 2nd order in mathematical term, and is no longer a linear model. 

It was also noted in \citet{Nadathur2019a} and \citet{Nadathur2020} that the GSM does not reduce to linear theory at the limit when the dispersion is small. However, it has been shown in \cite{Fisher1995} that the GSM naturally reduces to linear theory when allowing scale-dependent velocity dispersion. We have also verified numerically that when $\sigma_{{\parallel}}$ is very small, the GSM reduces to the case where there is only streaming velocity, i.e. Eq~(\ref{eq:full_Jacobian}). Further, if a linear coupling between the density and velocity is assumed, i.e. a pure linear system, then the GSM does get back to it.}.

All the above expressions account for only the streaming velocity. To include velocity dispersion, \cite{Nadathur2019a} convolve Eq.~(\ref{eq:full_Jacobian}) with the velocity dispersion as
\begin{align}
\label{eq:GSM_NP} \nonumber
    1 + \xi^s(s_{\perp}, s_{\parallel}) = &\int \left\{1 + \xi^s\left[s_{\perp}, s_{\parallel}-v_{\parallel}/(aH)\right]\right\}\\
     &\mathcal{P}(v_{\parallel}, \vec{r}) \mathrm{d} v_{\parallel} \ ,
\end{align}
where $\xi^s(\bf s)$ takes the expression from Eq.~(\ref{eq:full_Jacobian}), and $\mathcal{P}(v_{\parallel},\vec{r})$ is assumed to be Gaussian with a zero mean, i.e.
\begin{equation}
\label{eq:dispersion_only}
    \mathcal{P}(v_{\parallel},r, \mu) = \frac{1}{\sqrt{2\pi} \sigma_{\parallel}(r,\mu)}\exp \left[- \frac{v_{\parallel}^2}{2 \sigma_{\parallel}^{2}(r,\mu)}  \right]\ .
\end{equation}
This is similar to the well-known Gaussian streaming model (i.e. Eq.~\ref{eq:GSM_integral}), except that the streaming velocity is explicitly taken off from the exponential part, and being left to be accounted for by the Jacobian, i.e. the mapping from real to redshift-space CCF in Eq.~(\ref{eq:full_Jacobian}).\footnote{We notice that the $\mu$-dependence was missed out in the expression of $\mathcal{P}(v_{\parallel})$ in  \citep{Nadathur2019a, Nadathur2020}. Without the explicit $\mu$-dependence for the dispersion term, the impact of $\sigma_{\parallel}$ is slightly weakened.} 

We used our simulations to compare the performance of the this prescription (i.e. Eq.~\ref{eq:GSM_NP} with Eqs.~\ref{eq:full_Jacobian} \& \ref{eq:dispersion_only}) with the GSM for DS, taking all the ingredients of the model from our mock galaxies, i.e. $\xi(r |\Delta^i)$, $v_r(r|\Delta^i)$ and $\sigma_{{\parallel}}(r,\mu|\Delta^i)$. Fig.~\ref{fig:multipoles_GSM_vs_NP19} presents 
this comparison for DS1 and DS5. We can see that, although small deviations from the quadrupole can be spotted between 20-30 $h^{-1}$Mpc, and from the measured monopole in DS5 below $10\ h^{-1}$Mpc, their overall predictions are very similar across all scales for both DS1 and DS5.

Note that the solid and dashed lines are the most optimistic scenario where all the ingredients of the model are assumed to be known. Many studies have assumed linear coupling between the density and velocity (Eq.~\ref{eq:vel_den_coupling_1.5}) \citep[e.g.][]{Hamaus2016,Hamaus2017, Hamaus2020, Nadathur2019b, Nadathur2020}. To test the accuracy of this assumption, we show in dotted lines the GSM
with this approximation, labeled as `Linear GSM'. We can see that for DS1, Linear GSM shows larger deviation than GSM at small scales for the monopole, and slightly underpredicts the quadrupole at intermediate scales, but the overall agreement with the simulation is fairly good. This is expected, as DS1 is similar to the void-galaxy cross-correlation function, where non-linearity around voids is thought to be weaker. This does justify the linear coupling assumption for voids so far, as has been adopted in the literature, but at the per cent-level precision promised by future surveys, it is important to model this coupling relationship more accurately. 

For DS5, however, the deviations between the linear models and the measurements are much larger at a few tens of $h^{-1}$Mpc. This is because the density contrast for DS5 is larger, similar to the cluster-galaxy CCF, thus the growth of structure is expected to be more non-linear. The failure of the linear coupling model is also evident from Fig.~\ref{fig:velocity_profiles}, where the linear velocity profiles over predict the version from the simulations at small scales. It is also worth nothing that even though DS2-4 are transition stages between voids and clusters, thus having more moderate density constrasts, neither the linear nor the empirical density-velocity couplings are sufficient to correctly describe their radial velocity profiles in Fig.~\ref{fig:velocity_profiles}. After having calculated the radial velocity profiles directly from the dark matter distribution, we have found that the velocities sampled by galaxies around DS centres are biased against that of dark matter at the level of tens of km/s. We believe that this is due to the sparsity of our galaxy samples. While this velocity bias is sub-dominant for DS1 and DS5, it becomes relatively more severe for DS3 and DS4, which have smaller velocity magnitudes at small scales. A galaxy sample with a higher number density might be able to reduce this velocity bias and help reconcile the empirical model with the measurements. Generally speaking, we believe that there is substantial room for improving the prediction of velocity statistics for DS. In a sense, the free parameters from the empirical model we introduced are absorbing some of the uncertainties due to the assumptions in our modelling (e.g. the linear bias assumption). It is likely that many of the developments for the calculation of pairwise velocity statistics for the 2PCF \citep[e.g.][]{Vlah2016, Chen2020} will also be useful for DS. We plan to elaborate more on this problem in future work.

In summary, we have shown that:\\
 1. given the same conditions, and assuming that the model ingredients are fully known, the GSM and Eq.~(\ref{eq:GSM_NP}) lead to almost equivalent predictions. These results differs from what was reported in \cite{Nadathur2019a, Nadathur2020} for voids, where significant deviations for the GSM model were reported. \\
 2. the assumption of a linear coupling between density and velocity is not accurate for the statistical errors of our concern, i.e. a volume of (1.5$h^{-1}$ Gpc)$^3$. \\
We therefore will use the GSM model as default for our analysis. Eq.~(\ref{eq:GSM_NP}) should perform similarly, as demonstrated in this section.
 \\

\begin{figure*}
     \centering
     \begin{tabular}{cc}
     \hspace{-0.71 cm}
        \includegraphics[width=0.55\textwidth]{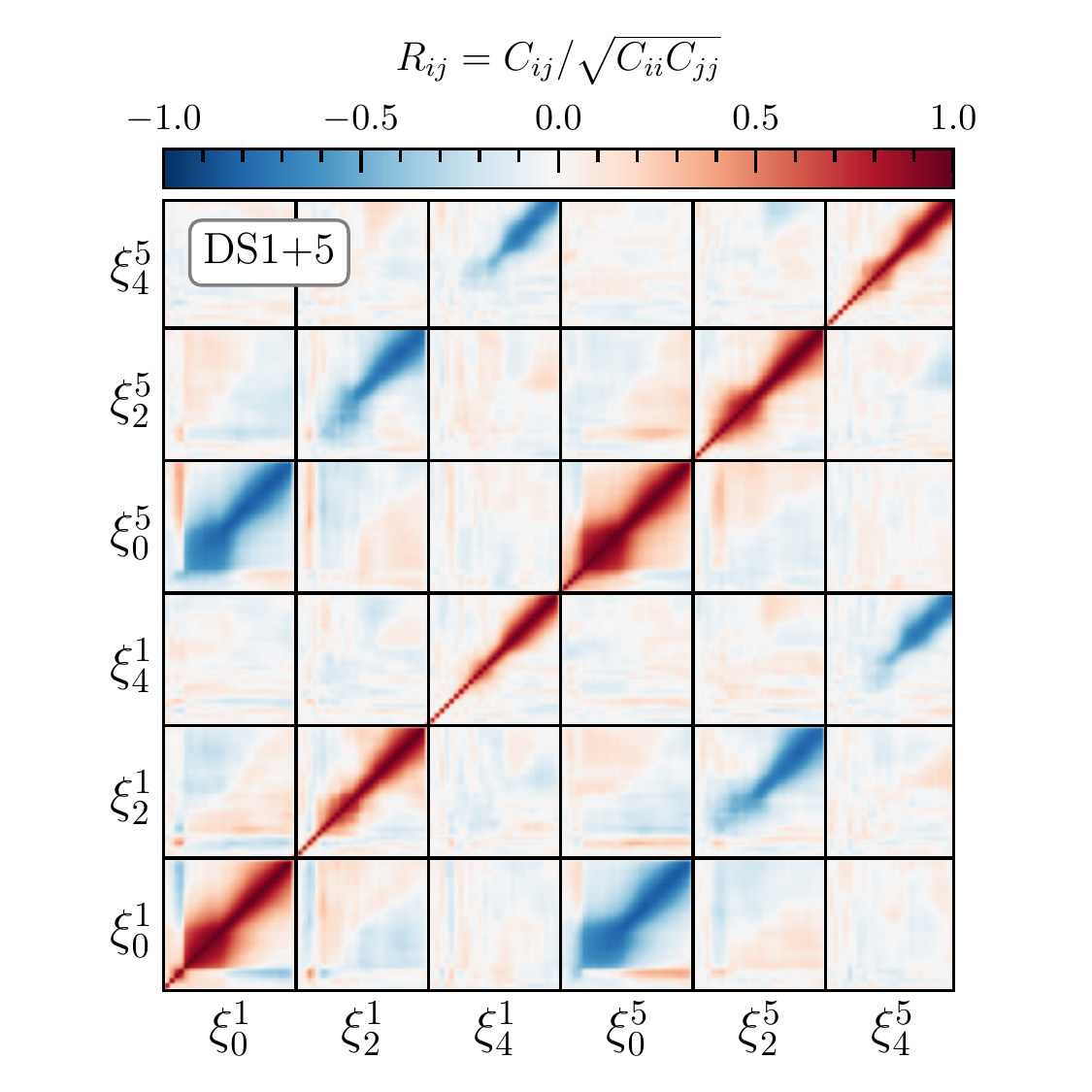} &
         \hspace{-1.4 cm}
        \includegraphics[width=0.55\textwidth]{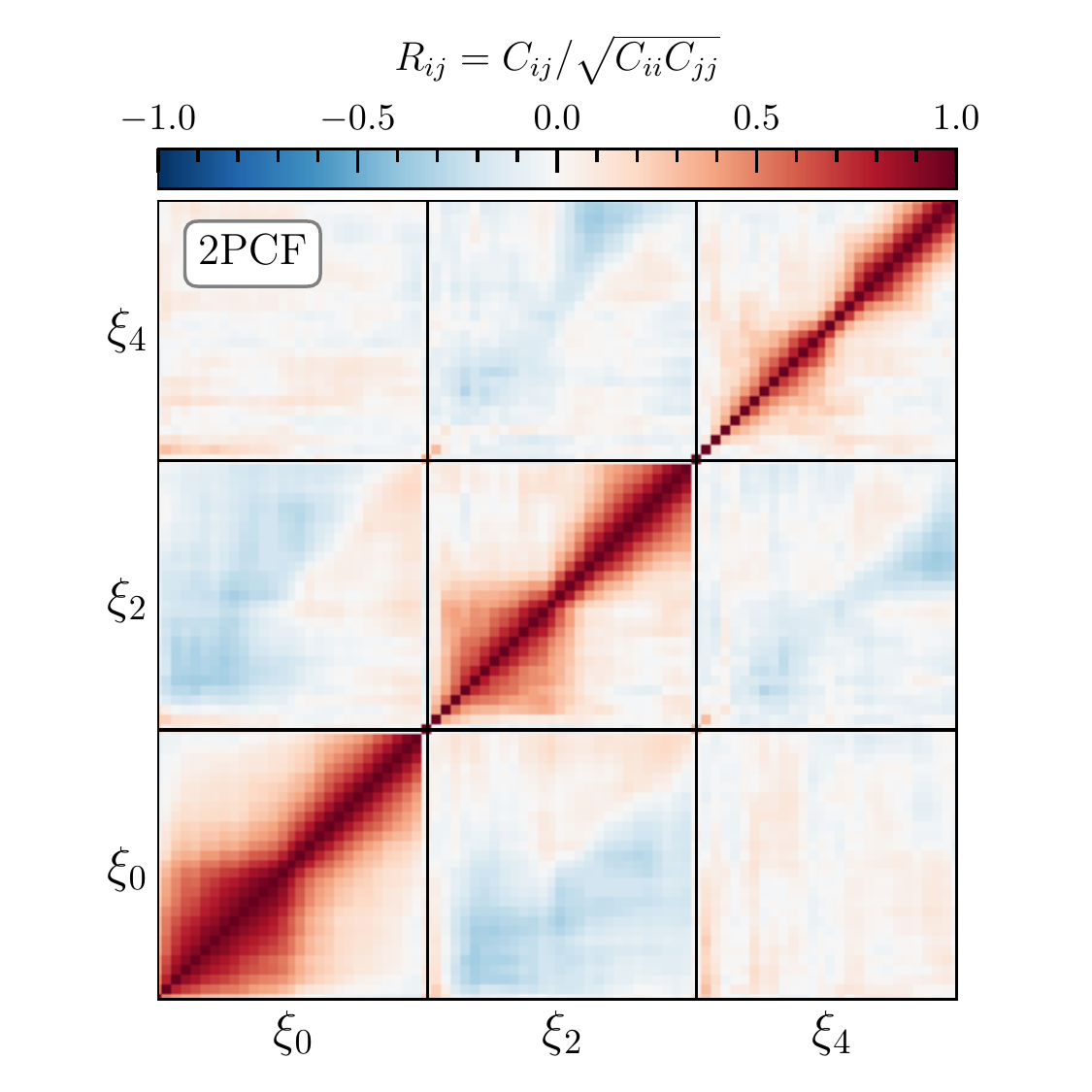}
     \end{tabular}
     \caption{Cross-correlation coefficients for the covariance matrix defined by Eq.~(\ref{eq:covariance}), for the Density Split (left) and the 2PCF (right). The solid lines divide the regions where each multipole contributes. The superindices in the DS covariance indicate the contributing quintile.  Significant contributions can be seen from the first $(\xi_{l}^{1})$ and fifth $(\xi_{l}^{5})$ quintiles, respectively. There are no strong correlations among $\xi_0$, $\xi_2$ and $\xi_4$ within each quintile, but there is significant anti-correlation between the multipoles from the first and fifth quintiles (voids and clusters) in DS.}
     \label{fig:covariance}
\end{figure*}

\section{Constraining cosmology with density split RSD}
\label{sec:constraints}

We have demonstrated in the previous section that the GSM works well for all CCFs in DS, and better than for the standard 2PCF. The main reason is that the key condition for the model -- the Gaussianity of the density and velocity field, is not valid at small scales for the 2PCF. This implies that the 2PCF is insufficient to capture all the information in this non-Gaussian regime. With DS, however, the Gaussianity condition continues to be valid. It is compelling to ask if the combination of the whole series of cross-correlations, i.e. $\xi(s,\mu|\Delta^{1,2,3,4,5})$, contains any more cosmological information than the 2PCF, $\xi(s, \mu)$. Using the same model (GSM) for both the DS and 2PCF sets a common ground for us to make a fair comparison for the cosmological information contained in these statistics.

Three ingredients are needed for extracting cosmological information with the GSM using redshift-space auto- and cross-correlation functions: 
\begin{itemize}
\item The real-space two-point correlation function  $\xi(r)$; or the conditioned correlation functions $\xi(r|\Delta^i)$. 

\item The real-space pairwise velocity profile $v_{r}(r)$; or conditioned (stacked) velocity profiles $v_r(r|\Delta^i)$.  
\item The real-space pairwise velocity dispersion $\sigma_{\parallel}(r,\mu)$; or the conditioned (stacked) velocity dispersions $\sigma_{\parallel}(r,\mu|\Delta^i)$. 
\end{itemize}

Significant efforts have been made to predict these three profiles. Among them, the real-space 2PCF $\xi(r)$ is the key. It can be predicted with the halo model, perturbation theory, or emulators with simulations \citep[e.g.][]{Peacock2000, Seljak2000, Zentner2005, Angulo2010, Zhai2019}. Once  $\xi(r)$ is known, it can be used to compute $\xi(r|\Delta^i)$ if the PDF of $\Delta$ is known. There are existing model frameworks that allow us to do this \citep{Abbas2005, Friedrich2018, Gruen2015, Neyrinck2018}. With spherical dynamics, \citet{Uhlemann2016} was able to predict the full PDF of $\Delta$ accurately at $10\,h^{-1}{\rm Mpc}$, which would be sufficient for our purpose. The velocity profiles can be predicted also from $\xi(r)$ and $\xi(r|\Delta^i)$, once the coupling relationship between density and velocity is known. In addition, when galaxies are used as tracers, a bias model is needed to connect the clustering of galaxies to the clustering of dark matter. 

In this paper, with the aim to compare the constraining power on cosmology of the 2PCF versus DS, we will use the real-space 2PCF and CCF measured from the simulations as our model inputs. We then employ the empirical model with one free parameter to model the coupling between the densities and velocities, i.e. Eqs.~(\ref{eq:vel_den_coupling_1}) \& (\ref{eq:vel_den_coupling_2}). We further adopt the shape of the velocity dispersion profiles from the simulations, and allow their overall amplitudes to vary with one single free parameter $\sigma_{v}$. We will also adopt a linear bias scheme, i.e. $\xi(r)=b^2\xi_m(r)$; and $\xi(r|\Delta^i)=b\xi_m(r|\Delta^i)$, where $\xi_m(r)$ is the two-point correlation function of matter (the condition on the spherical top-hat density is the same one we introduced in Sec.~\ref{sec:rsd_with_ds}). 

There may be considerations for another bias parameter related to the densities smoothed at 15~$h^{-1}$Mpc, which we will call $b_{\rm DS}$, and this is also relevant to the discussion for the possible peculiar motions of these 15~$h^{-1}$Mpc spheres \citep{Cai2016, Massara2018}. Indeed, if one measures the ratio between $\xi(r|\Delta^i)$ and the matter 2PCF $\xi_m(r)$, we find that on large scales, the result is consistent with the product of $b_{\rm DS}$ and the galaxy bias $b$, with $b_{\rm DS}$ varying from $\sim -3$ to $\sim$3 from DS1 to DS5. However, $b_{\rm DS}$ is irrelevant for our modelling, as it is only the ratio between $\xi(r|\Delta^i)$ and $\xi_m(r|\Delta^i)$, which is the linear galaxy bias $b$, that matters. Likewise, the possible peculiar motion of those 15~$h^{-1}$Mpc spheres does not seem to have any meaningful impact on the mapping between real and redshift-space CCFs, as evident by the success of the modelling for the redshift-space CCFs shown in Fig.~\ref{fig:sigma_pi_largescales} \& Fig.~\ref{fig:sigma_pi_smallscales}.

We note that the density splitting and model ingredients for the GSM have been obtained in real space. Even with the PDF of the density and all the ingredients of the GSM model being predictable analytically, as mentioned above, we still have to face the issue that density splitting may have to be done in redshift space in observation. As the ranks of the smoothed densities by a 15~$h^{-1}$Mpc top-hat filter may be affected by redshift-space distortions, splitting density in redshift space will inevitably cause mixing between quantiles from their real space version. This issue exists similarly for identifying voids in redshift space. There are at least three known approaches to tackle this. 1. \citet{Nadathur2019a} and their later work use reconstruction to re-install the positions of galaxies in real space, and perform void finding in the reconstructed galaxy field. This seems to work well and one can do the same for DS in principle. 2. \citet{Hamaus2020} uses the Abel transformation to reconstruct the real-space profiles of voids from the projected profiles along the transverse direction. Using this approach for DS, one may not need analytical predictions for the PDF and real-space profiles. 3. \citet{Repp2020} has also developed a method to map counts-in-cells between real and redshift space. This may also be applied to reinstall the real-space PDF of the smoothed galaxy number densities, and hence allow the subsequent density splitting in real space. We will leave it for a future work to incorporate some of the above approaches to be part of the analysis. 

Similarly, density splitting may also be affected by geometry distortions, the AP effect. In principle, for each $q_{\perp}$, $q_{\parallel}$ combination, the smoothing radius should be rescaled by $q = q_{\perp}^{2/3}q_{\parallel}^{1/3}$, and thus the density splitting would need to change at each iteration of the MCMC. However, we have explicitly tested this effect by performing the density splitting with different smoothing radii, using values that are within the range of our priors for the AP parameters. We have found that the inferred real-space profiles do not change significantly in shape or amplitude within the scale ranges that are used in the fits, which means that using a fixed smoothing radius is a good approximation under these conditions. One way around this issue for future analysis with observational data would consist in predicting the real-space density PDF, together with the real-space stacked density and velocity profiles, for any given cosmology and without having to rely on simulations, thus effectively generating the density splitting model ingredients on-the-fly for each step of the MCMC chain.

{Our observables/data vectors would be the combination of all five $\xi(s,\mu|\Delta^i)$ in redshift space for DS, and $\xi(s,\mu)$ for the 2PCF. In practice, we use Eq.~(\ref{eq:multipoles}) to extract the monopole, quadrupole and hexadecapole contributions, and concatenate the multipoles into a single vector:
\begin{equation}
{\bm \xi} = \left(\xi_0, \xi_2,\xi_4\right).  
\end{equation}
In this way, for DS our data vector is the concatenation of 3 multipoles and 5 quintiles, $\bm{\xi}^{1+2+3+4+5}(s)$}. For the 2PCF, it is simply the combination of 3 multipoles $\bm{\xi}(s)$.

The model parameters for the DS fits are 
\begin{equation}
{\bm \theta}_{\rm DS}=\left(f\sigma_{12}, b\sigma_{12}, \sigma_{v}, q_{\perp}, q_{\parallel}, \nu^1, ..., \nu^n\right),
\label{eq:params_ds}
\end{equation}
where $n$ is the number of quintiles to be used in the analysis, $f(z)$ is the growth rate of structure parameter; $\sigma_{12}$ is the rms mass fluctuation in spheres with a radius of $12\ {\rm Mpc}$, expressed in $h$-independent units
\citep[see][for a discussion on why $f\sigma_{12}$ is a more adequate parameter than the standard $f\sigma_8$ parameter combination to describe the information content of RSD]{Sanchez2020}, $b$ is the linear galaxy bias, $\sigma_{v}$ is the amplitude parameter for the velocity dispersion (one free parameter $\sigma_{v}$ is sufficient to fit all quintiles simultaneously, as we have explicitly verified that the individual DS quintiles show $\sigma_{\parallel}(r, \mu)$ functions that converge at approximately the same value on large scales),
$q_{\parallel}$ and $q_{\perp}$ are the Alcock-Paczynski scaling parameters that vary independently (Eq.~\ref{eq:AP_rescaling}) to re-scale the corresponding parallel and transverse distance separation vectors. As the ratio of these AP parameters is of cosmological interest, we will also show the derived parameter $q_{\perp}/q_{\parallel}$. $\nu^i$ are the parameters specifying the density-velocity couplings from Eqs.~(\ref{eq:vel_den_coupling_1}) \& (\ref{eq:vel_den_coupling_2}). For DS, we have one $\nu$ parameter per quintile, as the radial velocity profiles can have notoriously different shapes depending on the environment. {The analysis of the 2PCF 
requires the same parameters, with the exception of a single coupling parameter $\nu$, that is
\begin{equation}
{\bm \theta}_{\rm 2PCF}=\left(f\sigma_{12}, b\sigma_{12}, \sigma_{v}, q_{\perp}, q_{\parallel}, \nu\right).
\label{eq:params_2PCF}
\end{equation}
This means that the analysis of DS has additional degrees of freedom that could potentially degrade its constraining power for cosmology}. 

\begin{figure}
    \centering
    \includegraphics[width=1.0\columnwidth]{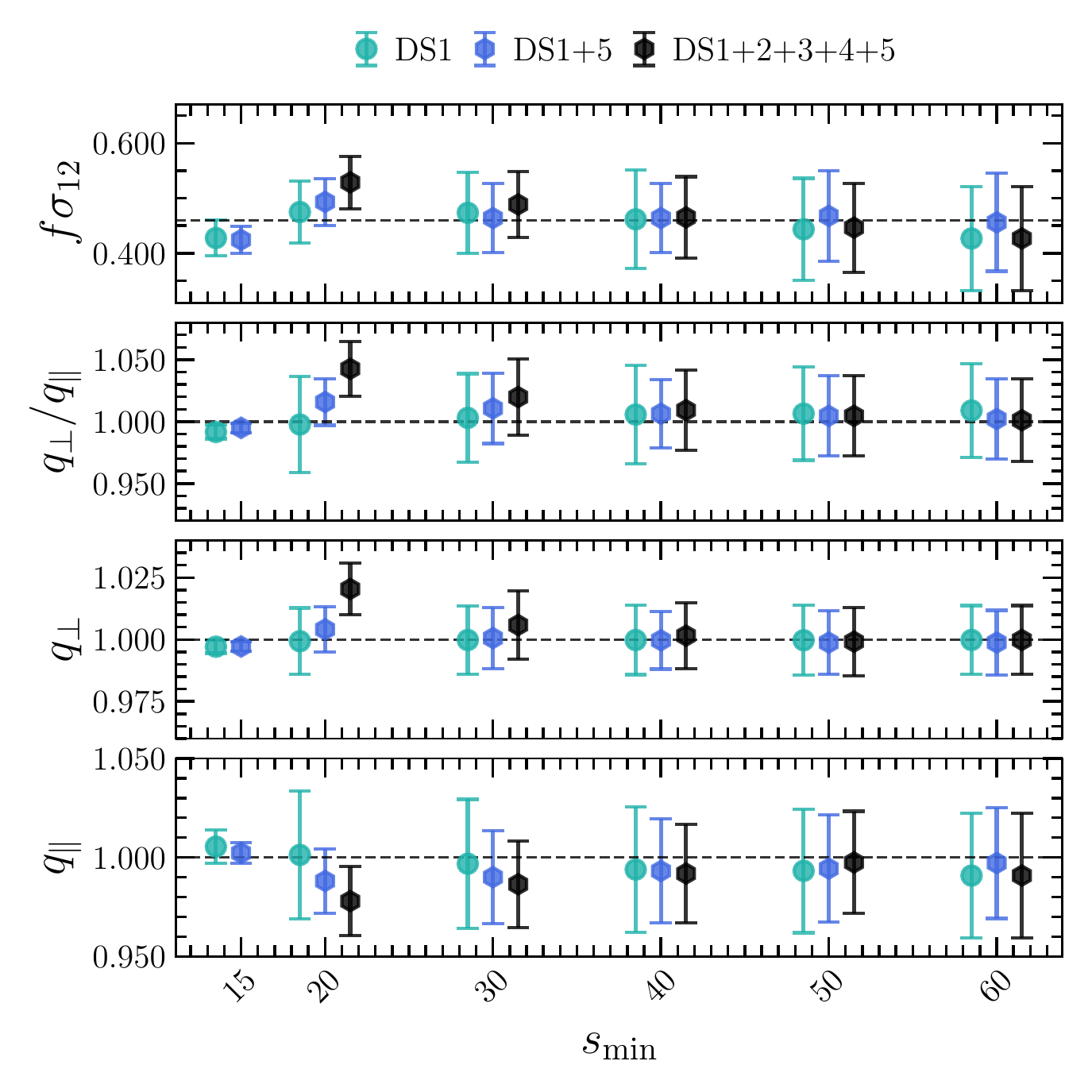}
    \caption{Cosmological parameter constraints obtained by combining CCFs from different DS quintiles (labeled in different colours and deliberately offset from each other for clarity). The markers show the median value of the marginalised posterior distribution for each parameter, while the error bars represent 68 per cent confidence levels. The horizontal dashed lines show the values from the cosmology of the simulations. Note that in order to reduce the size of the data vector, necessary to construct a reliable covariance matrix that includes all quintiles, given the limited number of simulation boxes, $\xi_4$'s are not used for the fitting in this figure. This only causes a minor increase of the errors.}
    \label{fig:constraints_scale_DS}
\end{figure}

\begin{figure*}
     \centering
     \begin{tabular}{cc}
         \hspace{-0.4 cm}
        \includegraphics[width=0.51\textwidth]{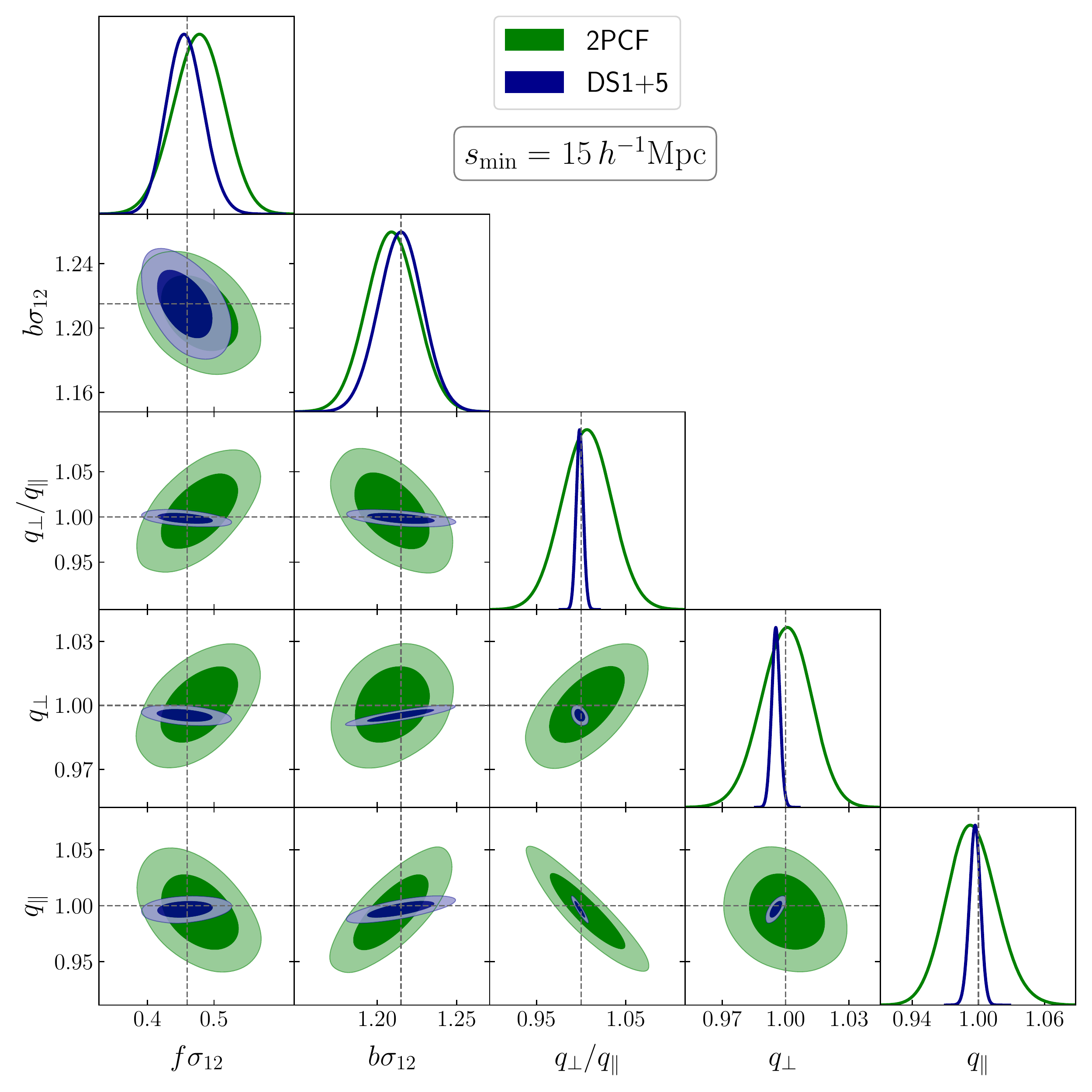} &  \hspace{-0.6 cm}
        \includegraphics[width=0.51\textwidth]{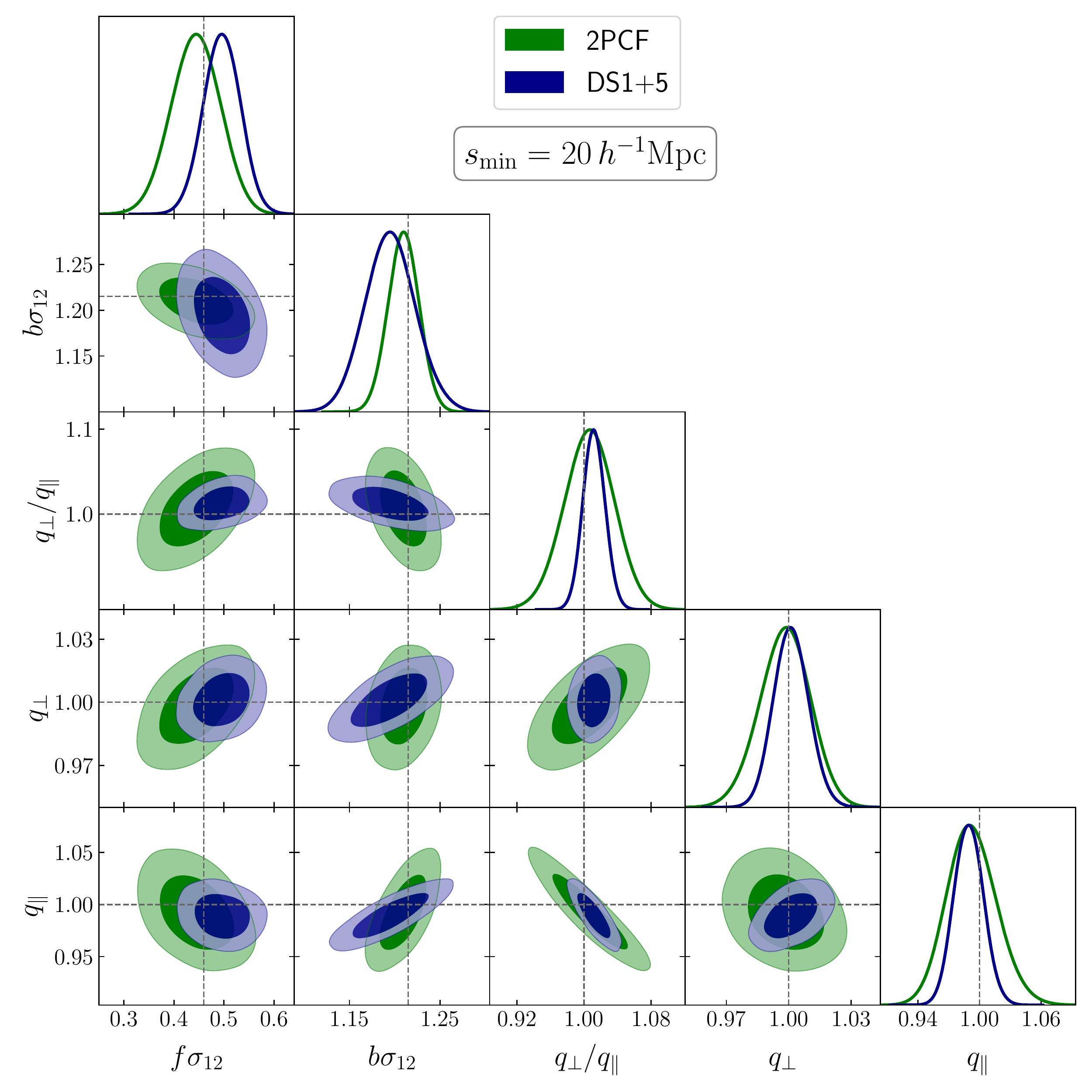}
     \end{tabular}
     \caption{Posterior distributions of the cosmological parameters inferred from two-point correlation functions (2PCF) and cross-correlation functions from density splitting (DS) (blue), with DS1+5 combined, using scales between 15-141\,$h^{-1}{\rm Mpc}$ (left-hand side side panel) and 20-141\,$h^{-1}{\rm Mpc}$ (right-hand panel). Darker and lighter shades show the 68 and 95 per cent confidence levels around the best-fit values, respectively. The dashed lines show values corresponding to the cosmology from the simulations.}
     \label{fig:posterior_15_20}
\end{figure*}

\begin{table*}
    \centering
    \caption{Comparing cosmological parameter constraints from two-point correlation functions (2PCF) and cross-correlation functions from density splitting (DS). All quoted errors (noted as $\sigma_{\rm stat}$) correspond to 68 per cent confidence levels around the best fit values $\mu$.
    }
    \label{table:HOD_parameters}
\begin{tabular}{@{}lcccccc}
\hline\hline
Parameter & Prior & 2PCF ($\mu \pm \sigma_{\mathrm{stat}}$) & 2PCF ($\sigma_{\mathrm{stat}} / \mu$) & DS ($\mu \pm \sigma_{\mathrm{stat}}$) & DS ($\sigma_{\mathrm{stat}} / \mu$) & $(\sigma_{\rm 2PCF} - \sigma_{\rm DS})/ \sigma_{\rm DS}$ \\
\hline
$s_{\mathrm{min}} = 20\,h^{-1}{\rm Mpc}$ & & &  & &  \\
$f\sigma_{12}$ & $[0.1,2.0]$ & $0.445 \pm 0.049$ & 0.110 & $0.496 \pm 0.037$ & 0.074 & 33.6\%\\ 
$b\sigma_{12}$ & $[0,3]$ & $1.210 \pm 0.017$ & 0.014 & $1.195 \pm 0.028$ & 0.023 & -39.6\%\\
$q_{\perp}/q_{\parallel}$ & &$1.007 \pm 0.029$ & 0.029 & $1.012 \pm 0.013$ & 0.013 & 121.2\% \\
$q_{\perp}$ & $[0.8,1.2]$  & $0.999 \pm 0.012$ & 0.012 & $1.001 \pm 0.009$ & 0.009 & 37.9\%\\
$q_{\parallel}$ & $[0.8, 1.2]$  & $0.991 \pm 0.025$ & 0.025 & $0.989 \pm 0.014$ & 0.014 & 76.1\%\\
\hline
$s_{\mathrm{min}} = 15\,h^{-1}{\rm Mpc}$ & & & & & \\
$f\sigma_{12}$ & $[0.1,2.0]$ & $0.477 \pm 0.036$ & $0.079$ & $0.456 \pm 0.028$ & $0.062$ & 33.7\%\\
$b\sigma_{12}$ & $[0,3]$ & $1.209 \pm 0.016$ & $0.013$ & $1.215 \pm 0.014$ & $0.011$ &  14.1\%\\
$q_{\perp}/q_{\parallel}$ & & $1.006 \pm 0.028$ & $0.027$ & $0.998 \pm 0.004$ & $0.004$ & 596.8\%\\
$q_{\perp}$ & $[0.8,1.2]$ & $1.000 \pm 0.012$ & $0.012$ & $0.995 \pm 0.002$ & $0.002$ & 502.2\%\\
$q_{\parallel}$ & $[0.8, 1.2]$ & $0.994 \pm 0.023$ & $0.023$ & $0.997 \pm 0.005$ & $0.005$ & 378.1\%\\
\hline\hline
\end{tabular}
\label{tab:constraints}
\end{table*}

Since the galaxy monopole was measured from a simulation with a fixed $b\sigma_{12}$ in the mock catalogues, we allow for DS a rescaling of its amplitude as
\begin{align}
\label{eq:DS_rescaling}
   \xi(r|\Delta^i) = b\sigma_{12}\frac{\xi(r|\Delta^i)^\mathrm{mock}}{(b\sigma_{12})^{\rm mock}},
\end{align}
and for the 2PCF as
\begin{align}
   \xi(r) = 
   (b\sigma_{12})^2\frac{\xi(r)^\mathrm{mock}}{[(b\sigma_{12})^2]^{\rm mock}},
\end{align}
where the subscript `$\mathrm{mock}$' denotes the quantity measured from the simulations.b Note that for DS, linear galaxy bias $b$ and the amplitude parameter $\sigma_{12}$ enter Eq.~(\ref{eq:DS_rescaling}) linearly, rather than quadratically, as $\xi(r|\Delta^i)$ is a measure of density profiles, and not a two-point statistic. This is also evident from Fig.~\ref{fig:velocity_profiles} where the density-velocity coupling in the linear regime for cases in DS follow Eq.~(\ref{eq:vel_den_coupling_1.5}) and not Eq.~(\ref{eq:vel_den_coupling_0}).

We explore the parameter space by means of a Markov Chain Monte Carlo (MCMC) procedure, using the \textsc{emcee} software\footnote{emcee.readthedocs.io/}. We set flat, non-informative priors for all  parameters, as specified in Table~\ref{tab:constraints}. At each step of the chain, we concatenate the model predictions for the monopole, quadrupole and hexadecapole from each DS quintile (or a single measurement in the case of the 2PCF) into a single theory vector $\bm{\xi}^ {\mathrm{theory}}$, and quantify its deviation from the measured data vector $\bm{\xi}^ {\mathrm{data}}$ by computing
\begin{align}
   \chi^2 = \left(\bm{\xi}^{\mathrm{theory}} - \bm{\xi}^{\mathrm{data}}\right) \mathbfss{C}^{-1} \left(\bm{\xi}^{\mathrm{theory}} - \bm{\xi}^{\mathrm{data}}\right)^{T} ,
\end{align}
where $\mathbfss{C}$ is the  covariance matrix of the data vector. Then, the log-likelihood can be expressed as 
\begin{equation}
    \log \mathcal{L} = -\frac{1}{2} \chi^2 \ .
\end{equation}
The covariance matrix is estimated by measuring $\bm{\xi}^{\mathrm{data}}$ on all the mocks, as
\begin{align} \label{eq:covariance}
    \mathbfss{C} = \frac{1}{N - 1} \sum_{k=1}^{N} \left({\bm{\xi}_{k}^ {\mathrm{data}}} - \overline{\bm{\xi}^ {\mathrm{data}}}\right)\left({\bm{\xi}_{k}^ {\mathrm{data}}} - \overline{\bm{\xi}^ {\mathrm{data}}}\right) \ ,
\end{align}
where $N=300$, and $\overline{\bm{\xi}^ {\mathrm{data}}}$ is the mean data vector. As this covariance is measured from a finite number of mocks, its inversion provides a biased estimator of the true precision matrix. In order to account for this effect, we multiply $\mathbfss{C}^{-1}$ by a correction factor \citep{Hartlap2007}, given by
\begin{equation}
    \alpha = \left(1 - \frac{N_{\mathrm{b}} + 1}{N - 1} \right) \ ,
\end{equation}
where $N_{\mathrm{b}}$ is the number of bins of the data vector, and $N$ is the number of mocks \citep[see also][for an alternative approach to robustly estimate the uncertainty in the covariance matrix]{Sellentin2016}. After including this correction factor, the estimate of the precision matrix is unbiased, although it can still be affected by noise due to the finite number of simulations that are used, which can propagate further into the cosmological constraints \citep{Percival2014}. However, we have explicitly verified that when using a theoretically-derived Gaussian covariance matrix that is free of noise \citep{Grieb2016}, our inferred cosmological parameters from the 2PCF are recovered to better than 1 per cent with respect to the case where the covariance is estimated from the mocks. Currently, it is not possible to perform such a test for DS, as we lack a method to generate a noise-free Gaussian covariance for the DS multipoles. Nevertheless, we expect such a correction to be of similar order to the 2PCF case. An explicit verification of this is left for future work. Examples for the full covariance matrices for DS1+5 and 2PCF are shown in Fig.~\ref{fig:covariance}. There are significant anti-correlations between the multipoles in DS1 and DS5, as these correspond to voids and peaks of the density field.

\begin{figure*}
     \centering
     \begin{tabular}{cc}
      \hspace{-0.7 cm}
        \includegraphics[width=0.52\textwidth]{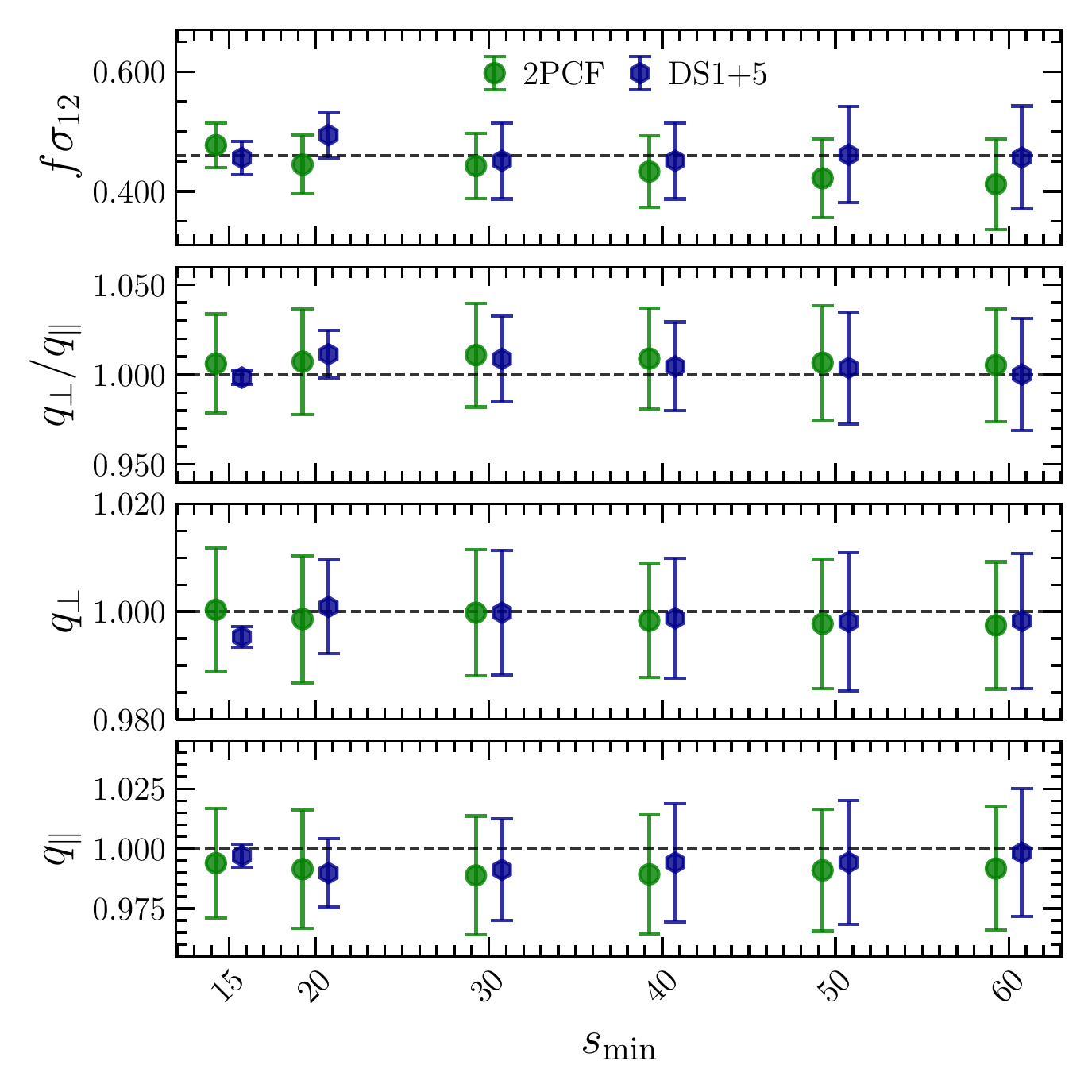} &
        \hspace{-0.5 cm}
        \includegraphics[width=0.52\textwidth]{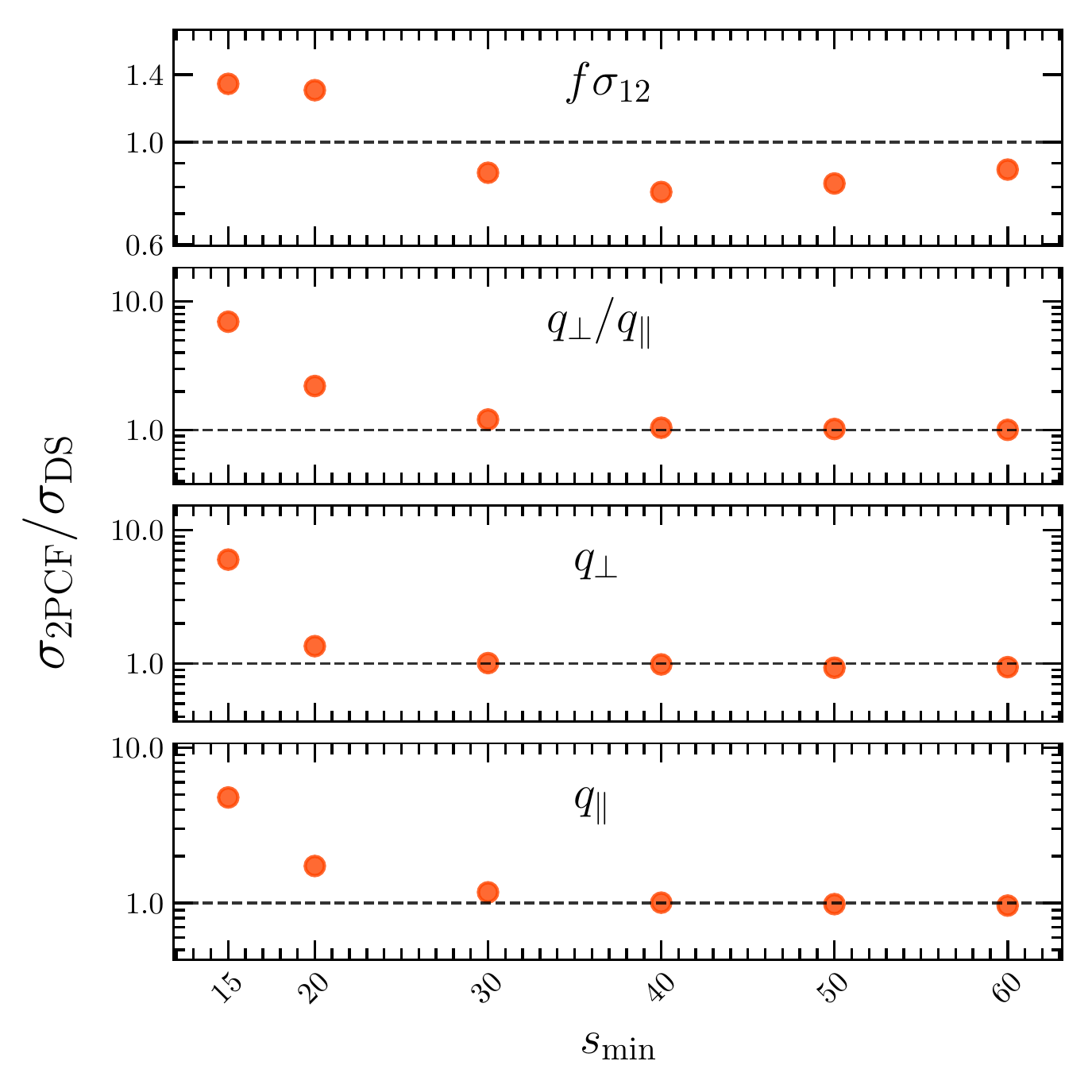}
     \end{tabular}
        \vspace{-0.4 cm}
     \caption{Left-hand side: parameter constraints with the 2PCF (green) and DS1+5 (blue) as a function of the minimum scale used in the fit. The error bars show the 1$\sigma$ spread around the best-fit values. The dashed-horizontal lines indicate values from the cosmology of the simulation. Right-hand side: The ratio of the 1$\sigma$ errors (from the left) between DS and the 2PCF. The y-axis is shown in a logarithmic scale.}
     \label{fig:constraints_scale_DS_vs_TPCF}
\end{figure*}

\subsection{Parameter constraints}

Before presenting our main cosmological constraints, we want to stress-test the numerical accuracy of the model and our simulation set up. There are at least two factors to consider: 1. with a large covariance matrix to invert, and a finite number of simulation boxes (300), we want to reduce the size of the covariance matrix where possible to make sure that it is invertible and accurate.
2. the precision of constraints increases rapidly when the minimum scale included in the fits, $s_{\rm min}$, is decreased. Despite the excellent agreement between the GSM {and the two-dimensional correlation functions measured from the simulations} (Figs.~\ref{fig:sigma_pi_largescales} \& \ref{fig:sigma_pi_smallscales}), we need to find a balance between precision and accuracy, which is likely to depend on $s_{\rm min}$.

We start by running MCMC for a wide range of $s_{\rm min}$, different combinations of density quintiles and a fixed maximum fitting scale $s_{\rm max} = 141\,h^{-1}{\rm Mpc}$. The constraints on cosmological parameters after marginalising over all the nuisance parameters (i.e. the velocity dispersion $\sigma_{v}$ and the couplings $\nu^i$) are shown in Fig.~\ref{fig:constraints_scale_DS}. We can see that on scales $s_{\rm min}\geq 50\,h^{-1}{\rm Mpc}$, using only DS1 (equivalent to the void-galaxy CCF) yields almost the same constraints as DS1+5 (equivalent to the void-galaxy CCF plus the cluster-galaxy CCF), or DS1+2+3+4+5. This is expected in the Gaussian linear regime. At $s_{\rm min}=30\ h^{-1}{\rm Mpc}$, the combination of DS1+5 starts to yield better constraints than DS1 alone, and DS1+2+3+4+5 is slightly better than DS1+5. At $s_{\rm min}=20\,h^{-1}{\rm Mpc}$, the combinaton DS1+5 leads to tighter constraints than DS1 alone, indicating the value of combining the void-galaxy and  cluster-galaxy CCFs. Although at these scales the uncertainties obtained from the joint analysis of DS1+2+3+4+5 (black points) are comparable to those obtained in the DS1+5 case, the results are clearly biased. Therefore, instead of the full combination, we will use DS1+5 as our default case for DS, knowing that they can already yield similar constraints as the case of DS1+2+3+4+5 in the range of scales where the model is unbiased. These tests also inform us that with the volume we are simulating, $(1.5\,h^{-1}{\rm Gpc})^3$, the GSM provides us with unbiased constraints down to scales $s\simeq 15\,h^{-1}{\rm Mpc}$. Therefore, we will we set $s_{\rm min}=15\,h^{-1}{\rm Mpc}$ as our default case and consider also the case of $s_{\rm min}=20\,h^{-1}{\rm Mpc}$ for comparison. Note that all data points in Fig.~\ref{fig:constraints_scale_DS} were obtained by fitting only the monopole and quadrupole. This is the case for DS1, DS1+5 and DS1+2+3+4+5. The motivation behind this is that we wanted to make a fair comparison between the different quantile combinations. Since the covariance matrix of DS1+2+3+4+5 is prohibitively large when using all 3 multipoles, we stick to using monopole and quadrupole for all combinations, which are thought to encapsulate most of the cosmological information of interest. This criterion was only applied for this figure. Throughout the rest of the paper, DS1+5 always uses all 3 multipoles.

The posterior distributions of the parameters that contain cosmological information are shown in Fig.~\ref{fig:posterior_15_20} with $s_{\rm min}=15\,h^{-1}{\rm Mpc}$ and $s_{\rm min}=20\,h^{-1}{\rm Mpc}$, respectively (see also Fig.~\ref{fig:full_posterior_15Mpc} \& \ref{fig:full_posterior_20Mpc} where all other parameters are presented). For $s_{\rm min}=20\,h^{-1}{\rm Mpc}$ (right-hand side panel), the marginalised constraints for the growth rate parameter at the 68 per cent CL are $f\sigma_{12} = 0.445 \pm 0.049$ for the 2PCF, and $f\sigma_{12} = 0.496 \pm 0.037$ for DS. The result recovered from the DS 1+5 combination represents a $\sim 33$ per cent improvement in precision over 2PCF, even though this case uses 60 per cent fewer sampling points than the full 2PCF. The advantage of DS is stronger for AP parameters. While the 2PCF yields a constraint for the AP ratio of $q_{\perp}/q_{\parallel} = 1.007 \pm 0.029$, DS has $q_{\perp}/q_{\parallel} = 1.012 \pm 0.013$, a $121$\% reduction of the error, and reaching a $\sim 1\%$ precision. Previous studies have already reported that the multipoles of the void-galaxy CCF are particularly good at measuring this ratio \citep{Nadathur2019b, Nadathur2020, Hamaus2020}, which is similar to having DS1. Here we can see that adding DS5, similar to the cluster-galaxy CCF, helps to beat down the errors for AP parameters (Fig.~\ref{fig:constraints_scale_DS}). The galaxy bias parameter encapsulated in $b \sigma_{12}$ is recovered at a precision of $\sim 2.3\%$ in DS, with $b\sigma_{12} = 1.195 \pm 0.028$. This is slightly worse than 1.4\% constraint from 2PCF. This perhaps indicates that on this scale, the constraints with DS are still partially penalised by having the extra free parameter. It is also interesting to see that the constraints on some parameter combinations from DS, such as $f \sigma_{12}$ against $q_{\perp} / q_{\parallel}$, show contours that are rotated with respect to the 2PCF, implying that there is possibility for complementing the information from these two probes in galaxy surveys, although this is beyond the scope of this work. A summary of the quoted constraints with their corresponding precision and figures of merit can be found in Table~\ref{tab:constraints}. 

At $s_{\rm min}=15\,h^{-1}{\rm Mpc}$, the parameters $f\sigma_{12}$ and $b\sigma_{12}$ are constrained to $\sim 6\%$ and $\sim 1\%$ respectively for DS. All the AP parameters are down at the sub-per cent level. The improvement for DS1+5 over 2PCF is much more significant (with the exception of the growth rate), as shown on the left of Fig.~\ref{fig:posterior_15_20} and in Table~\ref{tab:constraints}, with reductions of errors on the $f\sigma_{12}$ by $33\%$, $b\sigma_{12}$ by $14\%$, and AP parameters typically 400-600\% smaller. This is qualitatively expected. At smaller scales, the density and velocity fields become even more non-Gaussian. Less information is extracted from the two point statistics, and the cross-correlations between voids and peaks with the galaxy field (DS1+5) seem capable of uncovering it \citep{White1979, Saslaw1984, Fry1985, Fry1986}. However, we caution that the best-fit constraint for $q_{\perp}$ is 2.4$\sigma$ from the fiducial value, which is at the margin of being unbiased. Nevertheless, it is also important to realise that the error on $q_{\perp}$ is at the sub-per cent level, and is a factor of $\sim$6 smaller than that recovered from the 2PCF. Even if we take the model to be biased as the 3-$\sigma$ level, and add this as the systematic error, the total error is still significantly smaller (a factor of $\sim$2) than that from the 2PCF. It is likely that with decreasing $s_{\rm min}$, the advantage of using DS for cosmological constraints over the 2PCF will continue to increase. From the left-hand side panel of Fig.~\ref{fig:posterior_15_20}, the rotation of the contours for DS relative to the 2PCF becomes very obvious. This again indicates the potential of even better cosmological constraints with DS and the 2PCF combined.

The gain for the AP parameters $q_{\perp}$ and $q_{\parallel}$, as well as their ratio  $q_{\perp}/q_{\parallel}$ with DS over the 2PCF is more obvious than for the growth-rate parameter ($f\sigma_{12}$). This is likely because the CCFs in DS have relatively steep profiles around the scales of the top-hat filter. In particular, DS1 and DS5 have very different slopes, sometimes of the opposite sign from each other, while the 2PCF is relatively featureless at small scales. This makes those CCFs more sensitively to the change of AP parameters via Eq.~(\ref{eq:AP_rescaling}), i.e. the derivatives of the CCFs with respect to the re-scaling  parameters $q$ are larger. 

More generally, we also compare the constraining power between DS and the 2PCF for a wide range of minimal scales $s_{\rm min}$, shown in Fig.~\ref{fig:constraints_scale_DS_vs_TPCF}. On large scales, where the density field is nearly Gaussian, DS will not reveal information that is not already contained in the 2PCF. In fact, DS is expected to deliver slightly worse constraints, since the fit is penalised by the inclusion of an extra nuisance parameter with respect to the 2PCF ($\nu^1$ \& $\nu^5$ versus $\nu$). This is seen for the $f\sigma_{12}$ parameter for $s_{\rm min}>30\,h^{-1}{\rm Mpc}$, but the constraints for AP parameters are virtually the same between the two cases for those large scales.

As the minimum fitting scale progressively gets smaller, the density and velocity fields deviate from a Gaussian distribution, and the inclusion of the extra free parameter is compensated by the additional information that is captured by DS. This happens at around $s_{\mathrm{min}}=30\,h^{-1}{\rm Mpc}$. At $s_{\mathrm{min}}=20\,h^{-1}{\rm Mpc}$, the constraints for the growth parameter and AP parameters are all tighter with DS. This trend continues towards smaller scales and the improvement for DS over the 2PCF continues until $s_{\rm min}<15\,h^{-1}{\rm Mpc}$, where the model for DS starts to show systematic biases larger than the statistical errors for our sample.

\begin{figure}
    \centering
    \includegraphics[width=\columnwidth]{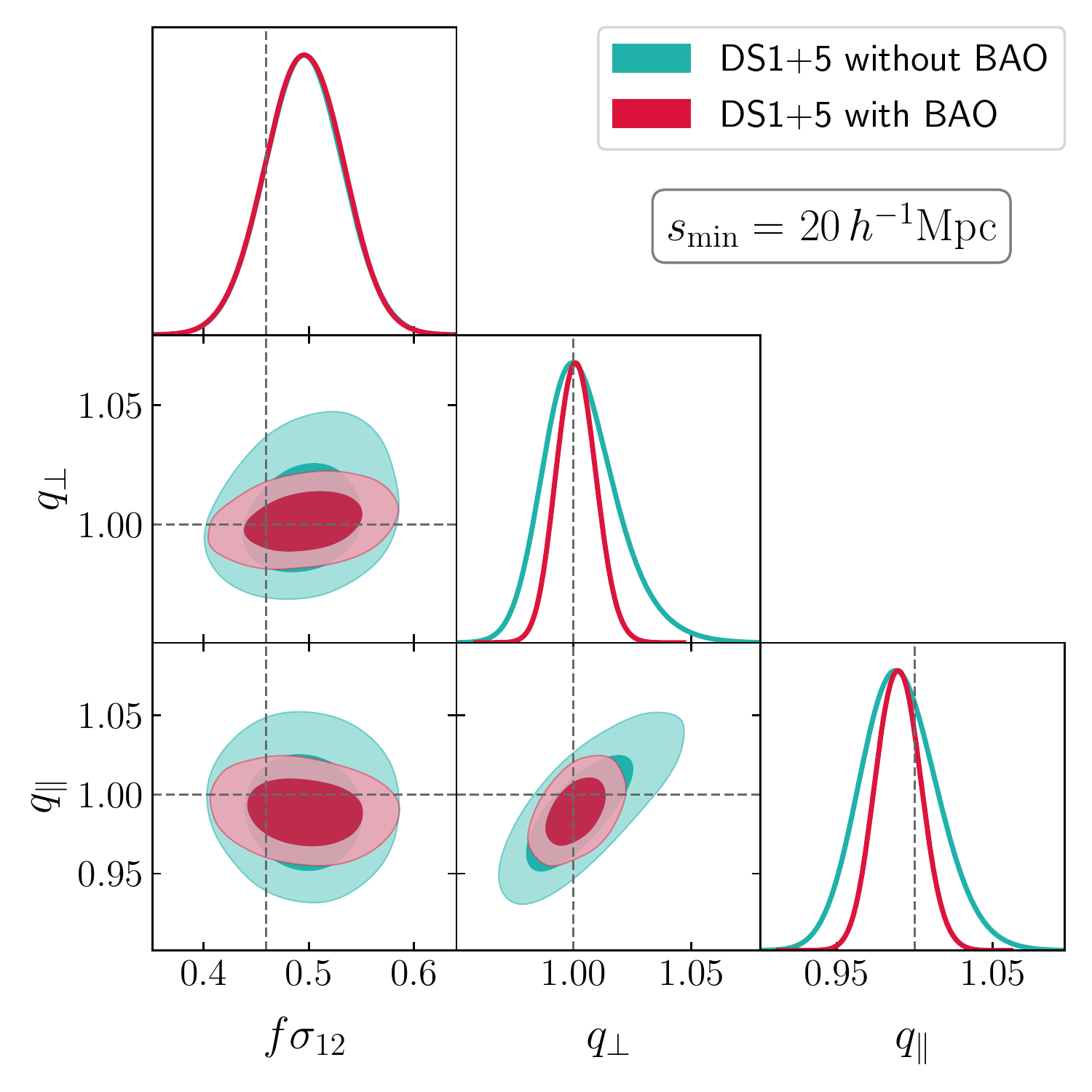}
    \caption{Comparing cosmological constraints with cross-correlation functions from DS, using scales between 20-80 $h^{-1}{\rm Mpc}$ (red) and 20-141 $h^{-1}{\rm Mpc}$ (turquoise). The latter case includes scale of the BAO. The BAO feature in the cross-correlation function helps to break degeneracies for the AP parameters ($q_{\perp}$ and $q_{\parallel}$) and improves their constraints, but it has negligible impact on the growth parameter $f\sigma_{12}$.}
    \label{fig:constraints_BAO_vs_noBAO}
\end{figure}

\subsection{Inclusion of BAO information in DS fits}
As discussed earlier, the BAO information encoded at very large scales allows us to constrain $q_{\perp}$ and $q_{\parallel}$ individually. This is demonstrated in Fig. \ref{fig:constraints_BAO_vs_noBAO}, where we compare the cases where the DS model fit is performed with and without including scales between $80$-$141\,h^{-1}{\rm Mpc}$, where the BAO feature is found. The exclusion of the BAO scales leads to a strong degeneracy between $q_{\perp}$ and $q_{\parallel}$. Since the correlation function in those intermediate scales is smooth and featureless, the model is not particularly sensitive to the angle-averaged parameter combination $q = q_{\perp}^{2/3} q_{\parallel}^{1/3}$, and can only constrain the AP ratio precisely. The inclusion of the BAO peak helps to break this degeneracy, allowing constraints at the $0.8\%$ and $1.4\%$ level for $q_{\perp}$ and $q_{\parallel}$, respectively. This information is valuable, as $q_{\perp}$ and $q_{\parallel}$ are defined in terms of the comoving angular diameter distance $D_{\rm M}(z)$ and the Hubble rate $H(z)$, respectively, which in turn allows us to put constraints on $\Omega_{\rm m}$. On the other hand, the constraint on $f\sigma_{12}$ is virtually unchanged with the addition of scales of $80$-$141\,h^{-1}{\rm Mpc}$. This is expected, since $f\sigma_{12}$ only benefits from the additional perturbation modes at those scales, which is a negligibly small fraction  compared to those at small scales.

In summary, we have shown that when applying the same GSM model to the same data with the same ranges of scale, DS tends to tighten the constraints for both the growth parameter ($f\sigma_{12}$) and AP parameters ($q_{\parallel}$ and $q_{\perp}$) over the 2PCF. The improvement of DS over the 2PCF is relatively more significant for AP parameters. We believe that there are two main reasons behind this: 
\\
\\
(1) the density PDF becomes non-Gaussian at small scales. The 2PCF, which is essentially a measure for the variance of the density field, becomes incomplete. The combination of a series of CCFs in DS allows us to sample the non-Gaussian distribution of the density PDF, hence recovering some information lost in the 2PCF. This appears to be effective even with the combination of only DS1 and DS5, the equivalent of voids plus clusters.
\\
\\
(2) the diverse and steep slopes of CCFs in DS near the top-hat smoothing scale makes the use of DS more sensitive to AP parameters at small scales than for the 2PCF. This allows DS to break the degeneracy between AP parameters and the growth parameter with just the small-scale information, and without employing the BAO. On the other hand, the 2PCF needs the BAO scale to be included to break such degeneracy. Therefore, when the BAO scale is not included in both cases, i.e. with $s<80\,h^{-1}{\rm Mpc}$, DS provides constraints on those cosmological parameters many times better than the 2PCF, as we have checked explicitly. This suggests that DS will be particularly powerful for RSD analysis with galaxy surveys covering a relatively small volume where the BAO peaks are not well constrained.

\section{Discussion and conclusions} \label{sec:conclusions}
We have presented a new method to analyse the signature of RSD in clustering measurements inferred from galaxy redshift surveys. Instead of the conventional 2PCF, we use the combination of a series of CCFs. These are CCFs 
obtained by splitting random positions according to the local galaxy density, and cross-correlating those positions with the entire galaxy field, thus using the exact same galaxy sample as in the 2PCF. This method builds upon the idea of DS statistics from weak lensing analysis \citep{Gruen2015,Gruen2018, Friedrich2018}, and generalises the modelling of RSD around voids to environments of different local densities. 
The algorithm can be summarised as follows:

\begin{enumerate}
    \item We smooth the real-space galaxy number density field with a spherical top-hat window in a number of randomly selected locations, and split the positions into quintiles according to the smoothed densities.
    
    \item We cross-correlate positions from each quintile with the galaxy number density field in redshift space, and model the RSD in their clustering pattern using the GSM.
    
    \item We use MCMC to jointly fit RSD and Alcock-Paczynski distortions for each quintile, and we combine their information to obtain cosmological constraints.

\end{enumerate}

We have tested and validated the model using a suite of mock galaxy catalogues that mimic the number density and clustering properties of the BOSS CMASS galaxy sample. Our main conclusions can be listed as follows:

\begin{itemize}

   \item[-] the velocity field within each quintile of split density is close to Gaussian, and this meets the key condition necessary for the GSM, providing the physical reason for it to perform well at small scales.

    \item[-] The GSM can fit the CCFs for every DS quintile with a remarkable accuracy. It appears to work well at small scales (below 20-30 $h^{-1}{\rm Mpc}$) where the same model, when applied to 2PCF, has stronger deviations from simulation. We have also shown that recent models for the void-galaxy CCF \citep{Cai2016, Nadathur2019a} have no fundamental difference from the Gaussian streaming model, and can work equally well for DS. 

    \item[-] By comparing the parameter constraints obtained with combining different DS quintiles, we find that most of the cosmological information is contained in the cross-correlations between the lowest and highest density quintiles (voids and clusters) with the galaxy field. Consequently, there is no major loss of constraining power when discarding the intermediate quintiles.
    
    \item[-] On large scales, where the density field is is nearly Gaussian, there is no additional gain of information when using DS, compared to the traditional RSD analysis using the galaxy 2PCF.
    
    \item[-] On scales below $\sim 30\,h^{-1}{\rm Mpc}$, the non-Gaussian features of the density and velocity field start to become important, and DS outperforms the constraints from the galaxy 2PCF for $f\sigma_{12}$, yielding a $\sim$7 per cent constraint with $s_{\rm min}=20\,h^{-1}{\rm Mpc}$. It is approximately $30$ per cent better in precision than the 2PCF. DS is most sensitive to geometrical distortions and can constrain the AP distortion parameters at the sub-per cent level, and up to $\sim 6$ times better than the 2PCF with $s_{\rm min}=15~h^{-1}$Mpc.
\end{itemize}

With two-point statistics, \cite{Alam2017} have reported a $\sim 6\%$ constraint on the growth parameter $f\sigma_8$ with a volume of 18.7 Gpc$^3$ using the combined samples from BOSS. The volume is 6.4~(Gpc/$h$)$^3$ with $h=0.7$, nearly a factor of two larger than the single box of our simulation. Part of their likelihood comes from \cite{Sanchez2016}, where a minimal scale of $s=20\,h^{-1}{\rm Mpc}$ was adopted. Our constraint for $f\sigma_{12}$ from the 2PCF with the same cutoff scale has a $\sim 10\%$ precision. Scaling our errors up to the same volume as BOSS, we expect to have $\sim 7\%$ constraint for $f\sigma_{12}$. This is broadly consistent with \cite{Alam2017}. We have shown that using the same data, a $\sim 7\%$ constraint can already be achieved using DS. This is equivalent to an increase of survey volume by nearly a factor 2.

The advantage of DS over 2PCF is most significant for AP parameters. DS can reduce the errors on $q_{\perp}$ and $q_{\parallel}$ by a few times compared to the 2PCF at $s_{\rm min}\sim 15\,h^{-1}{\rm Mpc}$. This is likely due to the diverse and steep slopes for the monopoles of CCFs in DS near the smoothing scale of the top-hat filter. The constraints on AP parameters can be translated into constraints  on parameters of the standard $\Lambda$CDM model. For example, varying only $\Omega_{\rm m}$ and $\sigma_{12}$, and all other cosmological parameters fixed in the model, we find the marginalised constraints with DS on $\Omega_{\rm m}$ and $\sigma_{12}$ under these somewhat strict assumptions to be a factor of $\sim 3$ and $\sim 40\%$ tighter over the 2PCF, respectively.

The sensitive response to cosmology of the tails of the density distribution has already been suggested in previous papers \citep[e.g.][]{White1979, Saslaw1984, Fry1985, Fry1986, Zorrilla2020}. Given that the most significant constraining power is coming from the extreme DS quintiles, it would be interesting to address whether the voids and clusters that are identified with the DS method provide tighter constraints than those identified with dedicated void-finding algorithms, such as the Spherical Void Finder \citep{Padilla2005} and the Watershed Void Finder \citep{vandeWeygaert2009}, or group-finding algorithms, such as redMaPPer \citep{Rykoff2014}. The selection criteria for these regions could play an important role in the modelling side. 

RSD with split densities is general. Although it is similar to the combination of void-galaxy and cluster cross-correlations, it does not depend on a specific void or cluster finding algorithm, and can be predicted from first principles. It is essentially a series of conditioned correlation functions, which can be predicted once the PDF of the density field and the real-space correlation function are known \citep{Abbas2005,Shi2018, Neyrinck2018, Friedrich2018}. Given the success of models that allow us to predict the PDF of a smoothed density field \citep{Uhlemann2016, Repp2018, Jamieson2020}, and the existing model framework for predicting CCFs, it is hopeful that the model prediction can be achieved with a high accuracy, which is our next task.

 At small scales, e.g. below $15\,h^{-1}{\rm Mpc}$, the GSM starts to be biased at the precision of our interest. We suspect that multiple reasons can lead to the complication at those small scales: 1. the linear galaxy bias model starts to break down; .2 higher order terms in the streaming model start to become important, especially for DS5 (cluster-galaxy CCF), where the velocity dispersion and its derivative may start to be significant \citep{Fisher1995,Reid2011}; 3. as a long positive tail in the density PDF is expected, splitting the density with 5 quantiles may not be sufficient, as the highest density quantile will have a wide range of densities, and this will violate the Gaussian condition for the velocities.

The linear coupling between the density and velocity is insufficient. Forthcoming large-area surveys, such as Euclid \citep{euclid} and DESI \citep{Levi2019} will generate data sets that will be several factors larger in volume than current surveys, requiring per cent-level precision on the modelling side. We certainly need to go beyond the linear coupling. 
We have tested employing spherical dynamics for modelling the velocity profiles, finding some improvements over the linear coupling model, as also elucidated by previous work for voids \citep[e.g.][]{Demchenko2016}, but the accuracy is still insufficient for our purpose. The empirical model for the coupling allows us to apply the GSM to a relatively small scale, at the cost that our cosmological constraints have been penalised by marginalising over the nuisance parameter. Given the potential gain of cosmological information at small scales, improving our modelling for the velocity and density coupling model may be very rewarding.

Throughout this work, we have not yet attempted the combination of DS CCFs with the 2PCF, nor have we involved the PDFs of the density field for cosmological constraints. Given the apparent different orientations for contours of the posteriors between DS and the 2PCF, e.g. Figs.~\ref{fig:full_posterior_15Mpc} \& \ref{fig:full_posterior_20Mpc}, it is hopeful that their combination would offer even tighter constraints for cosmology. This can be explored when a larger suit of simulations is employed, allowing a more reliable and accurate covariance matrix to be constructed; or having an accurate analytical covariance matrix \citep{Grieb2016}, . Likewise, the 3D PDFs of the density field, i.e. the counts-in-cells statistics, have been demonstrated to be complementary for parameter constraints \citep[e.g.][]{Uhlemann2020}, especially for breaking degeneracy between $b$ and $\sigma_{12}$ \citep[e.g.][]{Repp2020}. It is compelling to combine DS, the 2PCF, and counts in cells all together, which may one of the best ways to extract cosmological information from the initial conditions at small scales without explicitly employing higher order statistics. Developing the joint covariance for the three different cosmological probes would be a major target of our effort.

\section*{Data availability}

The data and codes used in this analysis will be made available upon publication at the following repository: \href{https://github.com/epaillas/densitysplit}{https://github.com/epaillas/densitysplit}.

\section*{Acknowledgements}
We would like to thank the anonymous referee for their helpful suggestions and remarks on our manuscript.
We thank Mark Neyrinck, Carolina Cuesta-Lazaro, Daniel Gruen, Joseph Kuruvilla and Seshadri Nadathur for useful comments. EP is supported by CONICYT-PCHA/Doctorado Nacional (2017- 21170093), EP and NP acknowledge support from CONICYT project Basal AFB-170002 and Fondecyt 1191813. YC acknowledges the support of the Royal Society through the award of a University Research Fellowship and an Enhancement Award. The Geryon cluster at the Centro de Astro-Ingenieria UC was extensively used for the calculations performed in this paper. BASAL CATA PFB-06, the Anillo ACT-86, FONDEQUIP AIC-57, and QUIMAL 130008 provided funding for several improvements to the Geryon cluster. This project received financial support from the European Union's Horizon 2020 Research and Innovation programme under 
the Marie Sklodowska-Curie grant agreement number 734374.   

\appendix

\renewcommand\thefigure{A\arabic{figure}} 
\setcounter{figure}{0} 

\begin{figure*}
    \centering
    \includegraphics[width=\textwidth]{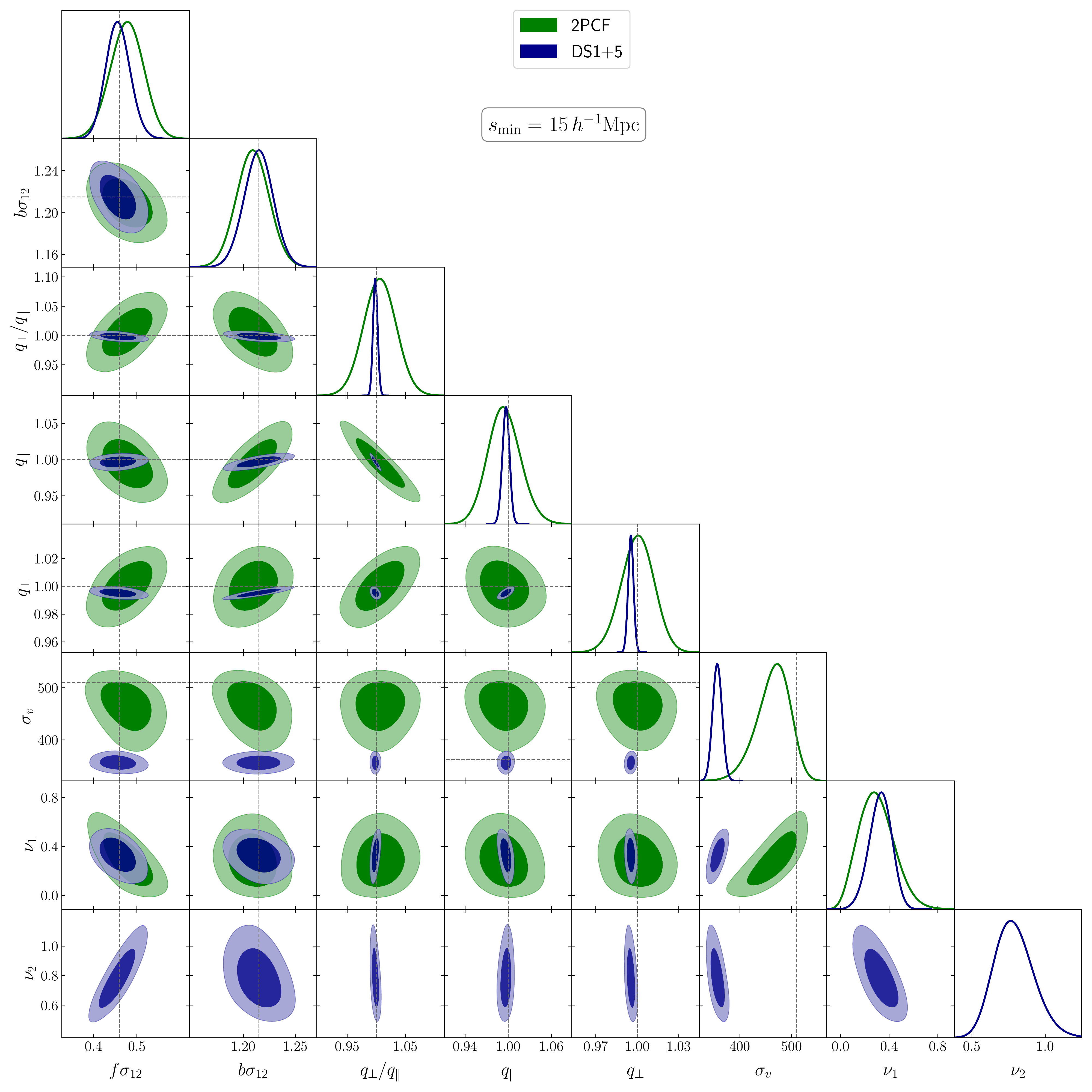}
    \caption{Similar to Fig. \ref{fig:posterior_15_20}, but showing the posterior distributions for all parameters included in the fit, with the exception of the second nuisance parameter $\nu_2$ in DS. The minimum fitting scale used was $s_{\mathrm{\min}} = 15\,h^{-1}{\rm Mpc}$.}
    \label{fig:full_posterior_15Mpc}
\end{figure*}

\begin{figure*}
    \centering
    \includegraphics[width=\textwidth]{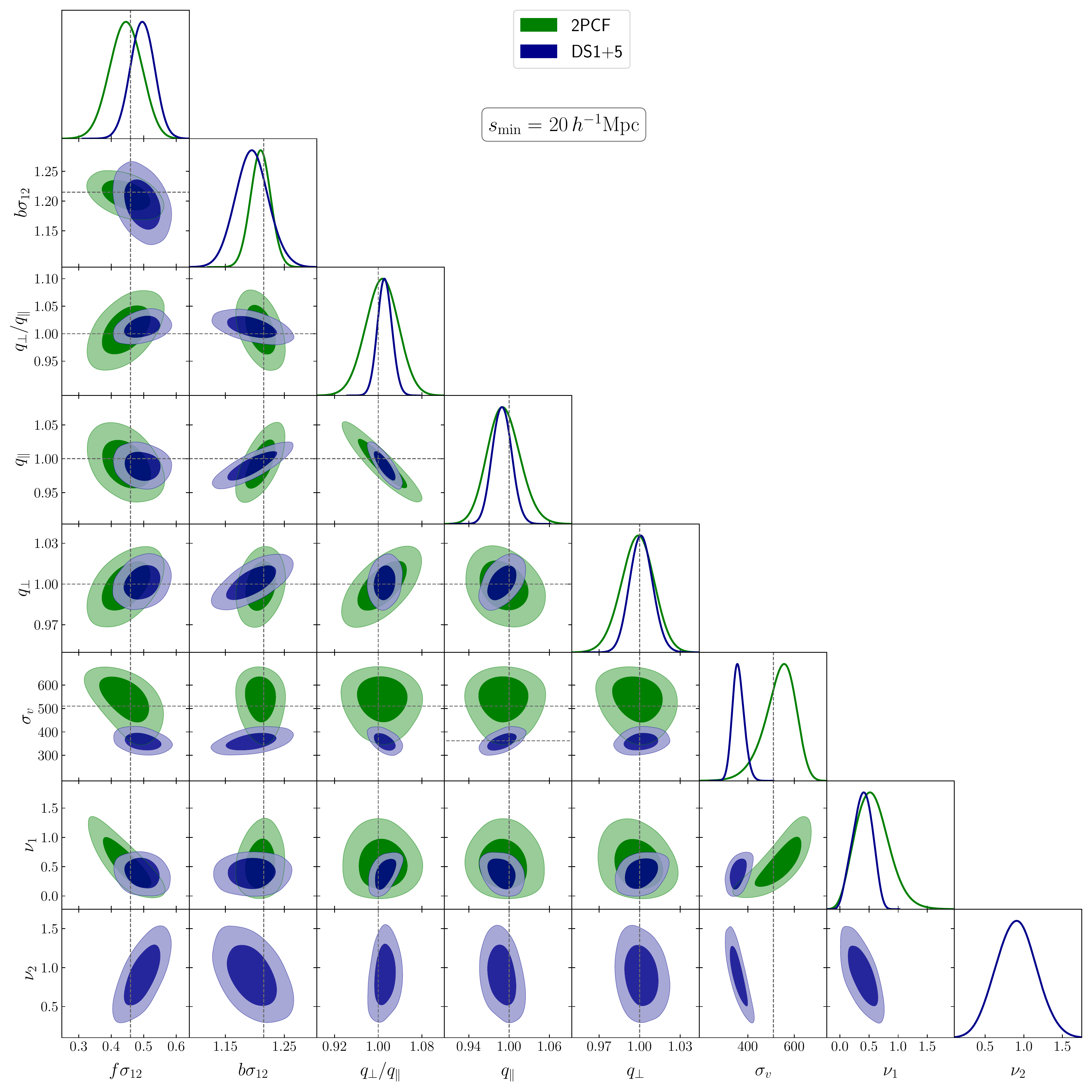}
    \caption{Similar to Fig. \ref{fig:full_posterior_15Mpc}, but showing the posterior distributions obtained using a minimum fitting scale of $s_{\mathrm{\min}} = 20\,h^{-1}{\rm Mpc}$.}
    \label{fig:full_posterior_20Mpc}
\end{figure*}

\bibliographystyle{mnras}
\bibliography{manuscript}

\begin{thebibliography}{}
\makeatletter
\relax
\def\mn@urlcharsother{\let\do\@makeother \do\$\do\&\do\#\do\^\do\_\do\%\do\~}
\def\mn@doi{\begingroup\mn@urlcharsother \@ifnextchar [ {\mn@doi@}
  {\mn@doi@[]}}
\def\mn@doi@[#1]#2{\def\@tempa{#1}\ifx\@tempa\@empty \href
  {http://dx.doi.org/#2} {doi:#2}\else \href {http://dx.doi.org/#2} {#1}\fi
  \endgroup}
\def\mn@eprint#1#2{\mn@eprint@#1:#2::\@nil}
\def\mn@eprint@arXiv#1{\href {http://arxiv.org/abs/#1} {{\tt arXiv:#1}}}
\def\mn@eprint@dblp#1{\href {http://dblp.uni-trier.de/rec/bibtex/#1.xml}
  {dblp:#1}}
\def\mn@eprint@#1:#2:#3:#4\@nil{\def\@tempa {#1}\def\@tempb {#2}\def\@tempc
  {#3}\ifx \@tempc \@empty \let \@tempc \@tempb \let \@tempb \@tempa \fi \ifx
  \@tempb \@empty \def\@tempb {arXiv}\fi \@ifundefined
  {mn@eprint@\@tempb}{\@tempb:\@tempc}{\expandafter \expandafter \csname
  mn@eprint@\@tempb\endcsname \expandafter{\@tempc}}}

\bibitem[\protect\citeauthoryear{{Abbas} \& {Sheth}}{{Abbas} \&
  {Sheth}}{2005}]{Abbas2005}
{Abbas} U.,  {Sheth} R.~K.,  2005, \mn@doi [\mnras]
  {10.1111/j.1365-2966.2005.09654.x}, \href
  {https://ui.adsabs.harvard.edu/abs/2005MNRAS.364.1327A} {364, 1327}

\bibitem[\protect\citeauthoryear{{Achitouv}}{{Achitouv}}{2019}]{Achitouv2019}
{Achitouv} I.,  2019, \mn@doi [\prd] {10.1103/PhysRevD.100.123513}, \href
  {https://ui.adsabs.harvard.edu/abs/2019PhRvD.100l3513A} {100, 123513}

\bibitem[\protect\citeauthoryear{{Achitouv}, {Blake}, {Carter}, {Koda}  \&
  {Beutler}}{{Achitouv} et~al.}{2017}]{Achitouv2017a}
{Achitouv} I.,  {Blake} C.,  {Carter} P.,  {Koda} J.,   {Beutler} F.,  2017,
  \mn@doi [\prd] {10.1103/PhysRevD.95.083502}, \href
  {https://ui.adsabs.harvard.edu/abs/2017PhRvD..95h3502A} {95, 083502}

\bibitem[\protect\citeauthoryear{{Alam} et~al.,}{{Alam}
  et~al.}{2017}]{Alam2017}
{Alam} S.,  et~al., 2017, \mn@doi [\mnras] {10.1093/mnras/stx721}, \href
  {https://ui.adsabs.harvard.edu/abs/2017MNRAS.470.2617A} {470, 2617}

\bibitem[\protect\citeauthoryear{{Alcock} \& {Paczynski}}{{Alcock} \&
  {Paczynski}}{1979}]{Alcock1979}
{Alcock} C.,  {Paczynski} B.,  1979, \mn@doi [\nat] {10.1038/281358a0}, \href
  {https://ui.adsabs.harvard.edu/abs/1979Natur.281..358A} {281, 358}

\bibitem[\protect\citeauthoryear{{Angulo} \& {White}}{{Angulo} \&
  {White}}{2010}]{Angulo2010}
{Angulo} R.~E.,  {White} S.~D.~M.,  2010, \mn@doi [\mnras]
  {10.1111/j.1365-2966.2010.16459.x}, \href
  {https://ui.adsabs.harvard.edu/abs/2010MNRAS.405..143A} {405, 143}

\bibitem[\protect\citeauthoryear{{Ballinger}, {Peacock}  \&
  {Heavens}}{{Ballinger} et~al.}{1996}]{Ballinger1996}
{Ballinger} W.~E.,  {Peacock} J.~A.,   {Heavens} A.~F.,  1996, \mn@doi [\mnras]
  {10.1093/mnras/282.3.877}, \href
  {https://ui.adsabs.harvard.edu/abs/1996MNRAS.282..877B} {282, 877}

\bibitem[\protect\citeauthoryear{{Bautista} et~al.,}{{Bautista}
  et~al.}{2021}]{Bautista2021}
{Bautista} J.~E.,  et~al., 2021, \mn@doi [\mnras] {10.1093/mnras/staa2800},
  \href {https://ui.adsabs.harvard.edu/abs/2021MNRAS.500..736B} {500, 736}

\bibitem[\protect\citeauthoryear{{Benson}, {Cole}, {Frenk}, {Baugh}  \&
  {Lacey}}{{Benson} et~al.}{2000}]{Benson2000}
{Benson} A.~J.,  {Cole} S.,  {Frenk} C.~S.,  {Baugh} C.~M.,   {Lacey} C.~G.,
  2000, \mn@doi [\mnras] {10.1046/j.1365-8711.2000.03101.x}, \href
  {http://adsabs.harvard.edu/abs/2000MNRAS.311..793B} {311, 793}

\bibitem[\protect\citeauthoryear{{Berlind} \& {Weinberg}}{{Berlind} \&
  {Weinberg}}{2002}]{Berlind2002}
{Berlind} A.~A.,  {Weinberg} D.~H.,  2002, \mn@doi [\apj] {10.1086/341469},
  \href {http://adsabs.harvard.edu/abs/2002ApJ...575..587B} {575, 587}

\bibitem[\protect\citeauthoryear{{Bernardeau} \& {Kofman}}{{Bernardeau} \&
  {Kofman}}{1995}]{Bernardeau1995}
{Bernardeau} F.,  {Kofman} L.,  1995, \mn@doi [\apj] {10.1086/175542}, \href
  {https://ui.adsabs.harvard.edu/abs/1995ApJ...443..479B} {443, 479}

\bibitem[\protect\citeauthoryear{{Beutler} et~al.,}{{Beutler}
  et~al.}{2012}]{Beutler2012}
{Beutler} F.,  et~al., 2012, \mn@doi [\mnras]
  {10.1111/j.1365-2966.2012.21136.x}, \href
  {https://ui.adsabs.harvard.edu/abs/2012MNRAS.423.3430B} {423, 3430}

\bibitem[\protect\citeauthoryear{{Beutler} et~al.,}{{Beutler}
  et~al.}{2017}]{Beutler2017}
{Beutler} F.,  et~al., 2017, \mn@doi [\mnras] {10.1093/mnras/stw3298}, \href
  {https://ui.adsabs.harvard.edu/abs/2017MNRAS.466.2242B} {466, 2242}

\bibitem[\protect\citeauthoryear{{Bianchi}, {Percival}  \& {Bel}}{{Bianchi}
  et~al.}{2016}]{Bianchi2016}
{Bianchi} D.,  {Percival} W.~J.,   {Bel} J.,  2016, \mn@doi [\mnras]
  {10.1093/mnras/stw2243}, \href
  {https://ui.adsabs.harvard.edu/abs/2016MNRAS.463.3783B} {463, 3783}

\bibitem[\protect\citeauthoryear{{Blake} et~al.,}{{Blake}
  et~al.}{2011}]{Blake2011}
{Blake} C.,  et~al., 2011, \mn@doi [\mnras] {10.1111/j.1365-2966.2011.18903.x},
  \href {https://ui.adsabs.harvard.edu/abs/2011MNRAS.415.2876B} {415, 2876}

\bibitem[\protect\citeauthoryear{{Blake} et~al.,}{{Blake}
  et~al.}{2013}]{Blake2013}
{Blake} C.,  et~al., 2013, \mn@doi [\mnras] {10.1093/mnras/stt1791}, \href
  {https://ui.adsabs.harvard.edu/abs/2013MNRAS.436.3089B} {436, 3089}

\bibitem[\protect\citeauthoryear{Cai, Taylor, Peacock  \& Padilla}{Cai
  et~al.}{2016}]{Cai2016}
Cai Y.-C.,  Taylor A.,  Peacock J.~A.,   Padilla N.,  2016, \mn@doi [\mnras]
  {10.1093/mnras/stw1809}, 462, 2465

\bibitem[\protect\citeauthoryear{{Cai}, {Neyrinck}, {Mao}, {Peacock}, {Szapudi}
   \& {Berlind}}{{Cai} et~al.}{2017}]{Cai2017}
{Cai} Y.-C.,  {Neyrinck} M.,  {Mao} Q.,  {Peacock} J.~A.,  {Szapudi} I.,
  {Berlind} A.~A.,  2017, \mn@doi [\mnras] {10.1093/mnras/stw3299}, \href
  {https://ui.adsabs.harvard.edu/abs/2017MNRAS.466.3364C} {466, 3364}

\bibitem[\protect\citeauthoryear{{Cautun}, {Paillas}, {Cai}, {Bose}, {Armijo},
  {Li}  \& {Padilla}}{{Cautun} et~al.}{2018}]{Cautun2017}
{Cautun} M.,  {Paillas} E.,  {Cai} Y.-C.,  {Bose} S.,  {Armijo} J.,  {Li} B.,
  {Padilla} N.,  2018, \mn@doi [\mnras] {10.1093/mnras/sty463}, \href
  {http://adsabs.harvard.edu/abs/2018MNRAS.476.3195C} {476, 3195}

\bibitem[\protect\citeauthoryear{{Chen}, {Vlah}, {Castorina}  \&
  {White}}{{Chen} et~al.}{2020}]{Chen2020}
{Chen} S.-F.,  {Vlah} Z.,  {Castorina} E.,   {White} M.,  2020, arXiv e-prints,
  \href {https://ui.adsabs.harvard.edu/abs/2020arXiv201204636C} {p.
  arXiv:2012.04636}

\bibitem[\protect\citeauthoryear{{Chiang}, {Wagner}, {S{\'a}nchez}, {Schmidt}
  \& {Komatsu}}{{Chiang} et~al.}{2015}]{Chiang2015}
{Chiang} C.-T.,  {Wagner} C.,  {S{\'a}nchez} A.~G.,  {Schmidt} F.,   {Komatsu}
  E.,  2015, \mn@doi [\jcap] {10.1088/1475-7516/2015/09/028}, \href
  {https://ui.adsabs.harvard.edu/abs/2015JCAP...09..028C} {2015, 028}

\bibitem[\protect\citeauthoryear{{Clampitt} \& {Jain}}{{Clampitt} \&
  {Jain}}{2015}]{Clampitt2015}
{Clampitt} J.,  {Jain} B.,  2015, \mn@doi [\mnras] {10.1093/mnras/stv2215},
  \href {https://ui.adsabs.harvard.edu/abs/2015MNRAS.454.3357C} {454, 3357}

\bibitem[\protect\citeauthoryear{{Coles} \& {Jones}}{{Coles} \&
  {Jones}}{1991}]{Coles1991}
{Coles} P.,  {Jones} B.,  1991, \mn@doi [\mnras] {10.1093/mnras/248.1.1}, \href
  {https://ui.adsabs.harvard.edu/abs/1991MNRAS.248....1C} {248, 1}

\bibitem[\protect\citeauthoryear{{Colombi}}{{Colombi}}{1994}]{Colombi1994}
{Colombi} S.,  1994, \mn@doi [\apj] {10.1086/174834}, \href
  {https://ui.adsabs.harvard.edu/abs/1994ApJ...435..536C} {435, 536}

\bibitem[\protect\citeauthoryear{{Copeland}, {Sami}  \& {Tsujikawa}}{{Copeland}
  et~al.}{2006}]{Copeland2006}
{Copeland} E.~J.,  {Sami} M.,   {Tsujikawa} S.,  2006, \mn@doi [International
  Journal of Modern Physics D] {10.1142/S021827180600942X}, \href
  {https://ui.adsabs.harvard.edu/abs/2006IJMPD..15.1753C} {15, 1753}

\bibitem[\protect\citeauthoryear{{Correa}, {Paz}, {Padilla}, {Ruiz}, {Angulo}
  \& {S{\'a}nchez}}{{Correa} et~al.}{2019}]{Correa2019}
{Correa} C.~M.,  {Paz} D.~J.,  {Padilla} N.~D.,  {Ruiz} A.~N.,  {Angulo} R.~E.,
    {S{\'a}nchez} A.~G.,  2019, \mn@doi [\mnras] {10.1093/mnras/stz821}, \href
  {https://ui.adsabs.harvard.edu/abs/2019MNRAS.485.5761C} {485, 5761}

\bibitem[\protect\citeauthoryear{{Correa}, {Paz}, {S{\'a}nchez}, {Ruiz},
  {Padilla}  \& {Angulo}}{{Correa} et~al.}{2020}]{Correa2020}
{Correa} C.~M.,  {Paz} D.~J.,  {S{\'a}nchez} A.~G.,  {Ruiz} A.~N.,  {Padilla}
  N.~D.,   {Angulo} R.~E.,  2020, \mn@doi [\mnras] {10.1093/mnras/staa3252},
  \href {https://ui.adsabs.harvard.edu/abs/2020MNRAS.tmp.3058C} {}

\bibitem[\protect\citeauthoryear{{Croft}, {Dalton}  \& {Efstathiou}}{{Croft}
  et~al.}{1999}]{Croft1999}
{Croft} R. A.~C.,  {Dalton} G.~B.,   {Efstathiou} G.,  1999, \mn@doi [\mnras]
  {10.1046/j.1365-8711.1999.02381.x}, \href
  {https://ui.adsabs.harvard.edu/abs/1999MNRAS.305..547C} {305, 547}

\bibitem[\protect\citeauthoryear{{Cuesta-Lazaro}, {Li}, {Eggemeier}, {Zarrouk},
  {Baugh}, {Nishimichi}  \& {Takada}}{{Cuesta-Lazaro}
  et~al.}{2020}]{Cuesta-Lazaro2020}
{Cuesta-Lazaro} C.,  {Li} B.,  {Eggemeier} A.,  {Zarrouk} P.,  {Baugh} C.~M.,
  {Nishimichi} T.,   {Takada} M.,  2020, arXiv e-prints, \href
  {https://ui.adsabs.harvard.edu/abs/2020arXiv200202683C} {p. arXiv:2002.02683}

\bibitem[\protect\citeauthoryear{{Demchenko}, {Cai}, {Heymans}  \&
  {Peacock}}{{Demchenko} et~al.}{2016}]{Demchenko2016}
{Demchenko} V.,  {Cai} Y.-C.,  {Heymans} C.,   {Peacock} J.~A.,  2016, \mn@doi
  [\mnras] {10.1093/mnras/stw2030}, \href
  {https://ui.adsabs.harvard.edu/abs/2016MNRAS.463..512D} {463, 512}

\bibitem[\protect\citeauthoryear{{Einasto}, {H{\"u}tsi}, {Liivam{\"a}gi}  \&
  {Einasto}}{{Einasto} et~al.}{2020}]{Einasto2020}
{Einasto} J.,  {H{\"u}tsi} G.,  {Liivam{\"a}gi} L.~J.,   {Einasto} M.,  2020,
  arXiv e-prints, \href {https://ui.adsabs.harvard.edu/abs/2020arXiv201113292E}
  {p. arXiv:2011.13292}

\bibitem[\protect\citeauthoryear{{Fisher}}{{Fisher}}{1995}]{Fisher1995}
{Fisher} K.~B.,  1995, \mn@doi [\apj] {10.1086/175980}, \href
  {https://ui.adsabs.harvard.edu/abs/1995ApJ...448..494F} {448, 494}

\bibitem[\protect\citeauthoryear{{Friedrich} et~al.,}{{Friedrich}
  et~al.}{2018}]{Friedrich2018}
{Friedrich} O.,  et~al., 2018, \mn@doi [\prd] {10.1103/PhysRevD.98.023508},
  \href {https://ui.adsabs.harvard.edu/abs/2018PhRvD..98b3508F} {98, 023508}

\bibitem[\protect\citeauthoryear{{Fry}}{{Fry}}{1985}]{Fry1985}
{Fry} J.~N.,  1985, \mn@doi [\apj] {10.1086/162859}, \href
  {https://ui.adsabs.harvard.edu/abs/1985ApJ...289...10F} {289, 10}

\bibitem[\protect\citeauthoryear{{Fry}}{{Fry}}{1986}]{Fry1986}
{Fry} J.~N.,  1986, \mn@doi [\apj] {10.1086/164348}, \href
  {https://ui.adsabs.harvard.edu/abs/1986ApJ...306..358F} {306, 358}

\bibitem[\protect\citeauthoryear{{Gil-Mar{\'\i}n}, {Wagner}, {Verde},
  {Porciani}  \& {Jimenez}}{{Gil-Mar{\'\i}n} et~al.}{2012}]{Gil-Marin2012}
{Gil-Mar{\'\i}n} H.,  {Wagner} C.,  {Verde} L.,  {Porciani} C.,   {Jimenez} R.,
   2012, \mn@doi [\jcap] {10.1088/1475-7516/2012/11/029}, \href
  {https://ui.adsabs.harvard.edu/abs/2012JCAP...11..029G} {2012, 029}

\bibitem[\protect\citeauthoryear{Grieb, Sánchez, Salazar-Albornoz  \&
  Dalla Vecchia}{Grieb et~al.}{2016}]{Grieb2016}
Grieb J.~N.,  Sánchez A.~G.,  Salazar-Albornoz S.,   Dalla Vecchia C.,  2016,
  \mn@doi [Monthly Notices of the Royal Astronomical Society]
  {10.1093/mnras/stw065}, 457, 1577–1592

\bibitem[\protect\citeauthoryear{{Gruen} et~al.,}{{Gruen}
  et~al.}{2016}]{Gruen2015}
{Gruen} D.,  et~al., 2016, \mn@doi [\mnras] {10.1093/mnras/stv2506}, \href
  {http://adsabs.harvard.edu/abs/2016MNRAS.455.3367G} {455, 3367}

\bibitem[\protect\citeauthoryear{{Gruen} et~al.,}{{Gruen}
  et~al.}{2018}]{Gruen2018}
{Gruen} D.,  et~al., 2018, \mn@doi [\prd] {10.1103/PhysRevD.98.023507}, \href
  {https://ui.adsabs.harvard.edu/abs/2018PhRvD..98b3507G} {98, 023507}

\bibitem[\protect\citeauthoryear{{Hamaus}, {Sutter}  \& {Wandelt}}{{Hamaus}
  et~al.}{2014}]{Hamaus2014}
{Hamaus} N.,  {Sutter} P.~M.,   {Wandelt} B.~D.,  2014, \mn@doi [\prl]
  {10.1103/PhysRevLett.112.251302}, \href
  {https://ui.adsabs.harvard.edu/abs/2014PhRvL.112y1302H} {112, 251302}

\bibitem[\protect\citeauthoryear{Hamaus, Sutter, Lavaux  \& Wandelt}{Hamaus
  et~al.}{2015}]{Hamaus2015}
Hamaus N.,  Sutter P.~M.,  Lavaux G.,   Wandelt B.~D.,  2015, \mn@doi [JCAP]
  {10.1088/1475-7516/2015/11/036}, 1511, 036

\bibitem[\protect\citeauthoryear{Hamaus, Pisani, Sutter, Lavaux, Escoffier,
  Wandelt  \& Weller}{Hamaus et~al.}{2016}]{Hamaus2016}
Hamaus N.,  Pisani A.,  Sutter P.~M.,  Lavaux G.,  Escoffier S.,  Wandelt
  B.~D.,   Weller J.,  2016, \mn@doi [Phys. Rev. Lett.]
  {10.1103/PhysRevLett.117.091302}, 117, 091302

\bibitem[\protect\citeauthoryear{{Hamaus}, {Cousinou}, {Pisani}, {Aubert},
  {Escoffier}  \& {Weller}}{{Hamaus} et~al.}{2017}]{Hamaus2017}
{Hamaus} N.,  {Cousinou} M.-C.,  {Pisani} A.,  {Aubert} M.,  {Escoffier} S.,
  {Weller} J.,  2017, \mn@doi [\jcap] {10.1088/1475-7516/2017/07/014}, \href
  {https://ui.adsabs.harvard.edu/abs/2017JCAP...07..014H} {2017, 014}

\bibitem[\protect\citeauthoryear{{Hamaus}, {Pisani}, {Choi}, {Lavaux},
  {Wandelt}  \& {Weller}}{{Hamaus} et~al.}{2020}]{Hamaus2020}
{Hamaus} N.,  {Pisani} A.,  {Choi} J.-A.,  {Lavaux} G.,  {Wandelt} B.~D.,
  {Weller} J.,  2020, arXiv e-prints, \href
  {https://ui.adsabs.harvard.edu/abs/2020arXiv200707895H} {p. arXiv:2007.07895}

\bibitem[\protect\citeauthoryear{{Hartlap}, {Simon}  \& {Schneider}}{{Hartlap}
  et~al.}{2007}]{Hartlap2007}
{Hartlap} J.,  {Simon} P.,   {Schneider} P.,  2007, \mn@doi [\aap]
  {10.1051/0004-6361:20066170}, \href
  {https://ui.adsabs.harvard.edu/abs/2007A&A...464..399H} {464, 399}

\bibitem[\protect\citeauthoryear{{Hawken}, {Aubert}, {Pisani}, {Cousinou},
  {Escoffier}, {Nadathur}, {Rossi}  \& {Schneider}}{{Hawken}
  et~al.}{2020}]{Hawken2020}
{Hawken} A.~J.,  {Aubert} M.,  {Pisani} A.,  {Cousinou} M.-C.,  {Escoffier} S.,
   {Nadathur} S.,  {Rossi} G.,   {Schneider} D.~P.,  2020, \mn@doi [\jcap]
  {10.1088/1475-7516/2020/06/012}, \href
  {https://ui.adsabs.harvard.edu/abs/2020JCAP...06..012H} {2020, 012}

\bibitem[\protect\citeauthoryear{{Higuchi}, {Oguri}  \& {Hamana}}{{Higuchi}
  et~al.}{2013}]{Higuchi2013}
{Higuchi} Y.,  {Oguri} M.,   {Hamana} T.,  2013, \mn@doi [\mnras]
  {10.1093/mnras/stt521}, \href
  {https://ui.adsabs.harvard.edu/abs/2013MNRAS.432.1021H} {432, 1021}

\bibitem[\protect\citeauthoryear{{Hou} et~al.,}{{Hou} et~al.}{2018}]{Hou2018}
{Hou} J.,  et~al., 2018, \mn@doi [\mnras] {10.1093/mnras/sty1984}, \href
  {https://ui.adsabs.harvard.edu/abs/2018MNRAS.480.2521H} {480, 2521}

\bibitem[\protect\citeauthoryear{{Jackson}}{{Jackson}}{1972}]{Jackson1972}
{Jackson} J.~C.,  1972, \mn@doi [\mnras] {10.1093/mnras/156.1.1P}, \href
  {https://ui.adsabs.harvard.edu/abs/1972MNRAS.156P...1J} {156, 1P}

\bibitem[\protect\citeauthoryear{{Jamieson} \& {Loverde}}{{Jamieson} \&
  {Loverde}}{2020}]{Jamieson2020}
{Jamieson} D.,  {Loverde} M.,  2020, arXiv e-prints, \href
  {https://ui.adsabs.harvard.edu/abs/2020arXiv201007235J} {p. arXiv:2010.07235}

\bibitem[\protect\citeauthoryear{{Jenkins}}{{Jenkins}}{2010}]{Jenkins2010}
{Jenkins} A.,  2010, \mn@doi [\mnras] {10.1111/j.1365-2966.2010.16259.x}, \href
  {https://ui.adsabs.harvard.edu/abs/2010MNRAS.403.1859J} {403, 1859}

\bibitem[\protect\citeauthoryear{{Jennings}, {Baugh}  \& {Pascoli}}{{Jennings}
  et~al.}{2011}]{Jennings2011}
{Jennings} E.,  {Baugh} C.~M.,   {Pascoli} S.,  2011, \mn@doi [\mnras]
  {10.1111/j.1365-2966.2010.17581.x}, \href
  {https://ui.adsabs.harvard.edu/abs/2011MNRAS.410.2081J} {410, 2081}

\bibitem[\protect\citeauthoryear{Joyce, Jain, Khoury  \& Trodden}{Joyce
  et~al.}{2015}]{Joyce2014}
Joyce A.,  Jain B.,  Khoury J.,   Trodden M.,  2015, \mn@doi [Phys. Rept.]
  {10.1016/j.physrep.2014.12.002}, 568, 1

\bibitem[\protect\citeauthoryear{{Juszkiewicz}, {Springel}  \&
  {Durrer}}{{Juszkiewicz} et~al.}{1999}]{Juszkiewicz1999}
{Juszkiewicz} R.,  {Springel} V.,   {Durrer} R.,  1999, \mn@doi [\apjl]
  {10.1086/312055}, \href
  {https://ui.adsabs.harvard.edu/abs/1999ApJ...518L..25J} {518, L25}

\bibitem[\protect\citeauthoryear{{Kaiser}}{{Kaiser}}{1987}]{Kaiser1987}
{Kaiser} N.,  1987, \mn@doi [\mnras] {10.1093/mnras/227.1.1}, \href
  {https://ui.adsabs.harvard.edu/abs/1987MNRAS.227....1K} {227, 1}

\bibitem[\protect\citeauthoryear{{Klypin}, {Prada}, {Betancort-Rijo}  \&
  {Albareti}}{{Klypin} et~al.}{2018}]{Klypin2018}
{Klypin} A.,  {Prada} F.,  {Betancort-Rijo} J.,   {Albareti} F.~D.,  2018,
  \mn@doi [\mnras] {10.1093/mnras/sty2613}, \href
  {https://ui.adsabs.harvard.edu/abs/2018MNRAS.481.4588K} {481, 4588}

\bibitem[\protect\citeauthoryear{{Koyama}}{{Koyama}}{2016}]{Koyama2016}
{Koyama} K.,  2016, \mn@doi [Rep. Prog. Phys.] {10.1088/0034-4885/79/4/046902},
  \href {http://adsabs.harvard.edu/abs/2016RPPh...79d6902K} {79, 046902}

\bibitem[\protect\citeauthoryear{{Krause}, {Chang}, {Dor{\'e}}  \&
  {Umetsu}}{{Krause} et~al.}{2013}]{Krause2013}
{Krause} E.,  {Chang} T.-C.,  {Dor{\'e}} O.,   {Umetsu} K.,  2013, \mn@doi
  [\apjl] {10.1088/2041-8205/762/2/L20}, \href
  {https://ui.adsabs.harvard.edu/abs/2013ApJ...762L..20K} {762, L20}

\bibitem[\protect\citeauthoryear{{Kravtsov}, {Berlind}, {Wechsler}, {Klypin},
  {Gottl{\"o}ber}, {Allgood}  \& {Primack}}{{Kravtsov}
  et~al.}{2004}]{Kravtsov2004}
{Kravtsov} A.~V.,  {Berlind} A.~A.,  {Wechsler} R.~H.,  {Klypin} A.~A.,
  {Gottl{\"o}ber} S.,  {Allgood} B.,   {Primack} J.~R.,  2004, \mn@doi [\apj]
  {10.1086/420959}, \href {http://adsabs.harvard.edu/abs/2004ApJ...609...35K}
  {609, 35}

\bibitem[\protect\citeauthoryear{{Kuruvilla} \& {Porciani}}{{Kuruvilla} \&
  {Porciani}}{2018}]{Kuruvilla2018}
{Kuruvilla} J.,  {Porciani} C.,  2018, \mn@doi [\mnras]
  {10.1093/mnras/sty1654}, \href
  {https://ui.adsabs.harvard.edu/abs/2018MNRAS.479.2256K} {479, 2256}

\bibitem[\protect\citeauthoryear{{Kuruvilla} \& {Porciani}}{{Kuruvilla} \&
  {Porciani}}{2020}]{Kuruvilla2020}
{Kuruvilla} J.,  {Porciani} C.,  2020, \mn@doi [\jcap]
  {10.1088/1475-7516/2020/07/043}, \href
  {https://ui.adsabs.harvard.edu/abs/2020JCAP...07..043K} {2020, 043}

\bibitem[\protect\citeauthoryear{Laureijs, Amiaux, Arduini, Auguères,
  Brinchmann, Cole  et~al.}{Laureijs et~al.}{2011}]{euclid}
Laureijs R.,  Amiaux J.,  Arduini S.,  Auguères J.~L.,  Brinchmann J.,  Cole
  R.,   et~al., 2011, preprint (\mn@eprint {arXiv} {1110.3193})

\bibitem[\protect\citeauthoryear{{Levi} et~al.,}{{Levi}
  et~al.}{2019}]{Levi2019}
{Levi} M.,  et~al., 2019, in Bulletin of the American Astronomical Society.
  p.~57 (\mn@eprint {arXiv} {1907.10688})

\bibitem[\protect\citeauthoryear{{Lilje} \& {Efstathiou}}{{Lilje} \&
  {Efstathiou}}{1988}]{Lilje1988}
{Lilje} P.~B.,  {Efstathiou} G.,  1988, \mn@doi [\mnras]
  {10.1093/mnras/231.3.635}, \href
  {https://ui.adsabs.harvard.edu/abs/1988MNRAS.231..635L} {231, 635}

\bibitem[\protect\citeauthoryear{{Lilje} \& {Efstathiou}}{{Lilje} \&
  {Efstathiou}}{1989}]{Lilje1989}
{Lilje} P.~B.,  {Efstathiou} G.,  1989, \mn@doi [\mnras]
  {10.1093/mnras/236.4.851}, \href
  {https://ui.adsabs.harvard.edu/abs/1989MNRAS.236..851L} {236, 851}

\bibitem[\protect\citeauthoryear{{Lilje} \& {Lahav}}{{Lilje} \&
  {Lahav}}{1991}]{Lilje1991}
{Lilje} P.~B.,  {Lahav} O.,  1991, \mn@doi [\apj] {10.1086/170094}, \href
  {https://ui.adsabs.harvard.edu/abs/1991ApJ...374...29L} {374, 29}

\bibitem[\protect\citeauthoryear{{Linder} \& {Cahn}}{{Linder} \&
  {Cahn}}{2007}]{Linder2007}
{Linder} E.~V.,  {Cahn} R.~N.,  2007, \mn@doi [Astroparticle Physics]
  {10.1016/j.astropartphys.2007.09.003}, \href
  {https://ui.adsabs.harvard.edu/abs/2007APh....28..481L} {28, 481}

\bibitem[\protect\citeauthoryear{Linder \& Polarski}{Linder \&
  Polarski}{2019}]{Linder2019}
Linder E.~V.,  Polarski D.,  2019, \mn@doi [Physical Review D]
  {10.1103/physrevd.99.023503}, 99

\bibitem[\protect\citeauthoryear{{Lippich} et~al.,}{{Lippich}
  et~al.}{2019}]{Lippich2019}
{Lippich} M.,  et~al., 2019, \mn@doi [\mnras] {10.1093/mnras/sty2757}, \href
  {https://ui.adsabs.harvard.edu/abs/2019MNRAS.482.1786L} {482, 1786}

\bibitem[\protect\citeauthoryear{{Mandal} \& {Nadkarni-Ghosh}}{{Mandal} \&
  {Nadkarni-Ghosh}}{2020}]{Mandal2020}
{Mandal} A.,  {Nadkarni-Ghosh} S.,  2020, \mn@doi [\mnras]
  {10.1093/mnras/staa2073}, \href
  {https://ui.adsabs.harvard.edu/abs/2020MNRAS.498..355M} {498, 355}

\bibitem[\protect\citeauthoryear{{Manera} et~al.,}{{Manera}
  et~al.}{2013}]{Manera2013}
{Manera} M.,  et~al., 2013, \mn@doi [\mnras] {10.1093/mnras/sts084}, \href
  {http://adsabs.harvard.edu/abs/2013MNRAS.428.1036M} {428, 1036}

\bibitem[\protect\citeauthoryear{{Massara} \& {Sheth}}{{Massara} \&
  {Sheth}}{2018}]{Massara2018}
{Massara} E.,  {Sheth} R.~K.,  2018, arXiv e-prints, \href
  {https://ui.adsabs.harvard.edu/abs/2018arXiv181103132M} {p. arXiv:1811.03132}

\bibitem[\protect\citeauthoryear{{Melchior}, {Sutter}, {Sheldon}, {Krause}  \&
  {Wandelt}}{{Melchior} et~al.}{2014}]{Melchior2014}
{Melchior} P.,  {Sutter} P.~M.,  {Sheldon} E.~S.,  {Krause} E.,   {Wandelt}
  B.~D.,  2014, \mn@doi [\mnras] {10.1093/mnras/stu456}, \href
  {https://ui.adsabs.harvard.edu/abs/2014MNRAS.440.2922M} {440, 2922}

\bibitem[\protect\citeauthoryear{{Mohammad}, {de la Torre}, {Bianchi}, {Guzzo}
  \& {Peacock}}{{Mohammad} et~al.}{2016}]{Mohammad2016}
{Mohammad} F.~G.,  {de la Torre} S.,  {Bianchi} D.,  {Guzzo} L.,   {Peacock}
  J.~A.,  2016, \mn@doi [\mnras] {10.1093/mnras/stw411}, \href
  {https://ui.adsabs.harvard.edu/abs/2016MNRAS.458.1948M} {458, 1948}

\bibitem[\protect\citeauthoryear{{Nadathur} \& {Percival}}{{Nadathur} \&
  {Percival}}{2019}]{Nadathur2019a}
{Nadathur} S.,  {Percival} W.~J.,  2019, \mn@doi [\mnras]
  {10.1093/mnras/sty3372}, \href
  {https://ui.adsabs.harvard.edu/abs/2019MNRAS.483.3472N} {483, 3472}

\bibitem[\protect\citeauthoryear{{Nadathur}, {Carter}, {Percival}, {Winther}
  \& {Bautista}}{{Nadathur} et~al.}{2019}]{Nadathur2019b}
{Nadathur} S.,  {Carter} P.~M.,  {Percival} W.~J.,  {Winther} H.~A.,
  {Bautista} J.~E.,  2019, \mn@doi [\prd] {10.1103/PhysRevD.100.023504}, \href
  {https://ui.adsabs.harvard.edu/abs/2019PhRvD.100b3504N} {100, 023504}

\bibitem[\protect\citeauthoryear{{Nadathur} et~al.,}{{Nadathur}
  et~al.}{2020}]{Nadathur2020}
{Nadathur} S.,  et~al., 2020, \mn@doi [\mnras] {10.1093/mnras/staa3074}, \href
  {https://ui.adsabs.harvard.edu/abs/2020MNRAS.499.4140N} {499, 4140}

\bibitem[\protect\citeauthoryear{{Neyrinck}}{{Neyrinck}}{2011}]{Neyrinck2011b}
{Neyrinck} M.~C.,  2011, \mn@doi [\apj] {10.1088/0004-637X/742/2/91}, \href
  {https://ui.adsabs.harvard.edu/abs/2011ApJ...742...91N} {742, 91}

\bibitem[\protect\citeauthoryear{{Neyrinck}, {Szapudi}  \& {Szalay}}{{Neyrinck}
  et~al.}{2009}]{Neyrinck2009}
{Neyrinck} M.~C.,  {Szapudi} I.,   {Szalay} A.~S.,  2009, \mn@doi [\apjl]
  {10.1088/0004-637X/698/2/L90}, \href
  {https://ui.adsabs.harvard.edu/abs/2009ApJ...698L..90N} {698, L90}

\bibitem[\protect\citeauthoryear{{Neyrinck}, {Szapudi}  \& {Szalay}}{{Neyrinck}
  et~al.}{2011}]{Neyrinck2011a}
{Neyrinck} M.~C.,  {Szapudi} I.,   {Szalay} A.~S.,  2011, \mn@doi [\apj]
  {10.1088/0004-637X/731/2/116}, \href
  {https://ui.adsabs.harvard.edu/abs/2011ApJ...731..116N} {731, 116}

\bibitem[\protect\citeauthoryear{{Neyrinck}, {Szapudi}, {McCullagh}, {Szalay},
  {Falck}  \& {Wang}}{{Neyrinck} et~al.}{2018}]{Neyrinck2018}
{Neyrinck} M.~C.,  {Szapudi} I.,  {McCullagh} N.,  {Szalay} A.~S.,  {Falck} B.,
    {Wang} J.,  2018, \mn@doi [\mnras] {10.1093/mnras/sty1074}, \href
  {https://ui.adsabs.harvard.edu/abs/2018MNRAS.478.2495N} {478, 2495}

\bibitem[\protect\citeauthoryear{{Okumura} et~al.,}{{Okumura}
  et~al.}{2016}]{Okumura2016}
{Okumura} T.,  et~al., 2016, \mn@doi [\pasj] {10.1093/pasj/psw029}, \href
  {https://ui.adsabs.harvard.edu/abs/2016PASJ...68...38O} {68, 38}

\bibitem[\protect\citeauthoryear{{Padilla}, {Ceccarelli}  \&
  {Lambas}}{{Padilla} et~al.}{2005}]{Padilla2005}
{Padilla} N.~D.,  {Ceccarelli} L.,   {Lambas} D.~G.,  2005, \mn@doi [\mnras]
  {10.1111/j.1365-2966.2005.09500.x}, \href
  {http://adsabs.harvard.edu/abs/2005MNRAS.363..977P} {363, 977}

\bibitem[\protect\citeauthoryear{{Paillas}, {Lagos}, {Padilla}, {Tissera},
  {Helly}  \& {Schaller}}{{Paillas} et~al.}{2017}]{Paillas2017}
{Paillas} E.,  {Lagos} C.~D.~P.,  {Padilla} N.,  {Tissera} P.,  {Helly} J.,
  {Schaller} M.,  2017, \mn@doi [\mnras] {10.1093/mnras/stx1514}, \href
  {http://adsabs.harvard.edu/abs/2017MNRAS.470.4434P} {470, 4434}

\bibitem[\protect\citeauthoryear{{Paz}, {Lares}, {Ceccarelli}, {Padilla}  \&
  {Lambas}}{{Paz} et~al.}{2013}]{Paz2013}
{Paz} D.,  {Lares} M.,  {Ceccarelli} L.,  {Padilla} N.,   {Lambas} D.~G.,
  2013, \mn@doi [\mnras] {10.1093/mnras/stt1836}, \href
  {https://ui.adsabs.harvard.edu/abs/2013MNRAS.436.3480P} {436, 3480}

\bibitem[\protect\citeauthoryear{{Peacock} \& {Smith}}{{Peacock} \&
  {Smith}}{2000}]{Peacock2000}
{Peacock} J.~A.,  {Smith} R.~E.,  2000, \mn@doi [\mnras]
  {10.1046/j.1365-8711.2000.03779.x}, \href
  {http://adsabs.harvard.edu/abs/2000MNRAS.318.1144P} {318, 1144}

\bibitem[\protect\citeauthoryear{{Peebles}}{{Peebles}}{1980}]{Peebles1980}
{Peebles} P.~J.~E.,  1980, {The large-scale structure of the universe}

\bibitem[\protect\citeauthoryear{{Percival} et~al.,}{{Percival}
  et~al.}{2014}]{Percival2014}
{Percival} W.~J.,  et~al., 2014, \mn@doi [\mnras] {10.1093/mnras/stu112}, \href
  {https://ui.adsabs.harvard.edu/abs/2014MNRAS.439.2531P} {439, 2531}

\bibitem[\protect\citeauthoryear{{Pezzotta} et~al.,}{{Pezzotta}
  et~al.}{2017}]{Pezzotta2017}
{Pezzotta} A.,  et~al., 2017, \mn@doi [\aap] {10.1051/0004-6361/201630295},
  \href {https://ui.adsabs.harvard.edu/abs/2017A&A...604A..33P} {604, A33}

\bibitem[\protect\citeauthoryear{{Planck Collaboration} et~al.,}{{Planck
  Collaboration} et~al.}{2016}]{Planck2016}
{Planck Collaboration} et~al., 2016, \mn@doi [\aap]
  {10.1051/0004-6361/201526926}, \href
  {http://adsabs.harvard.edu/abs/2016A%26A...594A..11P} {594, A11}

\bibitem[\protect\citeauthoryear{{Pollina}, {Hamaus}, {Dolag}, {Weller},
  {Baldi}  \& {Moscardini}}{{Pollina} et~al.}{2017}]{Pollina2017}
{Pollina} G.,  {Hamaus} N.,  {Dolag} K.,  {Weller} J.,  {Baldi} M.,
  {Moscardini} L.,  2017, \mn@doi [\mnras] {10.1093/mnras/stx785}, \href
  {https://ui.adsabs.harvard.edu/abs/2017MNRAS.469..787P} {469, 787}

\bibitem[\protect\citeauthoryear{{Raghunathan}, {Nadathur}, {Sherwin}  \&
  {Whitehorn}}{{Raghunathan} et~al.}{2020}]{Raghunathan2020}
{Raghunathan} S.,  {Nadathur} S.,  {Sherwin} B.~D.,   {Whitehorn} N.,  2020,
  \mn@doi [\apj] {10.3847/1538-4357/ab6f05}, \href
  {https://ui.adsabs.harvard.edu/abs/2020ApJ...890..168R} {890, 168}

\bibitem[\protect\citeauthoryear{{Reid} \& {White}}{{Reid} \&
  {White}}{2011}]{Reid2011}
{Reid} B.~A.,  {White} M.,  2011, \mn@doi [\mnras]
  {10.1111/j.1365-2966.2011.19379.x}, \href
  {https://ui.adsabs.harvard.edu/abs/2011MNRAS.417.1913R} {417, 1913}

\bibitem[\protect\citeauthoryear{{Repp} \& {Szapudi}}{{Repp} \&
  {Szapudi}}{2018}]{Repp2018}
{Repp} A.,  {Szapudi} I.,  2018, \mn@doi [\mnras] {10.1093/mnras/stx2615},
  \href {https://ui.adsabs.harvard.edu/abs/2018MNRAS.473.3598R} {473, 3598}

\bibitem[\protect\citeauthoryear{{Repp} \& {Szapudi}}{{Repp} \&
  {Szapudi}}{2020}]{Repp2020}
{Repp} A.,  {Szapudi} I.,  2020, \mn@doi [\mnras] {10.1093/mnrasl/slaa139},
  \href {https://ui.adsabs.harvard.edu/abs/2020MNRAS.498L.125R} {498, L125}

\bibitem[\protect\citeauthoryear{{Rykoff} et~al.,}{{Rykoff}
  et~al.}{2014}]{Rykoff2014}
{Rykoff} E.~S.,  et~al., 2014, \mn@doi [\apj] {10.1088/0004-637X/785/2/104},
  \href {https://ui.adsabs.harvard.edu/abs/2014ApJ...785..104R} {785, 104}

\bibitem[\protect\citeauthoryear{{S{\'a}nchez}}{{S{\'a}nchez}}{2020}]{Sanchez2020}
{S{\'a}nchez} A.~G.,  2020, \mn@doi [\prd] {10.1103/PhysRevD.102.123511}, \href
  {https://ui.adsabs.harvard.edu/abs/2020PhRvD.102l3511S} {102, 123511}

\bibitem[\protect\citeauthoryear{{S{\'a}nchez} et~al.,}{{S{\'a}nchez}
  et~al.}{2013}]{Sanchez2013}
{S{\'a}nchez} A.~G.,  et~al., 2013, \mn@doi [\mnras] {10.1093/mnras/stt799},
  \href {https://ui.adsabs.harvard.edu/abs/2013MNRAS.433.1202S} {433, 1202}

\bibitem[\protect\citeauthoryear{{S{\'a}nchez} et~al.,}{{S{\'a}nchez}
  et~al.}{2017a}]{Sanchez2016}
{S{\'a}nchez} A.~G.,  et~al., 2017a, \mn@doi [\mnras] {10.1093/mnras/stw2443},
  \href {https://ui.adsabs.harvard.edu/abs/2017MNRAS.464.1640S} {464, 1640}

\bibitem[\protect\citeauthoryear{{S{\'a}nchez} et~al.,}{{S{\'a}nchez}
  et~al.}{2017b}]{Sanchez2017}
{S{\'a}nchez} C.,  et~al., 2017b, \mn@doi [\mnras] {10.1093/mnras/stw2745},
  \href {http://adsabs.harvard.edu/abs/2017MNRAS.465..746S} {465, 746}

\bibitem[\protect\citeauthoryear{{Saslaw} \& {Hamilton}}{{Saslaw} \&
  {Hamilton}}{1984}]{Saslaw1984}
{Saslaw} W.~C.,  {Hamilton} A.~J.~S.,  1984, \mn@doi [\apj] {10.1086/161589},
  \href {https://ui.adsabs.harvard.edu/abs/1984ApJ...276...13S} {276, 13}

\bibitem[\protect\citeauthoryear{{Scoccimarro}}{{Scoccimarro}}{2004}]{Scoccimarro2004}
{Scoccimarro} R.,  2004, \mn@doi [\prd] {10.1103/PhysRevD.70.083007}, \href
  {https://ui.adsabs.harvard.edu/abs/2004PhRvD..70h3007S} {70, 083007}

\bibitem[\protect\citeauthoryear{{Sefusatti}, {Crocce}, {Pueblas}  \&
  {Scoccimarro}}{{Sefusatti} et~al.}{2006}]{Sefusatti2006}
{Sefusatti} E.,  {Crocce} M.,  {Pueblas} S.,   {Scoccimarro} R.,  2006, \mn@doi
  [\prd] {10.1103/PhysRevD.74.023522}, \href
  {https://ui.adsabs.harvard.edu/abs/2006PhRvD..74b3522S} {74, 023522}

\bibitem[\protect\citeauthoryear{{Seldner} \& {Peebles}}{{Seldner} \&
  {Peebles}}{1977}]{Seldner1977}
{Seldner} M.,  {Peebles} P.~J.~E.,  1977, \mn@doi [\apj] {10.1086/155404},
  \href {https://ui.adsabs.harvard.edu/abs/1977ApJ...215..703S} {215, 703}

\bibitem[\protect\citeauthoryear{{Seljak}}{{Seljak}}{2000}]{Seljak2000}
{Seljak} U.,  2000, \mn@doi [\mnras] {10.1046/j.1365-8711.2000.03715.x}, \href
  {https://ui.adsabs.harvard.edu/abs/2000MNRAS.318..203S} {318, 203}

\bibitem[\protect\citeauthoryear{{Seljak} \& {McDonald}}{{Seljak} \&
  {McDonald}}{2011}]{Seljak2011}
{Seljak} U.,  {McDonald} P.,  2011, \mn@doi [\jcap]
  {10.1088/1475-7516/2011/11/039}, \href
  {https://ui.adsabs.harvard.edu/abs/2011JCAP...11..039S} {2011, 039}

\bibitem[\protect\citeauthoryear{{Sellentin} \& {Heavens}}{{Sellentin} \&
  {Heavens}}{2016}]{Sellentin2016}
{Sellentin} E.,  {Heavens} A.~F.,  2016, \mn@doi [\mnras]
  {10.1093/mnrasl/slv190}, \href
  {https://ui.adsabs.harvard.edu/abs/2016MNRAS.456L.132S} {456, L132}

\bibitem[\protect\citeauthoryear{{Sheth}, {Hui}, {Diaferio}  \&
  {Scoccimarro}}{{Sheth} et~al.}{2001a}]{Sheth2001a}
{Sheth} R.~K.,  {Hui} L.,  {Diaferio} A.,   {Scoccimarro} R.,  2001a, \mn@doi
  [\mnras] {10.1046/j.1365-8711.2001.04222.x}, \href
  {https://ui.adsabs.harvard.edu/abs/2001MNRAS.325.1288S} {325, 1288}

\bibitem[\protect\citeauthoryear{{Sheth}, {Diaferio}, {Hui}  \&
  {Scoccimarro}}{{Sheth} et~al.}{2001b}]{Sheth2001b}
{Sheth} R.~K.,  {Diaferio} A.,  {Hui} L.,   {Scoccimarro} R.,  2001b, \mn@doi
  [\mnras] {10.1046/j.1365-8711.2001.04457.x}, \href
  {https://ui.adsabs.harvard.edu/abs/2001MNRAS.326..463S} {326, 463}

\bibitem[\protect\citeauthoryear{{Shi} \& {Sheth}}{{Shi} \&
  {Sheth}}{2018}]{Shi2018}
{Shi} J.,  {Sheth} R.~K.,  2018, \mn@doi [\mnras] {10.1093/mnras/stx2277},
  \href {https://ui.adsabs.harvard.edu/abs/2018MNRAS.473.2486S} {473, 2486}

\bibitem[\protect\citeauthoryear{{Shirasaki}, {Huff}, {Markovic}  \&
  {Rhodes}}{{Shirasaki} et~al.}{2021}]{Shirasaki2021}
{Shirasaki} M.,  {Huff} E.~M.,  {Markovic} K.,   {Rhodes} J.~D.,  2021, \mn@doi
  [\apj] {10.3847/1538-4357/abcc68}, \href
  {https://ui.adsabs.harvard.edu/abs/2021ApJ...907...38S} {907, 38}

\bibitem[\protect\citeauthoryear{{Slepian} \& {Eisenstein}}{{Slepian} \&
  {Eisenstein}}{2015}]{Slepian2015}
{Slepian} Z.,  {Eisenstein} D.~J.,  2015, \mn@doi [\mnras]
  {10.1093/mnras/stu2627}, \href
  {https://ui.adsabs.harvard.edu/abs/2015MNRAS.448....9S} {448, 9}

\bibitem[\protect\citeauthoryear{{Slepian} \& {Eisenstein}}{{Slepian} \&
  {Eisenstein}}{2017}]{Slepian2017}
{Slepian} Z.,  {Eisenstein} D.~J.,  2017, \mn@doi [\mnras]
  {10.1093/mnras/stx490}, \href
  {https://ui.adsabs.harvard.edu/abs/2017MNRAS.469.2059S} {469, 2059}

\bibitem[\protect\citeauthoryear{{Springel}}{{Springel}}{2005}]{Springel2005}
{Springel} V.,  2005, \mn@doi [\mnras] {10.1111/j.1365-2966.2005.09655.x},
  \href {https://ui.adsabs.harvard.edu/abs/2005MNRAS.364.1105S} {364, 1105}

\bibitem[\protect\citeauthoryear{{Springel}, {White}, {Tormen}  \&
  {Kauffmann}}{{Springel} et~al.}{2001}]{Springel2001}
{Springel} V.,  {White} S. D.~M.,  {Tormen} G.,   {Kauffmann} G.,  2001,
  \mn@doi [\mnras] {10.1046/j.1365-8711.2001.04912.x}, \href
  {https://ui.adsabs.harvard.edu/abs/2001MNRAS.328..726S} {328, 726}

\bibitem[\protect\citeauthoryear{{Sutter}, {Lavaux}, {Wandelt}  \&
  {Weinberg}}{{Sutter} et~al.}{2012}]{Sutter2012a}
{Sutter} P.~M.,  {Lavaux} G.,  {Wandelt} B.~D.,   {Weinberg} D.~H.,  2012,
  \mn@doi [\apj] {10.1088/0004-637X/761/2/187}, \href
  {https://ui.adsabs.harvard.edu/abs/2012ApJ...761..187S} {761, 187}

\bibitem[\protect\citeauthoryear{{Szapudi} \& {Pan}}{{Szapudi} \&
  {Pan}}{2004}]{Szapudi2004}
{Szapudi} I.,  {Pan} J.,  2004, \mn@doi [\apj] {10.1086/380920}, \href
  {https://ui.adsabs.harvard.edu/abs/2004ApJ...602...26S} {602, 26}

\bibitem[\protect\citeauthoryear{{Tinker}}{{Tinker}}{2007}]{Tinker2007}
{Tinker} J.~L.,  2007, \mn@doi [\mnras] {10.1111/j.1365-2966.2006.11157.x},
  \href {https://ui.adsabs.harvard.edu/abs/2007MNRAS.374..477T} {374, 477}

\bibitem[\protect\citeauthoryear{{Uhlemann}, {Codis}, {Pichon}, {Bernardeau}
  \& {Reimberg}}{{Uhlemann} et~al.}{2016}]{Uhlemann2016}
{Uhlemann} C.,  {Codis} S.,  {Pichon} C.,  {Bernardeau} F.,   {Reimberg} P.,
  2016, \mn@doi [\mnras] {10.1093/mnras/stw1074}, \href
  {https://ui.adsabs.harvard.edu/abs/2016MNRAS.460.1529U} {460, 1529}

\bibitem[\protect\citeauthoryear{{Uhlemann}, {Friedrich},
  {Villaescusa-Navarro}, {Banerjee}  \& {Codis}}{{Uhlemann}
  et~al.}{2020}]{Uhlemann2020}
{Uhlemann} C.,  {Friedrich} O.,  {Villaescusa-Navarro} F.,  {Banerjee} A.,
  {Codis} S.~r.,  2020, \mn@doi [\mnras] {10.1093/mnras/staa1155}, \href
  {https://ui.adsabs.harvard.edu/abs/2020MNRAS.495.4006U} {495, 4006}

\bibitem[\protect\citeauthoryear{{Vlah} \& {White}}{{Vlah} \&
  {White}}{2019}]{Vlah2019}
{Vlah} Z.,  {White} M.,  2019, \mn@doi [\jcap] {10.1088/1475-7516/2019/03/007},
  \href {https://ui.adsabs.harvard.edu/abs/2019JCAP...03..007V} {2019, 007}

\bibitem[\protect\citeauthoryear{{Vlah}, {Castorina}  \& {White}}{{Vlah}
  et~al.}{2016}]{Vlah2016}
{Vlah} Z.,  {Castorina} E.,   {White} M.,  2016, \mn@doi [\jcap]
  {10.1088/1475-7516/2016/12/007}, \href
  {https://ui.adsabs.harvard.edu/abs/2016JCAP...12..007V} {2016, 007}

\bibitem[\protect\citeauthoryear{{Wagner}, {Schmidt}, {Chiang}  \&
  {Komatsu}}{{Wagner} et~al.}{2015}]{Wagner2015}
{Wagner} C.,  {Schmidt} F.,  {Chiang} C.~T.,   {Komatsu} E.,  2015, \mn@doi
  [\mnras] {10.1093/mnrasl/slu187}, \href
  {https://ui.adsabs.harvard.edu/abs/2015MNRAS.448L..11W} {448, L11}

\bibitem[\protect\citeauthoryear{{Wang}, {Neyrinck}, {Szapudi}, {Szalay},
  {Chen}, {Lesgourgues}, {Riotto}  \& {Sloth}}{{Wang} et~al.}{2011}]{Wang2011}
{Wang} X.,  {Neyrinck} M.,  {Szapudi} I.,  {Szalay} A.,  {Chen} X.,
  {Lesgourgues} J.,  {Riotto} A.,   {Sloth} M.,  2011, \mn@doi [\apj]
  {10.1088/0004-637X/735/1/32}, \href
  {https://ui.adsabs.harvard.edu/abs/2011ApJ...735...32W} {735, 32}

\bibitem[\protect\citeauthoryear{{White}}{{White}}{1979}]{White1979}
{White} S.~D.~M.,  1979, \mn@doi [\mnras] {10.1093/mnras/186.2.145}, \href
  {https://ui.adsabs.harvard.edu/abs/1979MNRAS.186..145W} {186, 145}

\bibitem[\protect\citeauthoryear{{Yang}, {Mo}, {van den Bosch}, {Weinmann},
  {Li}  \& {Jing}}{{Yang} et~al.}{2005}]{Yang2005}
{Yang} X.,  {Mo} H.~J.,  {van den Bosch} F.~C.,  {Weinmann} S.~M.,  {Li} C.,
  {Jing} Y.~P.,  2005, \mn@doi [\mnras] {10.1111/j.1365-2966.2005.09351.x},
  \href {https://ui.adsabs.harvard.edu/abs/2005MNRAS.362..711Y} {362, 711}

\bibitem[\protect\citeauthoryear{{Zentner}, {Berlind}, {Bullock}, {Kravtsov}
  \& {Wechsler}}{{Zentner} et~al.}{2005}]{Zentner2005}
{Zentner} A.~R.,  {Berlind} A.~A.,  {Bullock} J.~S.,  {Kravtsov} A.~V.,
  {Wechsler} R.~H.,  2005, \mn@doi [\apj] {10.1086/428898}, \href
  {https://ui.adsabs.harvard.edu/abs/2005ApJ...624..505Z} {624, 505}

\bibitem[\protect\citeauthoryear{{Zhai} et~al.,}{{Zhai}
  et~al.}{2019}]{Zhai2019}
{Zhai} Z.,  et~al., 2019, \mn@doi [\apj] {10.3847/1538-4357/ab0d7b}, \href
  {https://ui.adsabs.harvard.edu/abs/2019ApJ...874...95Z} {874, 95}

\bibitem[\protect\citeauthoryear{{Zhao} et~al.,}{{Zhao}
  et~al.}{2020}]{Zhao2020}
{Zhao} C.,  et~al., 2020, \mn@doi [\mnras] {10.1093/mnras/stz3339}, \href
  {https://ui.adsabs.harvard.edu/abs/2020MNRAS.491.4554Z} {491, 4554}

\bibitem[\protect\citeauthoryear{{Zheng}, {Coil}  \& {Zehavi}}{{Zheng}
  et~al.}{2007}]{Zheng2007}
{Zheng} Z.,  {Coil} A.~L.,   {Zehavi} I.,  2007, \mn@doi [\apj]
  {10.1086/521074}, \href {http://adsabs.harvard.edu/abs/2007ApJ...667..760Z}
  {667, 760}

\bibitem[\protect\citeauthoryear{{Zorrilla Matilla}, {Sharma}, {Hsu}  \&
  {Haiman}}{{Zorrilla Matilla} et~al.}{2020}]{Zorrilla2020}
{Zorrilla Matilla} J.~M.,  {Sharma} M.,  {Hsu} D.,   {Haiman} Z.,  2020, arXiv
  e-prints, \href {https://ui.adsabs.harvard.edu/abs/2020arXiv200706529Z} {p.
  arXiv:2007.06529}

\bibitem[\protect\citeauthoryear{Zu \& Weinberg}{Zu \& Weinberg}{2013}]{Zu2013}
Zu Y.,  Weinberg D.~H.,  2013, \mn@doi [Monthly Notices of the Royal
  Astronomical Society] {10.1093/mnras/stt411}, 431, 3319–3337

\bibitem[\protect\citeauthoryear{{eBOSS Collaboration} et~al.,}{{eBOSS
  Collaboration} et~al.}{2020}]{eboss2020}
{eBOSS Collaboration} et~al., 2020, arXiv e-prints, \href
  {https://ui.adsabs.harvard.edu/abs/2020arXiv200708991E} {p. arXiv:2007.08991}

\bibitem[\protect\citeauthoryear{van~de Weygaert \& Platen}{van~de Weygaert \&
  Platen}{2011}]{vandeWeygaert2009}
van~de Weygaert R.,  Platen E.,  2011, \mn@doi [Int. J. Mod. Phys.]
  {10.1142/S2010194511000092}, 1, 41

\makeatother
\end{thebibliography}

\bsp	
\label{lastpage}
\end{document}